\RequirePackage{fixltx2e}
\documentclass[11pt,a4paper]{article}
\pdfoutput=1
\usepackage[utf8]{inputenc}
\usepackage{jheppub}
\makeatletter
\gdef\@fpheader{}
\makeatother
\hypersetup{unicode=true,bookmarksopen=true,breaklinks}

\usepackage{mathtools}
\usepackage{lmodern}
\usepackage[T1]{fontenc}
\usepackage{bbm}
\DeclareMathAlphabet{\mathbfi}{OML}{cmm}{b}{it}
\usepackage{textgreek}
\usepackage[scaled=0.8]{beramono}

\let\originalleft\left
\let\originalright\right
\renewcommand{\left}{\mathopen{}\mathclose\bgroup\originalleft}
\renewcommand{\right}{\aftergroup\egroup\originalright}

\makeatletter
\usepackage{alphalph}

\newenvironment{equations}[1][]{\subequations\ifx\relax#1\relax\else\label{#1}\fi\align\ignorespaces}{\endalign\ignorespacesafterend\endsubequations}
\def\@spliteq#1{\begin{equation}\begin{split}#1\end{split}\end{equation}}
\def\splitequation{\collect@body\@spliteq}

\makeatother

\renewcommand{\vec}[1]{{\ifnum9<1#1\mathbf{#1}\else\ifcat\noexpand#1\relax\boldsymbol{#1}\else\mathbfi{#1}\fi\fi}}

\newcommand{\mathi}{\mathrm{i}}
\let\oldre\Re
\let\oldim\Im
\renewcommand{\Re}{\oldre\mathfrak{e}\,}
\renewcommand{\Im}{\oldim\mathfrak{m}\,}
\newcommand{\total}{\mathop{}\!\mathrm{d}}

\newcommand{\abs}[1]{{\left\lvert{#1}\right\rvert}}

\newcommand{\unitmatrix}{\mathbbm{1}}
\newcommand{\tr}{\operatorname{tr}}
\newcommand{\eqend}[1]{\,#1}

\newcommand{\brst}{\mathop{}\!\mathsf{s}\hskip 0.05em\relax}

\allowdisplaybreaks

\usepackage{ytableau}
\ytableausetup{centertableaux,boxsize=1.3em}
\newcommand\ytableausmall[1]{{\,\ytableausetup{boxsize=0.7em}\ytableaushort[\scriptstyle]{#1}\ytableausetup{boxsize=1.3em}\,}{}}

\def\repdim#1{\repdimscan#1\relax}
\def\repdimscan#1{\ifx\relax#1\else\repdimel{#1}\expandafter\repdimscan\fi}
\def\repdimel#1{\ifnum9<1#1\mathbf{#1}\else\mathbfi{#1}\fi}

\usepackage[os=win]{menukeys}
\renewmenumacro{\keys}[+]{shadowedroundedkeys}

\newcommand{\xact}{\textsc{xAct}}
\newcommand{\fix}{\textsc{FieldsX}}

\usepackage{tabularx}
\usepackage{booktabs}
\usepackage{xcolor}

\usepackage{titletoc}
\newcommand{\ssubsection}[1]{\subsubsection{\texttt{#1}}}
\titlecontents*{subsubsection}
  [3.84em]
  {\small}
  {\thecontentspage~}
  {}
  {}
  []
  [\hspace{0.6em plus 0.2em minus 0.2em}]
  []

\newsavebox{\tabularbox}
\def\usagetable{\setbox\tabularbox=\hbox\bgroup\begin{tabular*}{\textwidth}{llp{\dimexpr \textwidth - 2cm}l}\toprule}
\def\endusagetable{\bottomrule\end{tabular*}\egroup\begingroup\fboxsep=0pt\colorbox{gray!15}{\box\tabularbox}\endgroup\vspace{6pt}}
\makeatletter
\def\usageline{\@ifstar\@usageline\@@usageline}
\def\@usageline#1#2{& \multicolumn{2}{l}{{\ttfamily #1}} & \\ & & #2 & \\}
\def\@@usageline#1#2{& \multicolumn{2}{l}{{\ttfamily #1}} & \\ & & #2 & \\ \midrule}
\makeatother
\newcommand{\cmdref}[2]{\hyperref[#1]{\nolinkurl{#2}}}
\makeatletter
\g@addto@macro\UrlSpecials{\do\!{\newline}}
\makeatother

\usepackage{upquote}
\usepackage{fancyvrb}
\newenvironment{longcode}{\Verbatim[xleftmargin=1.5em,listparameters=\setlength{\topsep}{2pt}\setlength{\partopsep}{2pt},codes={\catcode`$=3\catcode`^=7\catcode`_=8},commandchars=\\\{\},showspaces=true]}{\endVerbatim}
\newenvironment{mathout}{\Verbatim[fontsize=\footnotesize,listparameters=\setlength{\topsep}{2pt}\setlength{\partopsep}{2pt},codes={\catcode`$=3\catcode`^=7\catcode`_=8},commandchars=\\\{\},showspaces=true]}{\endVerbatim}
\makeatletter
\newenvironment{mathouteq}{\begingroup\setlength{\abovedisplayskip}{0pt}\start@align\tw@\st@rredtrue\m@ne}{\endalign\endgroup}
\makeatother

\usepackage{fonttable}

\begin{document}

\title{FieldsX --- An extension package for the xAct tensor computer algebra suite to include fermions, gauge fields and BRST cohomology}

\author{Markus B. Fr{\"o}b}
\affiliation{Institut f{\"u}r Theoretische Physik, Universit{\"a}t Leipzig, Br{\"u}derstra{\ss}e 16, 04103 Leipzig, Germany}

\emailAdd{mfroeb@itp.uni-leipzig.de}

\abstract{I present the tensor computer algebra package \fix, which extends the \xact\ suite of tensor algebra packages to perform computations in field theory with fermions and gauge fields. This includes the standard tools of curved-space $\gamma$ matrices, Fierz identities, invariant tensors on Lie algebras, arbitrary gradings and left and right variational derivatives, as well as the decomposition of spinor products into irreducible components following the approach of d'Auria, Fr{\'e}, Maina and Regge~[\href{https://doi.org/10.1016/0003-4916(82)90007-0}{\emph{Annals~Phys.} \textbf{139} (1982) 93}]. Lastly, it also includes functions to work with nilpotent differentials such as the BV--BRST differential for (supersymmetric) gauge theories and to compute their (relative) cohomologies, from which anomalies and gauge-invariant operators can be determined. I illustrate the use of the package with the example of $\mathcal{N}=1$ Super-Yang--Mills theory.\afterAbstractSpace
\noindent\textsc{Program summary:}

\begin{tabular}[t]{ll}
Program title: & \fix \\
Version: & 1.1.2 \\
Programming language: & \href{https://www.wolfram.com/mathematica/}{\textsc{Mathematica}} (version 8.0 or later) \\
Obtainable from: & Included with the arXiv submission, or GitHub \\
License: & \href{https://www.gnu.org/licenses/old-licenses/gpl-2.0}{GNU GPL 2.0} or \href{https://www.gnu.org/licenses/gpl}{later}
\end{tabular}}

\maketitle

\section{Introduction}
\label{sec_intro}

The free \xact\ suite of packages~\cite{xact1,xact2,xact3,xact4,xact5,xact6,xact7,xact8} for tensor algebra is a powerful tool to perform computations in general relativity and field theory which has found widespread use; the \href{http://xact.es/articles.html}{webpage} on papers that use \xact\ (which is certainly incomplete) lists about 630 papers to date. However, while functionality to work with the Newman-Penrose two-component spinor calculus in 3+1 dimensions is available~\cite{xact5}, working with $\gamma$ matrices, non-commutative spinors or Fierz identities requires additional and quite non-trivial effort from part of the user. Such functionality is available in the competitor computer algebra system Cadabra~\cite{cadabra1,cadabra2}, which however lacks other functionality present in \textsc{Mathematica} that is invaluable to deal with expressions containing tens of thousands of terms. \fix\ tries to fill this gap and make the \xact\ suite also useful for computations with fermions and gauge theories. For an efficient generation of suitable ans{\"a}tze for the computation of BRST cohomology, \fix\ relies on the \href{https://library.wolfram.com/infocenter/MathSource/8115/}{\textsc{Multisets}} package which is \textcopyright\ 2011 David Bevan and distributed under the Wolfram Library Archive License. \fix\ is fully integrated with other \xact\ packages, for example the \textsc{xPert} package to compute perturbations (including perturbations of curved-space $\gamma$ matrices and spin connections) or the \textsc{TexAct} package for \TeX\ output.

This paper is organised as follows: section~\ref{sec_inst} describes how to install and load the package, section~\ref{sec_sym} illustrates the main parts of \fix\ by verifying the supersymmetry and conformal symmetry of (classical) $\mathcal{N}=1$ Super-Yang--Mills theory and by computing its possible anomalies, and section~\ref{sec_doc} contains a complete list of all functions and variables (with short descriptions) contained in \fix. It is assumed that the reader is already familiar with the basic functionality of \xact; commands that the user has to execute in \textsc{Mathematica} are given in typewriter face (\texttt{Command[]}), while output from \textsc{Mathematica} is given in smaller size.

\section{Installation}
\label{sec_inst}

\fix\ requires a working installation of \xact, at least version 1.1.4 released on 16 February 2020. \fix\ is then installed by downloading the package file \texttt{FieldsX.m}, which is included as ancillary file with the arXiv submission of this paper and published on GitHub at~\href{https://github.com/mfroeb/FieldsX}{\nolinkurl{github.com/mfroeb/FieldsX}}, and placing it into the \xact\ installation directory, which for a per-user installation of \xact\ is obtained from \texttt{FileNameJoin[{\$UserBaseDirectory, "Applications/xAct"}]}, or for a system-wide installation from \texttt{FileNameJoin[{\$BaseDirec\-tory, "Applications/xAct"}]}. To install the required \textsc{Multisets} package, the package file \texttt{Multisets.m} must be downloaded from \href{https://library.wolfram.com/infocenter/MathSource/8115/}{\nolinkurl{library.wolfram.com/infocenter/MathSource/8115/}} and placed into the application directory \texttt{FileNameJoin[{\$UserBaseDirectory, "Appli\-cations"}]} (for a per-user install) or \texttt{FileNameJoin[{\$BaseDirectory, "Applications"}]} (for a system-wide install).\footnote{If the cohomology functionality of \fix\ is not needed, installation of the \textsc{Multisets} package is not necessary since it is loaded dynamically with \texttt{DeclarePackage}.}

The package can then be loaded in the standard way with \texttt{Get["xAct`FieldsX`"]} or \texttt{Needs["xAct`FieldsX`"]}, and will itself load the required packages from the \xact\ suite. Short information for any function is displayed in the same manner as for all Mathematica functions by typing \texttt{?FunctionName}.

\subsection{Compatibility with other packages}
\label{sec_inst_compat}

The \xact\ package \textsc{Spinors} to work with two-component spinors in four dimensions also defines the commands \cmdref{sec_doc_spingamma_defspinstructure}{DefSpinStructure}, \cmdref{sec_doc_spingamma_undefspinstructure}{UndefSpinStructure}, \cmdref{sec_doc_spinors_defspinor}{DefSpinor} and \cmdref{sec_doc_spinors_undefspinor}{UndefSpinor}.\footnote{I thank Thomas Bäckdahl for pointing this out to me.} If both packages are to be loaded at the same time, one could use the fully qualified names \texttt{xAct`Spinors`DefSpinStructure}, \texttt{xAct`FieldsX`DefSpinStructure} and so on to refer to the respective commands. Alternatively, if \fix\ is loaded \emph{after} \textsc{Spinors}, the commands \texttt{DefSpinStructure}, \texttt{UndefSpinStructure}, \texttt{DefSpinor} and \texttt{UndefSpinor} from the \textsc{Spinors} package are redefined by \fix\ to \texttt{Def2CompSpinStructure}, \texttt{Undef2CompSpin\-Structure}, \texttt{Def2CompSpinor} and \texttt{Undef2CompSpinor} if the variable \texttt{\$Keep2CompDefs} is not set or \texttt{False}, while the commands \texttt{DefSpinStructure}, \texttt{UndefSpinStructure}, \texttt{DefSpinor} and \texttt{UndefSpinor} from the \fix\ package are redefined to \texttt{DefGenSpinStructure}, \texttt{UndefGen\-SpinStructure}, \texttt{DefGenSpinor} and \texttt{UndefGenSpinor} if the variable \texttt{\$Keep2CompDefs} is \texttt{True}. Note that \texttt{\$Keep2CompDefs} must be set \emph{before} loading \fix.

\section{\texorpdfstring{$\mathcal{N}=1$}{N=1} Super-Yang--Mills theory}
\label{sec_sym}

To show how this package can be used, let us consider $\mathcal{N} = 1$ Super-Yang--Mills theory~\cite{ferrarazumino1974,salamstrathdee1974,dewitfreedman1975}, which in addition to the Yang--Mills vector boson contains a Majorana spinor $\chi^a$ and an auxiliary field $D^a$ in the adjoint representation. The action reads
\begin{equation}
S = - \frac{1}{4} \int F_{\mu\nu}^a F^{\mu\nu a} \total x - \frac{1}{2} \int \bar{\chi}^a \gamma^\mu \left( D_\mu \chi \right)^a \total x + \frac{1}{2} \int D^a D_a \total x \eqend{,}
\end{equation}
where $\total x \equiv \sqrt{-g} \total^4 x$, with the field strength
\begin{equation}
F_{\mu\nu}^a \equiv \nabla_\mu A_\nu^a - \nabla_\nu A_\mu^a + \mathi g f_{abc} A_\mu^b A_\nu^c
\end{equation}
and the gauge-covariant derivative
\begin{equation}
\left( D_\mu \chi \right)^a \equiv \nabla_\mu \chi^a + \mathi g f_{abc} A_\mu^b \chi^c \eqend{,}
\end{equation}
and is invariant under the gauge transformation
\begin{equations}
\delta^\text{gauge}_\xi A_\mu^a &= \left( D_\mu \xi \right)^a \eqend{,} & \delta^\text{gauge}_\xi D^a &= - \mathi g f_{abc} \xi^b D^c \eqend{,} \\
\delta^\text{gauge}_\xi \chi^a &= - \mathi g f_{abc} \xi^b \chi^c \eqend{,} & \delta^\text{gauge}_\xi \bar{\chi}^a &= - \mathi g f_{abc} \xi^b \bar{\chi}^c \eqend{,}
\end{equations}
and the supersymmetry transformation
\begin{equations}
\delta^\text{susy}_\epsilon A_\mu^a &= - \bar{\epsilon} \gamma_\mu \chi^a \eqend{,} & \delta^\text{susy}_\epsilon D^a &= \mathi \bar{\epsilon} \gamma_* \gamma^\mu \left( D_\mu \chi \right)^a \eqend{,} \\
\delta^\text{susy}_\epsilon \chi^a &= \frac{1}{2} \gamma^{\mu\nu} F_{\mu\nu}^a \epsilon + \mathi D^a \gamma_* \epsilon \eqend{,} & \delta^\text{susy}_\epsilon \bar{\chi}^a &= - \frac{1}{2} \bar{\epsilon} \gamma^{\mu\nu} F_{\mu\nu}^a + \mathi D^a \bar{\epsilon} \gamma_* \eqend{,}
\end{equations}
where $\epsilon$ is a constant Grassmann-odd Majorana spinor. To check invariance under the supersymmetry transformation, one needs to use Fierz rearrangement identities~\cite{fierz1937,freedmanvanproeyen}\footnote{Attributed to Pauli by Fierz himself: ``The content of this first section originates from Prof. W. Pauli and I am indebted to him for ceding me his calculations.''~\cite{fierz1937}, footnote 2.} and the Bianchi identities for the field strength tensor. Two supersymmetry transformations close into a gauge transformation and a translation,
\begin{equations}[sec_sym_susyalgebra]
\left[ \delta^\text{susy}_2, \delta^\text{susy}_1 \right] A_\mu^a &= - b^\nu F_{\mu\nu}^a = b^\nu \nabla_\nu A_\mu^a + \delta^\text{gauge}_\xi A_\mu^a \eqend{,} \\
\left[ \delta^\text{susy}_2, \delta^\text{susy}_1 \right] \chi^a &= b^\nu \nabla_\nu \chi^a + \delta^\text{gauge}_\xi \chi^a \eqend{,} \\
\left[ \delta^\text{susy}_2, \delta^\text{susy}_1 \right] D^a &= b^\nu \nabla_\nu D^a + \delta^\text{gauge}_\xi D^a \eqend{,}
\end{equations}
(Fierz rearrangement identities are needed again), with the translation parameter $b^\nu$ and the field-dependent gauge transformation parameter $\xi^a$ defined by
\begin{equation}
\label{sec_sym_susy_bxi_def}
b^\nu \equiv 2 \bar{\epsilon}_1 \gamma^\nu \epsilon_2 \eqend{,} \qquad \xi^a \equiv - b^\nu A_\nu^a \eqend{.}
\end{equation}

\subsection{Gauge and supersymmetry invariance}

To check (or obtain) these results using \fix, we first need to load the package (which automatically loads all required \xact\ packages). For demonstration purposes, we also set the \TeX\ output symbol for the non-commutative product to a dot; all output shown later on was copied using \texttt{CopyToClipboard@TexPrint[expr]} and not modified further.
\begin{longcode}
<< xAct`FieldsX`
\$CenterDotTexSymbol = "$\cdot$";
\end{longcode}
\begin{mathout}
------------------------------------------------------------
Package xAct`xPerm`  version 1.2.3, \{2015,8,23\}
CopyRight (C) 2003-2018, Jose M. Martin-Garcia, under the General Public License.

... (many lines of output not displayed)
\end{mathout}
We then define the manifold $M$, the metric $gg$ and covariant derivative $\nabla$, the spin structure ($\gamma$ matrices) and the inner bundle (abstract Lie algebra in the adjoint representation of dimension \textit{liedim}, with metric \textit{cartankilling}). The option \texttt{SymCovDQ} from the \textsc{xTras} package makes it possible to use symmetrised covariant derivatives $\nabla_{\mu\cdots\nu} \equiv \nabla_{(\mu} \cdots \nabla_{\nu)}$, which is extremely useful to obtain canonical forms of expressions, required for the computation of cohomologies.
\begin{longcode}
DefManifold[M, 4, \{$\alpha$, $\beta$, $\gamma$, $\delta$, $\mu$, $\nu$, $\rho$, $\sigma$\}];
DefMetric[-1, gg[-$\mu$, -$\nu$], CD, \{";", "$\nabla$"\}, PrintAs $\to$ "g", SymCovDQ $\to$ True];
DefSpinStructure[gg, \{A, B, F, G, H, J, K, L, P, Q\}];
DefConstantSymbol[liedim];
DefVBundleWithMetric[lie, M, liedim, \{a, b, c, d, e, f, i, j, k, l, m, n, p, q\},
    cartankilling];
\end{longcode}
Next we define the fields of the theory and the coupling $g$. Note that the spinors $\chi$ and $\epsilon$ have a spin-bundle index $-A$, which must always be the last index.
\begin{longcode}
DefEvenTensor[AA[-$\mu$, a], M, PrintAs $\to$ "A"];
DefOddSpinor[chi[a, -A], M, SpinorType $\to$ Majorana, PrintAs $\to$ "$\chi$"];
DefEvenTensor[auxD[a], M, PrintAs $\to$ "D"];
DefEvenTensor[xi[a], M, PrintAs $\to$ "$\xi$"];
DefOddSpinor[eps[-A], M, SpinorType $\to$ Majorana, PrintAs $\to$ "$\epsilon$"];
DefConstantSymbol[g];
\end{longcode}
The vector boson $A_\mu^a$, the auxiliary field $D^a$ and the gauge parameter $\xi^a$ are Grassmann even, while the spinor $\chi^a_A$ and supersymmetry parameter $\epsilon_A$ are Grassmann odd. Together with $\chi$ and $\epsilon$, the charge conjugate spinors $\bar{\chi}$ and $\bar{\epsilon}$ are automatically defined, and read \texttt{barchi[a, A]} and \texttt{bareps[A]}. We also have to ensure that the supersymmetry parameter is covariantly constant; a space-time dependent supersymmetry parameter leads to supergravity theories.
\begin{longcode}
CD[\_\_]@eps[\_] \textasciicircum:= 0;
CD[\_\_]@bareps[\_] \textasciicircum:= 0;
\end{longcode}

The field strength and gauge-covariant derivative are given in terms of the vector boson, but we first need to obtain the structure constants of the Lie algebra as the antisymmetric invariant tensor of rank 3 (and define a pretty output):
\begin{longcode}
lief = InvariantTraceTensor[lie, 3, Antisymmetric];
Tex[lief] = "f";
\end{longcode}
\begin{mathout}
** DefTensor: Defining tensor Invlief[a,b,c].
\end{mathout}
We then define the field strength and functions to convert it into derivatives of the vector boson and back. For the transformation back, we replace derivatives $\nabla_\mu A_\nu^a$ by the sum of its symmetric part and the antisymmetric part, and rewrite the antisymmetric part in terms of $F_{\mu\nu}^a$:
\begin{longcode}
DefEvenTensor[FF[-$\mu$, -$\nu$, a], M, Antisymmetric[\{-$\mu$, -$\nu$\}], PrintAs $\to$ "F"];
FToGradA[expr\_] := expr /. \{FF[$\mu$\_, $\nu$\_, a\_] :> CD[$\mu$]@AA[$\nu$, a] - CD[$\nu$]@AA[$\mu$, a]
    + I g With[\{b = DummyIn[lie], c = DummyIn[lie]\},
    lief[a, -b, -c] AA[$\mu$, b] AA[$\nu$, c]]\}
GradAToF[expr\_] := expr /. \{CD[$\mu$\_]@AA[$\nu$\_, a\_] :> 1/2 (CD[$\mu$]@AA[$\nu$, a]
    + CD[$\nu$]@AA[$\mu$, a]) + 1/2 (FF[$\mu$, $\nu$, a] - I g With[\{b = DummyIn[lie],
    c = DummyIn[lie]\}, lief[a, -b, -c] AA[$\mu$, b] AA[$\nu$, c]])\}
\end{longcode}
\begin{mathout}
** DefTensor: Defining tensor FF[-$\mu$,-$\nu$,a].
\end{mathout}
While \xact\ provides a function to define the gauge-covariant derivative, we need to specify that it acts on the lie bundle (and the spin bundle). We then need to give the correct values to the relevant generalised Christoffel symbols, where we have to check the bundle to which the indices belong using the condition operator \texttt{/;}, and we check the results by converting the gauge-covariant derivative into the covariant derivative $\nabla$. Since the gauge-covariant derivative does not act on space-time or spin indices, the corresponding Christoffel symbols are the same as for $\nabla$.
\begin{longcode}
DefCovD[DD[-$\mu$], \{lie, SpinM\}, \{":", "D"\}];
\end{longcode}
\begin{mathout}
** DefCovD: Defining covariant derivative DD[-$\mu$].
** DefTensor: Defining vanishing torsion tensor TorsionDD[$\alpha$,-$\beta$,-$\gamma$].
** DefTensor: Defining symmetric Christoffel tensor ChristoffelDD[$\alpha$,-$\beta$,-$\gamma$].
...
** DefTensor: Defining nonsymmetric AChristoffel tensor AChristoffelDD[a,-$\beta$,-c].
\end{mathout}
\begin{longcode}
AChristoffelDD[a\_, $\mu$\_, b\_] /; (VBundleOfIndex[a] == lie) && (VBundleOfIndex[b]
    == lie) := - I g With[\{c = DummyIn[lie]\}, lief[a, b, c] AA[$\mu$, -c]]
AChristoffelDD[a\_, $\mu$\_, b\_] /; (VBundleOfIndex[a] == SpinM) &&
    (VBundleOfIndex[b] == SpinM) := AChristoffelCD[a, $\mu$, b]
ChristoffelDD[$\alpha$\_, $\mu$\_, $\nu$\_] := ChristoffelCD[$\alpha$, $\mu$, $\nu$]
ChangeCovD[ChangeCovD[DD[-$\mu$]@AA[-$\nu$, a], DD, PD], PD, CD]
ChangeCovD[ChangeCovD[DD[-$\mu$]@chi[a, -A], DD, PD], PD, CD]
\end{longcode}
\begin{mathouteq}
& -i g A_{\mu c} A_{\nu }{}^{b} f^{a}{}_{b}{}^{c} + \nabla_{\mu }A_{\nu }{}^{a} \\
& -i g A_{\mu c} \chi^{b}{}_{A} f^{a}{}_{b}{}^{c} + \nabla_{\mu }\chi^{a}{}_{A} &
\end{mathouteq}

We can now write down the action (or rather the Lagrangian), where the non-commutative product of spinors is written $\cdot$ and can be entered with \keys{Esc + . + Esc}. The $\gamma$ matrices can be entered with \texttt{Gammagg}, which automatically selects the right one (\texttt{Gammagg1}, \dots) depending on the number of indices. We also define a pretty output for them:
\begin{longcode}
Tex[Gammagg1] \textasciicircum= "{\textbackslash\textbackslash}gamma"; Tex[Gammagg2] \textasciicircum= "{\textbackslash\textbackslash}gamma";
Tex[Gammagg3] \textasciicircum= "{\textbackslash\textbackslash}gamma"; Tex[Gammagg4] \textasciicircum= "{\textbackslash\textbackslash}gamma";
action = -1/4 FF[$\mu$, $\nu$, a] FF[-$\mu$, -$\nu$, -a] + 1/2 auxD[a] auxD[-a]
    - 1/2 barchi[a, A] $\cdot$ Gammagg[$\mu$, -A, B] $\cdot$ DD[-$\mu$]@chi[-a, -B]
\end{longcode}
\vspace*{2mm}
\begin{mathouteq}
&\tfrac{1}{2} D_{a} D^{a} -  \tfrac{1}{4} F_{\mu \nu a} F^{\mu \nu a} -  \tfrac{1}{2} \bar{\chi}^{aA}\cdot D_{\mu }\chi_{aB} \, \gamma^{\mu }{}_{A}{}^{B} &
\end{mathouteq}
We now expand the action, writing everything in terms of the basic fields and covariant derivative:
\begin{longcode}
action2 = CollectTensors@ChangeCovD[ChangeCovD[FToGradA[action], DD, PD], PD, CD]
\end{longcode}
\begin{mathouteq}
&\tfrac{1}{2} D_{a} D^{a} -  \tfrac{1}{2} \bar{\chi}^{aA}\cdot \nabla_{\mu }\chi_{aB} \gamma^{\mu }{}_{A}{}^{B} + \tfrac{1}{2}i g A^{\mu a} \bar{\chi}_{b}{}^{A}\cdot \chi_{cB} \gamma_{\mu A}{}^{B} f_{a}{}^{bc} + \tfrac{1}{4} g^2 A_{\mu }{}^{b} A^{\mu a} A_{\nu }{}^{d} A^{\nu c} f_{ac}{}^{e} f_{bde} \\
&+ i g A^{\mu a} A^{\nu b} f_{abc} \nabla_{\nu }A_{\mu }{}^{c} + \tfrac{1}{2} \nabla_{\mu }A_{\nu a} \nabla^{\nu }A^{\mu a} -  \tfrac{1}{2} \nabla_{\nu }A_{\mu a} \nabla^{\nu }A^{\mu a} &
\end{mathouteq}
Expanding the action, we obtain a term with the product of two structure constants $f_{ade} f^a{}_{bc}$. Since the structure constants satisfy the multiterm Jacobi identity $f_{ab[c} f^a{}_{de]} = 0$, it is more difficult to bring such terms into a canonical form. One way to deal with multiterm identities are Young projectors (as used in the \textsc{xTras} package to deal with the Bianchi identities of the Riemann tensor), where one has to project $f_{abc} f^a{}_{de}$ on the tableau \ytableausmall{bd,ce} --- antisymmetric in each pair $bc$ and $de$, and symmetric under the interchange of the two pairs. This results in
\begin{longcode}
YoungProject[lief[-a, -b, -c] lief[a, -d, -e], \{\{-b, -d\}, \{-c, -e\}\}]
\end{longcode}
\vspace*{2mm}
\begin{mathouteq}
&- \tfrac{1}{3} f_{be}{}^{a} f_{cda} + \tfrac{1}{3} f_{bd}{}^{a} f_{cea} + \tfrac{2}{3} f_{bc}{}^{a} f_{dea} \eqend{,}&
\end{mathouteq}
and one checks that the Jacobi identity is automatically fulfilled. The disadvantage of this approach is that one has to find the required Young tableau by hand, and \fix\ thus takes a different approach, inspired by the approach to Fierz identities of d'Auria, Fr{\'e}, Maina and Regge~\cite{dauriafremainaregge1982} based on the decomposition of the product of group representations into irreducible components. For the invariant tensors of the Lie algebra, the corresponding group is the permutation group, but it is not necessary to perform a full decomposition into irreducibles. Instead, it is enough to use the formula for the commutator of two generators $\mathfrak{t}_a$ of the Lie algebra, multiply by other generators and take the trace to obtain
\begin{equation*}
\mathi f_{ab}{}^c \, \tr\left[ \mathfrak{t}_c \mathfrak{t}_d \cdots \mathfrak{t}_e \right] = \tr\left[ \left[ \mathfrak{t}_a, \mathfrak{t}_b \right] \mathfrak{t}_d \cdots \mathfrak{t}_e \right] = \tr\left[ \mathfrak{t}_a \mathfrak{t}_b \mathfrak{t}_d \cdots \mathfrak{t}_e \right] - \tr\left[ \mathfrak{t}_b \mathfrak{t}_a \mathfrak{t}_d \cdots \mathfrak{t}_e \right] \eqend{,}
\end{equation*}
expressing the product of a structure constant with an invariant tensor of the Lie algebra by the sum of two invariant tensors of higher rank. This gives
\begin{longcode}
ReduceInvariantTraceTensors[lief[-a, -b, -c] lief[a, -d, -e]]
\end{longcode}
\begin{mathout}
** DefTensor: Defining tensor Invlie4[a,b,c,d].
\end{mathout}
\begin{mathouteq}
&i (tr[lie]_{bcde} -  tr[lie]_{bced} -  tr[lie]_{bdec} + tr[lie]_{bedc}) &
\end{mathouteq}
and the action
\begin{longcode}
action3 = CollectTensors@ReduceInvariantTraceTensors@Expand[action2]
\end{longcode}
\begin{mathouteq}
&\tfrac{1}{2} D_{a} D^{a} - \tfrac{1}{2} \bar{\chi}^{aA}\cdot \nabla_{\mu }\chi_{aB} \gamma^{\mu }{}_{A}{}^{B} -  \tfrac{1}{2}i g^2 A_{\mu }{}^{b} A^{\mu a} A_{\nu }{}^{d} A^{\nu c} tr[lie]_{abcd} + \tfrac{1}{2}i g^2 A_{\mu }{}^{b} A^{\mu a} A_{\nu }{}^{d} A^{\nu c} tr[lie]_{acbd} \\
&+ \tfrac{1}{2}i g A^{\mu a} \bar{\chi}_{b}{}^{A}\cdot \chi_{cB} \gamma_{\mu A}{}^{B} f_{a}{}^{bc} + i g A^{\mu a} A^{\nu b} f_{abc} \nabla_{\nu }A_{\mu }{}^{c} + \tfrac{1}{2} \nabla_{\mu }A_{\nu a} \nabla^{\nu }A^{\mu a} -  \tfrac{1}{2} \nabla_{\nu }A_{\mu a} \nabla^{\nu }A^{\mu a} &
\end{mathouteq}

Now gauge and supersymmetry invariance can be checked. We define the gauge transformations
\begin{longcode}
gauge[xi\_][AA[$\mu$\_, a\_]] := DD[$\mu$]@xi[a]
gauge[xi\_][FF[$\mu$\_, $\nu$\_, a\_]] := - I g With[\{b = DummyIn[lie], c = DummyIn[lie]\},
    lief[a, -b, -c] xi[b] FF[$\mu$, $\nu$, c]]
gauge[xi\_][chi[a\_, B\_]] := - I g With[\{b = DummyIn[lie], c = DummyIn[lie]\},
    lief[a, -b, -c] xi[b] chi[c, B]]
gauge[xi\_][barchi[a\_, B\_]] := - I g With[\{b = DummyIn[lie], c = DummyIn[lie]\},
    lief[a, -b, -c] xi[b] barchi[c, B]]
gauge[xi\_][auxD[a\_] := - I g With[\{b = DummyIn[lie], c = DummyIn[lie]\},
    lief[a, -b, -c] xi[b] auxD[c]]
gauge[xi\_][\_] := 0
\end{longcode}
and the supersymmetry transformations
\begin{longcode}
susy[eps\_][AA[$\mu$\_, a\_]] := With[\{A = DummyIn[SpinM], B = DummyIn[SpinM]\},
    - ConjugateSpinor[eps][A] $\cdot$ Gammagg[$\mu$, -A, B] $\cdot$ chi[a, -B]]
susy[eps\_][FF[$\mu$\_, $\nu$\_, a\_]] := With[\{A = DummyIn[SpinM], B = DummyIn[SpinM]\},
    ConjugateSpinor[eps][A] $\cdot$ Gammagg[$\mu$, -A, B] $\cdot$ DD[$\nu$]@chi[a, -B]
    - ConjugateSpinor[eps][A] $\cdot$ Gammagg[$\nu$, -A, B] $\cdot$ DD[$\mu$]@chi[a, -B]]
susy[eps\_][chi[a\_, B\_]] := With[\{$\mu$ = DummyIn[TangentM], $\nu$ = DummyIn[TangentM],
    A = DummyIn[SpinM]\}, 1/2 Gammagg[$\mu$, $\nu$, B, A] FF[-$\mu$, -$\mu$, a] eps[-A]
    + I auxD[a] GammaggStar[B, A] eps[-A]]
susy[eps\_][barchi[a\_, B\_]] := With[\{$\mu$ = DummyIn[TangentM], $\nu$ =
    DummyIn[TangentM], A = DummyIn[SpinM]\}, - 1/2 ConjugateSpinor[eps][A]
    Gammagg[$\mu$, $\nu$, -A, B] FF[-$\mu$, -$\nu$, a]
    + I auxD[a] ConjugateSpinor[eps][A] GammaggStar[-A, B]]
susy[eps\_][auxD[a\_]] := With[\{$\mu$ = DummyIn[TangentM], A = DummyIn[SpinM],
    B = DummyIn[SpinM], C = DummyIn[SpinM]\}, I ConjugateSpinor[eps][A] $\cdot$
    GammaggStar[-A, B] $\cdot$ Gammagg[$\mu$, -B, C] $\cdot$ DD[-$\mu$]@chi[a, -C]]
susy[eps\_][\_] := 0
\end{longcode}
where the last definitions ensure that the gauge or supersymmetry transformation of any other tensor vanishes. Using the functionality of the \textsc{xPert} package, we then consider a generic first-order perturbation, and set the perturbation of the metric to zero and the perturbation of any other tensor or spinor to its gauge or supersymmetry transformation:
\begin{longcode}
gaugetrafo[expr\_, xi\_] := Expand[ExpandPerturbation[Perturbation[expr, 1]]
    /. \{Perturbationgg[\_\_] :> 0, Perturbation[term\_] :> gauge[xi][term]\}]
susytrafo[expr\_, eps\_] := Expand[ExpandPerturbation[Perturbation[expr, 1]]
    /. \{Perturbationgg[\_\_] :> 0, Perturbation[term\_] :> susy[eps][term]\}]
\end{longcode}
Let us check that the transformations of the field strength are the correct ones:
\begin{longcode}
CollectTensors@ReduceInvariantTraceTensors@GradAToF@SymmetrizeCovDs[
    ChangeCovD[ChangeCovD[gaugetrafo[FToGradA[FF[-$\mu$, -$\nu$, a]], xi]
    - gaugetrafo[FF[-$\mu$, -$\nu$, a], xi], DD, PD], PD, CD]]
CollectTensors@GradAToF@SymmetrizeCovDs[ChangeCovD[ChangeCovD[
    susytrafo[FToGradA[FF[-$\mu$, -$\nu$, a]], eps]
    - susytrafo[FF[-$\mu$, -$\nu$, a], eps], DD, PD], PD, CD]]
\end{longcode}
\begin{mathouteq}
& 0 \\
& 0 \eqend{.}&
\end{mathouteq}
For the action, a gauge transformation gives
\begin{longcode}
gtf = gaugetrafo[action3, xi]
\end{longcode}
\begin{mathouteq}
&- \tfrac{1}{2}i g D_{a} D^{c} f^{a}{}_{bc} \xi^{b} + \tfrac{1}{2}i g \bar{\chi}^{cA}\cdot \nabla_{\mu }\chi_{aB} \gamma^{\mu }{}_{A}{}^{B} f^{a}{}_{bc} \xi^{b} -  \tfrac{1}{2}i g D^{a} D^{d} cartankilling_{ba} f^{b}{}_{cd} \xi^{c} - \ldots&
\end{mathouteq}
To show that this vanishes, we have to bring each term into canonical form: change the gauge-covariant derivatives into $\nabla$, symmetrise covariant derivatives, contract all metrics, reduce the invariant tensors of the Lie algebra and collect all terms. We then obtain
\begin{longcode}
mycanon[expr\_] := CollectTensors@ReduceInvariantTraceTensors@ContractMetric@
    SymmetrizeCovDs@Expand[ChangeCovD[ChangeCovD[expr, DD, PD], PD, CD]]
mycanon[gtf]
\end{longcode}
\begin{mathout}
** DefTensor: Defining tensor Invlie5[a,b,c,d,e].
\end{mathout}
\begin{mathouteq}
&0 \eqend{,}&
\end{mathouteq}
which shows that the action (in this case, even the Lagrangian) is gauge invariant. To obtain this result, \cmdref{sec_doc_inner_reduceinvarianttracetensors}{ReduceInvariantTraceTensors} expanded a product of a rank-4 invariant tensor with structure constants coming from the gauge transformation into rank-5 invariant tensors, which then cancelled.

For supersymmetry invariance, we have to work a bit harder. First perform again a perturbation of the action, insert the supersymmetry transformations, and use the same simplifications:
\begin{longcode}
stf = susytrafo[action3, eps]
\end{longcode}
\begin{mathouteq}
&\tfrac{1}{4} \bar{\epsilon}^{F}\cdot \nabla_{\mu }\chi_{aB} F_{\alpha \beta }{}^{a} \gamma^{\mu }{}_{A}{}^{B} \gamma^{\alpha \beta }{}_{F}{}^{A} + \tfrac{1}{2}i D_{a} \bar{\epsilon}^{A}\cdot D_{\alpha }\chi^{a}{}_{F} \gamma^{\alpha }{}_{B}{}^{F} \gamma^{*}{}[g]_{A}{}^{B} \\
&+ \tfrac{1}{2}i D^{a} cartankilling_{ba} \bar{\epsilon}^{A}\cdot D_{\alpha }\chi^{b}{}_{F} \gamma^{\alpha }{}_{B}{}^{F} \gamma^{*}{}[g]_{A}{}^{B} - \tfrac{1}{2}i D^{a} \bar{\epsilon}^{F}\cdot \nabla_{\mu }\chi_{aB} \gamma^{\mu }{}_{A}{}^{B} \gamma^{*}{}[g]_{F}{}^{A} + \dots&
\end{mathouteq}
We see that there are products of (generalised) $\gamma$ matrices (the $\gamma^\mu{}_A{}^B \gamma^{\alpha\beta}{}_B{}^F$ in the first term) that can be expressed in terms of generalised $\gamma$ matrices of higher order, and that there are both $\bar{\epsilon} \chi$ and $\bar{\chi} \epsilon$ terms. The first issue can be solved using \cmdref{sec_doc_spingamma_joingammamatrices}{JoinGammaMatrices}, and the second with \cmdref{sec_doc_flipfierz_sortspinor}{SortSpinor}. This gives
\begin{longcode}
stf2 = mycanon@SortSpinor[JoinGammaMatrices[stf], eps $\to$ bareps]
\end{longcode}
\begin{mathouteq}
&\tfrac{1}{2} \bar{\epsilon}^{A}\cdot \nabla_{\nu }\chi_{aB} F_{\mu }{}^{\nu a} \gamma^{\mu }{}_{A}{}^{B} + \tfrac{1}{4} \bar{\epsilon}^{A}\cdot \nabla_{\alpha }\chi_{aB} F^{\mu \nu a} \gamma_{\mu \nu }{}^{\alpha }{}_{A}{}^{B} + \tfrac{1}{12} D^{a} \bar{\epsilon}^{A}\cdot \nabla_{\beta }\chi_{aB} \epsilon g_{\mu \nu \alpha }{}^{\beta } \gamma^{\mu \nu \alpha }{}_{A}{}^{B} \\
&+ i g^2 A_{\mu }{}^{b} A^{\mu a} A^{\nu c} \bar{\epsilon}^{A}\cdot \chi_{dB} \gamma_{\nu A}{}^{B} tr[lie]_{abc}{}^{d} + i g^2 A_{\mu }{}^{b} A^{\mu a} A^{\nu c} \bar{\epsilon}^{A}\cdot \chi_{dB} \gamma_{\nu A}{}^{B} tr[lie]_{ab}{}^{d}{}_{c} - \dots \eqend{,}&
\end{mathouteq}
and the Lagrangian is not invariant under supersymmetry. However, the result is a surface term (total derivative) such that the action is invariant:
\begin{longcode}
surface = 1/4 mycanon@CD[-$\mu$][ bareps[A] $\cdot$ ( 2 Gammagg[-$\nu$, -A, B] gg[$\mu$, -$\rho$]
    - Gammagg[-$\nu$, -$\rho$, $\mu$, -A, B] ) $\cdot$ chi[-a, -B] FF[$\rho$, $\nu$, a] - 1/3 auxD[a]
    epsilongg[$\mu$, $\nu$, $\rho$, $\sigma$] bareps[A] $\cdot$ Gammagg[-$\nu$, -$\rho$, -$\sigma$, -A, B] $\cdot$ chi[-a, -B] ];
stf3 = mycanon@FToGradA[stf2 - surface]
\end{longcode}
\begin{mathouteq}
&- \tfrac{1}{2}i g \bar{\chi}_{a}{}^{A}\cdot \chi_{bB} \bar{\epsilon}^{F}\cdot \chi_{cG} \gamma_{\mu F}{}^{G} \gamma^{\mu }{}_{A}{}^{B} f^{abc} + \tfrac{1}{2} A^{\mu a} \bar{\epsilon}^{A}\cdot \chi_{aB} \gamma^{\nu \rho \sigma }{}_{A}{}^{B} R[\nabla ]_{\mu \nu \rho \sigma } \eqend{.}&
\end{mathouteq}
The second term vanishes because $\gamma^{\nu\rho\sigma}$ is totally antisymmetric, and by the Bianchi identity $R_{\mu[\nu\rho\sigma]} = 0$, which is taken care of by the function \texttt{RiemannYoungProject} from the \textsc{xTras} package. On the other hand, the first term vanishes after using Fierz identities and symmetrising using the Majorana flip relations:
\begin{longcode}
mycanon@SpinorFlipSymmetrize@JoinGammaMatrices@Expand[RiemannYoungProject@stf3
    /. \{HoldPattern[x\_\_ $\cdot$ chi[i1\_\_] barchi[i2\_\_] $\cdot$ y\_\_] :> 1/3 x $\cdot$ chi[i1]
    barchi[i2] $\cdot$ y - 2/3 x $\cdot$ y FierzExpand[barchi[i2],chi[i1]]\}]
\end{longcode}
\begin{mathouteq}
&0 \eqend{.}&
\end{mathouteq}
Because of the automatic rules set up for the noncommutative product, in pattern matching \texttt{HoldPattern} or \texttt{Verbatim} must be used. The function \cmdref{sec_doc_flipfierz_fierzexpand}{FierzExpand} gives the basic two-spinor Fierz identity (valid in this form for both commuting and anticommuting spinors)
\begin{equation}
\label{sec_sym_fierz}
\bar{\psi}^A \chi_B = 2^{-[d/2]} \sum_{k=0}^d \frac{1}{k!} (-1)^{k (k-1)/2} \gamma_{\mu_1 \cdots \mu_k}{}_B{}^A \left( \bar{\psi}^C \gamma^{\mu_1 \cdots \mu_k}{}_C{}^D \chi_D \right) \eqend{,}
\end{equation}
which is nothing else than a completeness relation in the space of $[d/2]\times[d/2]$ matrices~\cite{freedmanvanproeyen}.

However, finding a suitable Fierz rearrangement in each case might not be easy, and a more systematic approach is to be preferred. This is the one of d'Auria, Fr{\'e}, Maina and Regge~\cite{dauriafremainaregge1982} based on the decomposition of the product of group representations into irreducible components, applied to the Lorentz group. Consider two Dirac spinors in 4 dimensions, each transforming in the $(\tfrac{1}{2},0) \oplus (0,\tfrac{1}{2})$ representation of $\textrm{SL}(2,\mathbb{C})$, the universal cover of the Lorentz group. The tensor product of two decomposes according to the well-known rules as
\begin{equation}
\left[ (\tfrac{1}{2},0) \oplus (0,\tfrac{1}{2}) \right] \otimes \left[ (\tfrac{1}{2},0) \oplus (0,\tfrac{1}{2}) \right] = (0,0) \oplus (\tfrac{1}{2},\tfrac{1}{2}) \oplus \left[ (1,0) \oplus (0,1) \right] \oplus (\tfrac{1}{2},\tfrac{1}{2}) \oplus (0,0) \eqend{,}
\end{equation}
which are two scalars, two vectors and an antisymmetric tensor. This is exactly the result obtained using the Fierz identity~\eqref{sec_sym_fierz} in four dimensions:
\begin{splitequation}
\bar{\psi}^A \psi_B &= \frac{1}{4} \delta_B{}^A \left( \bar{\psi} \psi \right) + \frac{1}{4} \gamma_\mu{}_B{}^A \left( \bar{\psi} \gamma^\mu \psi \right) - \frac{1}{8} \gamma_{\mu\nu}{}_B{}^A \left( \bar{\psi} \gamma^{\mu\nu} \psi \right) - \frac{1}{24} \gamma_{\mu\nu\rho}{}_B{}^A \left( \bar{\psi} \gamma^{\mu\nu\rho} \psi \right) \\
&\quad+ \frac{1}{96} \gamma_{\mu\nu\rho\sigma}{}_B{}^A \left( \bar{\psi} \gamma^{\mu\nu\rho\sigma} \psi \right) \\
&= \frac{1}{4} \delta_B{}^A \Psi^{(2,\repdim{1})} + \frac{1}{4} \gamma^\mu{}_B{}^A \Psi^{(2,\repdim{4}')}_\mu - \frac{1}{8} \gamma^{\mu\nu}{}_B{}^A \Psi^{(2,\repdim{6})}_{\mu\nu} - \frac{\mathi}{4} \left( \gamma_* \gamma^\mu \right)_B{}^A \Psi^{(2,\repdim{4})}_\mu - \frac{\mathi}{4} \gamma_*{}_B{}^A \Psi^{(2,\repdim{1}')} \eqend{,}
\end{splitequation}
where the scalars are $\Psi^{(2,\repdim{1})} = \bar{\psi} \psi$ and $\Psi^{(2,\repdim{1}')} = \frac{1}{24} \epsilon^{\mu\nu\rho\sigma} \bar{\psi} \gamma_{\mu\nu\rho\sigma} \psi = \mathi \bar{\psi} \gamma_* \psi$, the vectors are $\Psi^{(2,\repdim{4})}_\mu = \frac{1}{6} \epsilon_{\mu\nu\rho\sigma} \bar{\psi} \gamma^{\nu\rho\sigma} \psi = - \mathi \bar{\psi} \gamma_* \gamma_\mu \psi$ and $\Psi^{(2,\repdim{4}')}_\mu = \bar{\psi} \gamma_\mu \psi$, and the antisymmetric tensor is $\Psi^{(2,\repdim{6})}_{\mu\nu} = \bar{\psi}^C \gamma_{\mu\nu} \psi$. Here, we used the alternative convention of labeling objects transforming in irreducible representations by their dimension $\repdim{d}$ instead of the rank\footnote{In the work of d'Auria, Fr{\'e}, Maina and Regge~\cite{dauriafremainaregge1982}, yet another labeling is used, where instead of $(j,k)$ one writes $[j+k,\abs{j-k}]$.}, and it is easily checked that applying the Fierz identity~\eqref{sec_sym_fierz} again to each tensor $\Psi^{(2,\repdim{d})}$ the same tensor is obtained, such that they indeed transform irreducibly. For a Majorana spinor, Majorana flip relations show that $\Psi^{(2,\repdim{4}')}_\mu = 0 = \Psi^{(2,\repdim{6})}_{\mu\nu}$, which is consistent with dimension counting: a Majorana spinor in four dimensions has four degrees of freedom, such that the $\tfrac{4 \cdot 3}{2} = 6$ degrees of freedom in a tensor product of anticommuting spinors are decomposed into $1+1+4$ of two scalars and a vector. In the following, we only consider Majorana spinors.

For the product of three spinors, the tensors transforming in irreducible representations have a free spinor index, and any trace and their contraction on the spinor and any tensor index with a $\gamma$ matrix (spinor trace) vanishes. For example, we have the decomposition
\begin{equation}
\Psi^{(2,\repdim{4})}_\mu \psi_A = \Psi^{(3,\repdim{12})}_{\mu A} + \frac{1}{4} \gamma_{\mu A}{}^B \Psi^{(3,\repdim{4}')}_B \eqend{,}
\end{equation}
where $\gamma^\mu{}_A{}^B \Psi^{(3,\repdim{12})}_{\mu B} = 0$, analogous to the decomposition of a symmetric 2-tensor into a trace and a traceless part. Applying the Fierz identity~\eqref{sec_sym_fierz}, one actually obtains $\Psi^{(3,\repdim{12})}_{\mu B} = 0$ and $\Psi^{(3,\repdim{4}')}_A = - \frac{1}{6} \epsilon^{\mu\nu\rho\sigma} \gamma_{\mu\nu\rho\sigma A}{}^B \Psi^{(3,\repdim{4})}_B$, where
\begin{equation}
\Psi^{(3,\repdim{4})}_A = \Psi^{(2,\repdim{1})} \psi_A \eqend{,}
\end{equation}
and in addition the decomposition
\begin{equation}
\Psi^{(2,\repdim{1}')} \psi_A = - \frac{1}{24} \epsilon^{\mu\nu\rho\sigma} \gamma_{\mu\nu\rho\sigma A}{}^B \Psi^{(3,\repdim{4})}_B \eqend{.}
\end{equation}
Taking all together, one obtains
\begin{equation}
\psi_A \bar{\psi}^B \psi_C = - \frac{1}{4} \left[ \delta_{[A}^B \delta_{C]}^D + \frac{1}{6} \gamma_{\mu\nu\rho [A}{}^B \gamma^{\mu\nu\rho}{}_{C]}{}^D - \frac{1}{24} \gamma_{\mu\nu\rho\sigma [A}{}^B \gamma^{\mu\nu\rho\sigma}_{C]}{}^D \right] \Psi^{(3,\repdim{4})}_D \eqend{,}
\end{equation}
which agrees with the counting of degrees of freedom for anticommuting Majorana spinors: $4 (4-1) (4-2)/(2\cdot 3) = 4$. Lastly, for four spinors it follows that
\begin{equation}
\psi_A \bar{\psi}^B \psi_C \bar{\psi}^D = - \frac{1}{16} \left[ \delta_{[A}^B \delta_{C]}^D + \frac{1}{6} \gamma_{\mu\nu\rho [A}{}^B \gamma^{\mu\nu\rho}{}_{C]}{}^D - \frac{1}{24} \gamma_{\mu\nu\rho\sigma [A}{}^B \gamma^{\mu\nu\rho\sigma}_{C]}{}^D \right] \Psi^{(4,\repdim{1})}
\end{equation}
with
\begin{equation}
\Psi^{(4,\repdim{1})} = \left( \bar{\psi}^A \psi_A \right)^2 \eqend{,}
\end{equation}
and the product of five or more anticommuting spinors vanishes since at least two components will be equal.

For Lie-algebra valued spinors, one has to decompose representations of the product of Lorentz group and symmetric group (for permutations of the Lie algebra indices), where the representations of the symmetric group are labeled by Young tableaux; see appendix~\ref{app_decomp} for details. The decomposition of a product of spinors into tensors transforming in irreducible representations is done using \cmdref{sec_doc_irreducible_irreduciblespindecompose}{IrreducibleSpinDecompose}, and the $\gamma$ matrix orthogonality conditions and projections on the Young tableau (for Lie-algebra valued spinors) are imposed with \cmdref{sec_doc_irreducible_irreduciblespinproject}{IrreducibleSpinProject}. Invariance of the action under supersymmetry transformation then follows straightforwardly:
\begin{longcode}
IrreducibleSpinProject[mycanon@JoinGammaMatrices@mycanon@
    IrreducibleSpinDecompose[RiemannYoungProject@stf3, chi], chi]
\end{longcode}
\begin{mathouteq}
&0 \eqend{.}&
\end{mathouteq}

Lastly, we also want to check the closure~\eqref{sec_sym_susyalgebra} of the supersymmetry transformations on the fields. To reduce clutter, we now suppress informational output.
\begin{longcode}
\$DefInfoQ = False;
\$UndefInfoQ = False;
\end{longcode}
We declare two constant Grassmann-odd Majorana spinors $\epsilon_i$ and define the auxiliary vector $b^\mu$~\eqref{sec_sym_susy_bxi_def}:
\begin{longcode}
DefOddSpinor[eps1[-A], M, PrintAs $\to$ "$\epsilon1$"];
DefOddSpinor[eps2[-A], M, PrintAs $\to$ "$\epsilon2$"];
eps1 /: CD[\_]@eps1[\_] := 0
bareps1 /: CD[\_]@bareps1[\_] := 0
eps2 /: CD[\_]@eps2[\_] := 0
bareps2 /: CD[\_]@bareps2[\_] := 0
bb[mu\_] := With[{A = DummyIn[SpinM], B = DummyIn[SpinM]}, 2 bareps1[A] $\cdot$
    Gammagg[mu, -A, B] $\cdot$ eps2[-B]]
\end{longcode}
We then compute the commutator of two supersymmetry transformations and obtain for the gauge boson the required result~\eqref{sec_sym_susyalgebra}
\begin{longcode}
mycanon[SortSpinor[JoinGammaMatrices@mycanon[susytrafo[susytrafo[AA[-$\mu$, a],
    eps1], eps2] - susytrafo[susytrafo[AA[-$\mu$, a], eps2], eps1]],
    eps1 $\to$ bareps1] + bb[$\nu$] FF[-$\mu$, -$\nu$, a]]
\end{longcode}
\begin{mathouteq}
&0 \eqend{.}&
\end{mathouteq}
To bring all terms into a form where only the conjugate spinor $\bar{\epsilon}_1$ appears, we have used Majorana flip relations~\cite{freedmanvanproeyen}
\begin{equation}
\bar{\psi}^A \gamma_{\mu_1 \cdots \mu_k}{}_A{}^B \chi_B = s_k \bar{\chi}^A \gamma_{\mu_1 \cdots \mu_k}{}_A{}^B \psi_B \eqend{,}
\end{equation}
where the sign $s_k = \pm 1$ depends on the number of space-time dimensions and is given in~\cite{freedmanvanproeyen}, table 3.1 (the choices in boldface). They are implemented in \fix\ by the functions~\cmdref{sec_doc_flipfierz_flipspinor}{FlipSpinor}, \cmdref{sec_doc_flipfierz_spinorflipsymmetrize}{SpinorFlipSymmetrize} and~\cmdref{sec_doc_flipfierz_sortspinor}{SortSpinor}, with the last one used above.

For the gaugino $\chi^a$, we compute
\begin{longcode}
erg = mycanon@TimesToCenterDot@JoinGammaMatrices@mycanon[mycanon@
    JoinGammaMatrices@mycanon[susytrafo[susytrafo[chi[a, -A], eps1], eps2]
    - susytrafo[susytrafo[chi[a, -A], eps2], eps1]]]
\end{longcode}
\begin{mathouteq}
&\overline{\epsilon_{1}{}}^{B}\cdot \epsilon_{2}{}_{G}\cdot \nabla_{\beta }\chi^{a}{}_{F} \gamma^{\alpha }{}_{B}{}^{F} \gamma_{\alpha }{}^{\beta }{}_{A}{}^{G} -  \overline{\epsilon_{2}{}}^{B}\cdot \epsilon_{1}{}_{G}\cdot \nabla_{\beta }\chi^{a}{}_{F} \gamma^{\alpha }{}_{B}{}^{F} \gamma_{\alpha }{}^{\beta }{}_{A}{}^{G} + \dots \eqend{.}&
\end{mathouteq}
Here we have used \cmdref{sec_doc_grassmann_timestocenterdot}{TimesToCenterDot} to ensure that all non-commutative objects are properly multiplied with the non-commutative product. We then perform a Fierz rearrangement to ensure that the two parameters $\epsilon_i$ are contracted together, and use \cmdref{sec_doc_spingamma_gammamatricestodual}{GammaMatricesToDual} which replaces $\gamma$ matrices with three and four indices by their duals to shorten the required $\gamma$ matrix algebra.
\begin{longcode}
erg2 = CollectTensors@JoinGammaMatrices@CollectTensors@JoinGammaMatrices@
    GammaMatricesToDual[erg /. \{HoldPattern[bareps1[A\_] $\cdot$ eps2[B\_] $\cdot$ t\_] :>
    FierzExpand[bareps1[A], eps2[B]] $\cdot$ t, HoldPattern[bareps2[A\_] $\cdot$ eps1[B\_] $\cdot$ t\_]
    :> FierzExpand[bareps2[A], eps1[B]] $\cdot$ t, HoldPattern[bareps1[A\_] $\cdot$ t\_
    $\cdot$ eps2[B\_]] :> -FierzExpand[bareps1[A], eps2[B]] $\cdot$ t, HoldPattern[bareps2[A\_]
    $\cdot$ t\_ $\cdot$ eps1[B\_]] :> -FierzExpand[bareps2[A], eps1[B]] $\cdot$ t\}];
\end{longcode}
Lastly, we bring all into a form where only the conjugate spinor $\bar\epsilon_1$ appears, using again Majorana flip relations.
\begin{longcode}
erg3 = CollectTensors@JoinGammaMatrices@SortSpinor[erg2, eps1 $\to$ bareps1];
mycanon@JoinGammaMatrices[mycanon[erg3 - (bb[$\nu$] CD[-$\nu$]@chi[a, -A]
    + gauge[xi][chi[a, -A]] /. \{xi[a\_] :> With[\{$\nu$ = DummyIn@TangentM\},
    - bb[$\nu$] AA[-$\nu$, a]]\})]]
\end{longcode}
\begin{mathouteq}
&0 \eqend{,}&
\end{mathouteq}
which is the required result~\eqref{sec_sym_susyalgebra}. Lastly, we check the susy algebra on the auxiliary field $D^a$:
\begin{longcode}
erg = mycanon@JoinGammaMatrices@SortSpinor[mycanon[susytrafo[susytrafo[
    auxD[a], eps1], eps2] - susytrafo[susytrafo[auxD[a], eps2], eps1]],
    eps1 $\to$ bareps1];
\end{longcode}
We again use the decomposition of a product of spinors into tensors transforming in irreducible representations, this time for the gaugino $\chi^a$:
\begin{longcode}
erg2 = mycanon@FToGradA@JoinGammaMatrices@mycanon@
    IrreducibleSpinDecompose[erg, chi];
\end{longcode}
and obtain the required result~\eqref{sec_sym_susyalgebra}
\begin{longcode}
mycanon@RiemannYoungProject@mycanon[erg2 - (bb[$\nu$] CD[-$\nu$]@auxD[a]
    + gauge[xi][auxD[a]] /. \{xi[a\_] :> With[\{$\nu$ = DummyIn@TangentM\},
    - bb[$\nu$] AA[-$\nu$, a]]\})]
\end{longcode}
\begin{mathouteq}
&0 \eqend{.}&
\end{mathouteq}

\fix\ is also fully integrated with the \textsc{xPert} package for perturbative expansions, as we have already seen previously for computing the gauge and supersymmetry variations. In particular, one can easily obtain the stress tensor by performing a general perturbation of the action (where the variation of the $\gamma$ matrices and the spin connection follows~\cite{forgerroemer2004}), setting all perturbations but the metric one to zero. Using \cmdref{sec_doc_varder_leftvard}{LeftVarD} then performs the needed integration by parts, and it follows that
\begin{longcode}
actionvar = mycanon[2/Sqrt[-Detgg[]] (ExpandPerturbation@Perturbation[action
    Sqrt[-Detgg[]], 1] /. \{Perturbation[\_] :> 0\})];
stress = mycanon[LeftVarD[Perturbationgg[LI[1], $\mu$, $\nu$], CD]@actionvar
    /. \{delta[-LI[1], LI[1]] $\to$ 1\}]
\end{longcode}
\begin{mathouteq}
&F_{\mu }{}^{\alpha a} F_{\nu \alpha a} + \tfrac{1}{4} \bar{\chi}^{aA}\cdot \nabla_{\nu }\chi_{aB} \gamma_{\mu A}{}^{B} + \dots \eqend{,}&
\end{mathouteq}
which becomes the standard expression after expressing everything in terms of the gauge-covariant derivative:
\begin{longcode}
stress2 = CollectTensors@JoinGammaMatrices@CollectTensors@
    ChangeCovD[ChangeCovD[stress, CD, PD], PD, DD]
\end{longcode}
\begin{mathouteq}
&F_{\mu }{}^{\alpha a} F_{\nu \alpha a} + \tfrac{1}{8} \bar{\chi}^{aA}\cdot D_{\nu }\chi_{aB} \gamma_{\mu A}{}^{B} + \tfrac{1}{8} \chi^{a}{}_{B}\cdot D_{\nu }\bar{\chi}_{a}{}^{A} \gamma_{\mu A}{}^{B} + \tfrac{1}{8} \bar{\chi}^{aA}\cdot D_{\mu }\chi_{aB} \gamma_{\nu A}{}^{B} \\
&+ \tfrac{1}{8} \chi^{a}{}_{B}\cdot D_{\mu }\bar{\chi}_{a}{}^{A} \gamma_{\nu A}{}^{B} + \tfrac{1}{2} D_{a} D^{a} g_{\mu \nu } -  \tfrac{1}{4} F_{\alpha \beta a} F^{\alpha \beta a} g_{\mu \nu } -  \tfrac{1}{4} \bar{\chi}^{aA}\cdot D_{\alpha }\chi_{aB} \gamma^{\alpha }{}_{A}{}^{B} g_{\mu \nu } \\
&-  \tfrac{1}{4} \chi^{a}{}_{B}\cdot D_{\alpha }\bar{\chi}_{a}{}^{A} \gamma^{\alpha }{}_{A}{}^{B} g_{\mu \nu } \eqend{.}&
\end{mathouteq}

\subsection{BRST formalism}

A very general approach to the quantisation of gauge theories is the BRST (Becchi--Rouet--Stora--Tyutin) formalism~\cite{becchirouetstora1976,tyutin1975}, which is sufficient for theories with closed gauge algebras (such as supersymmetric theories with auxiliary fields). For theories with open gauge algebras (that only close modulo the equations of motion), the BV or field--antifield formalism~\cite{batalinvilkovisky1981,batalinvilkovisky1983,batalinvilkovisky1984} is necessary, which of course can also be treated with \fix. Because extensive reviews of the BRST(-BV) formalism exist~\cite{henneaux1990,gomisparissamuel1995,barnichetal2000}, I only list the essential steps: For each symmetry transformation $\delta^{(s)}_\xi$ with parameter $\xi$, one introduces a ghost field $c^{(s)}$ (of ghost number 1), an antighost field $\bar{c}^{(s)}$ (of ghost number -1) and an auxiliary field $B^{(s)}$ (of ghost number 0), where the ghost and antighost fields have opposite Grassmann parity to that of the parameter $\xi$, while $B^{(s)}$ has the same Grassmann parity; for global symmetries, the non-minimal fields (antighost and auxiliary field) are not needed~\cite{brandthenneauxwilch1996,brandthenneauxwilch1998}.\footnote{For reducible symmetries (which become symmetries of the ghost fields) one has to repeat the procedure, leading to ``ghosts for ghosts'' or even higher ``$n$-th order ghosts''~\cite{townsend1979,namaziestorey1980,thierrymieg1990,siegel1980,kimura1980,kimura1981}; the ghost number of the $n$-th order (anti-)ghost is $\pm n$.} One then defines the (Grassmann-odd) BRST differential $\brst$, which acts on the original fields of the theory as the sum of all symmetry transformations, with the parameters replaced by the corresponding ghosts. The action of $\brst$ on the ghosts is determined by requiring that $\brst^2 = 0$ when acting on the original fields; the closure of the gauge algebra ensures that this also entails $\brst^2 = 0$ on the ghosts. For the non-minimal fields, one sets $\brst \bar{c}^{(s)} = B^{(s)}$ and $\brst B^{(s)} = 0$, which ensures that the BRST differential is nilpotent, $\brst^2 = 0$, when acting on anything.

The relevance of the BRST(-BV) formalism lies in the fact that all quantities of interest can be obtained as (representatives of) the cohomology of the BRST differential $\brst$ in different gradings, and that Slavnov-Taylor-Ward-Takahashi identities can be straightforwardly derived~\cite{barnichetal2000,piguetsorella,dragonbrandt2012}. In particular, the Wess-Zumino consistency conditions~\cite{wesszumino1971} are nothing else but the statement of $\brst$ invariance for the integrated anomaly, and the problem of determining all consistent interactions for a given free theory is reduced to finding all elements of the cohomology at ghost number $0$~\cite{barnichhenneaux1993,barnichhenneauxtatar1994}; see~\cite{piguetsibold1982a,piguetsibold1982b} for the case of $\mathcal{N} = 1$ super-Yang--Mills theory. Gauge-fixing terms are incorporated by choosing a gauge-fixing fermion $\Psi$ of ghost number $-1$ and adding $\brst \Psi$ to the action. Since the BRST differential incorporates all the symmetries under which the action $S$ is invariant, one has $\brst S = 0$, and since $\brst^2 = 0$ adding the gauge-fixing term does not change this. For $\mathcal{N} = 1$ super-Yang--Mills theory, one needs the ghost for gauge transformations $c^a$ (a Grassmann-odd Lie-algebra valued scalar), the ghost for supersymmetry transformations $\theta$ (a constant Grassmann-even Majorana spinor), and the ghost for translations $\alpha^\rho$ (a constant Grassmann-odd vector), which is necessary to close the supersymmetry transformations~\eqref{sec_sym_susyalgebra}.

We thus define all the required gradings and fields:\footnote{The given choice of engineering dimensions of the (anti-)ghosts and auxiliary fields is such that $\brst$ augments the engineering dimension by $1$.}
\begin{longcode}
DefGrading[\{Dimension, GhostNumber\}];
DefOddTensor[cc[a], M, PrintAs $\to$ "c"];
DefEvenSpinor[theta[-A], M, SpinorType $\to$ Majorana, PrintAs $\to$ "$\theta$"];
DefOddTensor[alpha[$\mu$], M, PrintAs $\to$ "$\alpha$"];
CD[\_\_]@theta[\_\_] \textasciicircum:= 0;
CD[\_\_]@bartheta[\_\_] \textasciicircum:= 0;
CD[\_\_]@alpha[\_\_] \textasciicircum:= 0;
SetGrading[AA, \{Dimension $\to$ 1, GhostNumber $\to$ 0\}];
SetGrading[FF, \{Dimension $\to$ 2, GhostNumber $\to$ 0\}];
SetGrading[auxD, \{Dimension $\to$ 2, GhostNumber $\to$ 0\}];
SetGrading[cc, \{Dimension $\to$ 1, GhostNumber $\to$ 1\}];
SetGrading[chi, \{Dimension $\to$ 3/2, GhostNumber $\to$ 0\}];
SetGrading[theta, \{Dimension $\to$ 1/2, GhostNumber $\to$ 1\}];
SetGrading[alpha, \{Dimension $\to$ 0, GhostNumber $\to$ 1\}];
\end{longcode}
We then define the BRST transformations using the predefined \texttt{BRST} operator:
\begin{longcode}
AA /: BRST[AA[$\mu$\_, a\_]] := gauge[cc]@AA[$\mu$, a] + susy[theta]@AA[$\mu$, a]
    + With[\{$\rho$ = DummyIn@TangentM\}, alpha[$\rho$] CD[-$\rho$]@AA[$\mu$, a]]
FF /: BRST[FF[$\mu$\_, $\nu$\_, a\_]] := gauge[cc]@FF[$\mu$, $\nu$, a] + susy[theta]@FF[$\mu$, $\nu$, a]
    + With[\{$\rho$ = DummyIn@TangentM\}, alpha[$\rho$] CD[-$\rho$]@FF[$\mu$, $\nu$, a]]
chi /: BRST[chi[a\_, -A\_]] := gauge[cc]@chi[a, -A] + susy[theta]@chi[a, -A]
    + With[\{$\rho$ = DummyIn@TangentM\}, alpha[$\rho$] $\cdot$ CD[-$\rho$]@chi[a, -A]]
barchi /: BRST[barchi[a\_, A\_]] := gauge[cc]@barchi[a, A] + susy[theta]@
    barchi[a, A] + With[\{$\rho$ = DummyIn@TangentM\}, alpha[$\rho$] $\cdot$ CD[-$\rho$]@barchi[a, A]]
auxD /: BRST[auxD[a\_]] := gauge[cc]@auxD[a] + susy[theta]@auxD[a]
    + With[\{$\rho$ = DummyIn@TangentM\}, alpha[$\rho$] CD[-$\rho$]@auxD[a]]
cc /: BRST[cc[a\_]] := With[\{A = DummyIn@SpinM, B = DummyIn@SpinM, mu =
    DummyIn@TangentM\}, AA[-mu, a] bartheta[A] Gammagg[mu, -A, B] theta[-B]]
    - I/2 g With[\{b = DummyIn@lie, c = DummyIn@lie\}, lief[a, b, c] cc[-b] $\cdot$ cc[-c]]
    + With[\{mu = DummyIn@TangentM\}, alpha[mu] $\cdot$ CD[-mu]@cc[a]]
theta /: BRST[theta[A\_]] := 0
bartheta /: BRST[bartheta[A\_]] := 0
alpha /: BRST[alpha[mu\_]] := With[\{A = DummyIn@SpinM, B = DummyIn@SpinM\},
    - bartheta[A] Gammagg[mu, -A, B] theta[-B]]
\end{longcode}
To check BRST invariance of the action, we also have to define the BRST transformations of $\gamma$ matrices and the $\epsilon$ tensor, which are not predefined (since they change under diffeomorphisms, which are needed for gravity theories), and to impose that it commutes with covariant derivatives:
\begin{longcode}
Map[Function[BRST[#[i\_\_\_]] := 0], \$GammaMatrices];
BRST[epsilongg[\_\_]] := 0;
BRST[CD[i\_\_][expr\_]] := CD[i]@BRST[expr];
BRST[DD[i\_\_][expr\_]] := BRST@CollectTensors@ChangeCovD[
    ChangeCovD[DD[i]@expr, DD, PD], PD, CD];
\end{longcode}

Let us check that with these definitions the BRST operator $\brst$ is nilpotent:
\begin{longcode}
IrreducibleSpinProject[mycanon@TimesToCenterDot@JoinGammaMatrices@mycanon@
    IrreducibleSpinDecompose[mycanon@JoinGammaMatrices@mycanon@
    BRST[BRST[cc[a]]],theta],theta]
BRST[BRST[alpha[$\mu$]]]
mycanon@FToGradA@JoinGammaMatrices@mycanon@IrreducibleSpinDecompose[mycanon@
    JoinGammaMatrices@mycanon@BRST[BRST[AA[$\mu$, a]]], theta]
mycanon@TimesToCenterDot@JoinGammaMatrices@mycanon@IrreducibleSpinDecompose[
    mycanon@JoinGammaMatrices@mycanon@BRST[BRST[chi[a, -A]]], theta]
CollectTensors@RiemannYoungProject@JoinGammaMatrices@mycanon@FToGradA@
     IrreducibleSpinDecompose[mycanon@JoinGammaMatrices@mycanon@
     IrreducibleSpinDecompose[mycanon@JoinGammaMatrices@mycanon@
     BRST[BRST[auxD[a]]], theta], chi]    
\end{longcode}
\begin{mathouteq}
&0 \\
&0 \\
&- \tfrac{1}{2} A^{\alpha a} \alpha_{\beta }\cdot \alpha_{\gamma } R[\nabla ]^{\mu }{}_{\alpha }{}^{\beta \gamma } \\
&- \tfrac{1}{8} \alpha_{\gamma }\cdot \alpha_{\delta }\cdot \chi^{a}{}_{B} \gamma^{\alpha \beta }{}_{A}{}^{B} R[\nabla ]_{\alpha \beta }{}^{\gamma \delta } \\
&- \tfrac{1}{12} \bar{\theta}^{A} \alpha_{\beta }\cdot \chi^{a}{}_{B} \epsilon g_{\alpha \gamma \delta \mu } \gamma^{\gamma \delta \mu }{}_{A}{}^{B} R[\nabla ]^{\alpha \beta } \eqend{,}&
\end{mathouteq}
which quite obviously only vanishes in a flat background, as is well known, such that for the following we set all curvature tensors to vanish:
\begin{longcode}
RiemannCD[\_\_] := 0
RicciCD[\_\_] := 0
RicciScalarCD[] := 0
\end{longcode}
For the action, we obtain as before that it is invariant up to a surface term:
\begin{longcode}
surface = - I/2 bartheta[A] GammaggStar[-A, B] Gammagg[$\mu$, -B, F] chi[-a, -F]
    auxD[a] - 1/2 bartheta[A] Gammagg[-$\nu$, -A, B] chi[a, -B] FF[$\mu$, $\nu$, -a]
    - 1/4 bartheta[A] Gammagg[$\mu$, $\nu$, $\rho$, -A, B] chi[-a, -B] FF[-$\nu$, -$\rho$, a]
    - alpha[$\mu$] $\cdot$ ReplaceDummies[action];
sn = CollectTensors@GammaMatricesToDual@IrreducibleSpinDecompose[
    CollectTensors@SpinorFlipSymmetrize@RiemannYoungProject@mycanon@
    FToGradA@CollectTensors@JoinGammaMatrices@mycanon@
    TimesToCenterDot[BRST[action] + CD[-$\mu$]@surface], chi];
IrreducibleSpinProject[CollectTensors@JoinGammaMatrices@sn, chi]
\end{longcode}
\begin{mathouteq}
&0 \eqend{.}&
\end{mathouteq}

We now turn to cohomologies (and thus in particular anomalies). To compute them, besides a brute-force approach (which for any but the most trivial cases is too lengthy) one can choose between various methods such as descent equations~\cite{piguetsorella,barnichetal2000} or spectral sequences~\cite{dixon1991}. A very useful concept are filtrations~\cite{piguetsorella}, defined by a filtration operator $\mathcal{F}$ with non-negative integer eigenvalues. $\mathcal{F}$ can be defined by assigning each field of the theory a certain weight, and defining the action of $\mathcal{F}$ on a monomial by the sum of the weights of the fields in the monomial. Assume further that the BRST differential $\brst$ can be written as a sum of differentials $\brst_k$ with integer $k \geq 0$, such that $\left[ \mathcal{F}, \brst_k \right] = k \brst_k$. Then it follows that the lowest-order term $\brst_0$ is itself nilpotent, and that the cohomology of $\brst$ is isomorphic to a subset of the cohomology of $\brst_0$~\cite{piguetsorella}. That is, one can determine the cohomology of $\brst$ by first computing the cohomology of $\brst_0$ and then checking whether each element can be extended. A simple filtration operator is given by assigning each field wright 1, such that $\mathcal{F}$ simply counts the number of fields, and in this case $\brst_0$ is the BRST differential of the free theory. In general, the choice of weights is obviously restricted by the fact that $\brst = \sum_{k \geq 0} \brst_k$, and that $\brst_0$ is non-trivial. If $\mathcal{F}$ commutes with derivatives, the same result applies to the relative cohomologies. Functions to work with filtrations (and check whether a choice for the weights leads to a viable filtration) are defined by \fix, see section~\ref{sec_doc_brst}.

We would like to verify that the well-known super-chiral anomalies~\cite{clarkpiguetsibold1978,piguetsibold1984,konishishizuya1985,guadagninimintchev1985} satisfy the consistency criterion. The gauge part of the anomaly reads
\begin{equation}
\mathcal{A}^\text{gauge} = \int c^a \epsilon^{\mu\nu\rho\sigma} d_{abc} \left[ \nabla_\mu A_\nu^b \nabla_\rho A_\sigma^e + \frac{\mathi}{4} g f_{cde} \nabla_\mu \left( A_\nu^b A_\rho^d A_\sigma^e \right) \right] \total x \eqend{,}
\end{equation}
while the supersymmetric part of the anomaly is\footnote{This differs from the expression given in~\cite{guadagninimintchev1985}, since they use Weyl instead of Majorana fermions.}
\begin{equation}
\mathcal{A}^\text{susy} = \int \epsilon^{\mu\nu\rho\sigma} d_{abc} \, \bar{\theta} \left[ \frac{1}{8} \gamma_{\mu\nu\rho\sigma} \chi^a \left( \bar{\chi}^b \chi^c \right) - \gamma_\mu \chi^a \left( 2 A_\nu^b \nabla_\rho A_\sigma^c + \frac{3}{4} \mathi g f_{bde} A_\nu^c A_\rho^d A_\sigma^e \right) \right] \total x \eqend{.}
\end{equation}
Since gauge transformations close among themselves, the gauge transformation of $\mathcal{A}^\text{gauge}$ must vanish which is its consistency condition. On the other hand, because supersymmetry transformations close into a gauge transformation and a translation, the consistency condition for $\mathcal{A}^\text{susy}$ is~\cite{guadagninimintchev1985}
\begin{equation}
\delta^\text{susy}_\theta \mathcal{A}^\text{susy} = \mathcal{A}^\text{gauge} \big\rvert_{c^a \to - \bar{\theta} \gamma^\mu \theta A_\mu^a} \eqend{.}
\end{equation}

We thus first define the anomalies together with the needed surface terms:
\begin{longcode}
lied = InvariantTraceTensor[lie, 3, Symmetric];
lied4 = InvariantTraceTensor[lie, 4];
gaugeanomaly = mycanon@mycanon[epsilongg[-$\mu$, -$\nu$, -$\rho$, -$\sigma$] cc[a] lied[-a, -b, -c]
    (CD[$\mu$]@AA[$\nu$, b] CD[$\rho$]]@AA[$\sigma$, c] + I/4 g lief[c, -d, -e]
    CD[$\mu$]@(AA[$\nu$, b] AA[$\rho$, d] AA[$\sigma$, e]))];
gaugeanomalysurface = mycanon[I g epsilongg[$\mu$, $\nu$, $\rho$, $\sigma$] (CD[-$\sigma$]@AA[-$\nu$, a]
    AA[-$\rho$, b] cc[c] $\cdot$ cc[d] lied4[-a, -b, -c, -d] - AA[-$\nu$, a] CD[-$\sigma$]@AA[-$\rho$, b]
    cc[c] $\cdot$ cc[d] lied4[-a, -b, -c, -d] + AA[-$\nu$, a] AA[-$\rho$, b] cc[c] $\cdot$
    CD[-$\sigma$]@cc[d] lied4[-a, -c, -b, -d] + g AA[-$\nu$, a] AA[-$\rho$, b] AA[-$\sigma$, c]
    cc[d] $\cdot$ cc[e] Invlie5[-a, -b, -c, -d, -e])];
susyanomaly = mycanon[lied[-a, -b, -c] epsilongg[-$\mu$, -$\nu$, -$\rho$, -$\sigma$] (1/8
    bartheta[A] Gammagg[$\mu$, $\nu$, $\rho$, $\sigma$, -A, B] chi[a,-B] $\cdot$ barchi[b,F] $\cdot$ chi[c,-F]
    - bartheta[A] $\cdot$ Gammagg[$\nu$, -A, B] $\cdot$ chi[a,-B] (2 AA[$\nu$, b] CD[$\rho$]@AA[$\sigma$, c]
    + 3/4 I g lief[b, d, e] AA[$\nu$, c] AA[$\rho$, -d] AA[$\sigma$, -e]))];
susyanomalysurface = mycanon[lied[-a, -b, -c] epsilongg[-$\nu$, -$\rho$, -$\sigma$, -$\alpha$] (1/48
    AA[-$\beta$, a] barchi[b,A] $\cdot$ Gammagg[$\nu$, $\rho$, $\sigma$, $\alpha$, -A, B] $\cdot$ chi[c, -B] bartheta[F]
    Gammagg[$\mu$, $\beta$, -F, G] theta[-G] - 1/4 AA[$\nu$, a] barchi[b, A] $\cdot$ chi[c, -A]
    gg[$\mu$, $\alpha$] bartheta[F] Gammagg[$\rho$, $\sigma$, -F, G] theta[-G] - 1/12 AA[$\nu$, a]
    barchi[c, A] $\cdot$ Gammagg[$\rho$, $\sigma$, $\alpha$, -A, B] $\cdot$ chi[b, -B] bartheta[F]
    Gammagg[$\mu$, -F, G] theta[-G] - 1/12 AA[-$\beta$, a] barchi[b, A] $\cdot$ Gammagg[$\nu$, $\rho$, $\sigma$,
    -A, B] $\cdot$ chi[c, -B] gg[$\mu$, $\alpha$] bartheta[F] Gammagg[$\beta$, -F, G] theta[-G])];
\end{longcode}

Since gauge transformations close among themselves, we can define a consistent filtration of the BRST differential:
\begin{longcode}
CheckFiltration[\{AA $\to$ 0, FF $\to $ 0, chi $\to$ 0, barchi $\to$ 0, cc $\to$ 0, auxD $\to$ 0,
    theta $\to$ 2, bartheta $\to$ 1, alpha $\to$ 2\}, BRST, Display $\to$ Full]
\end{longcode}
\begin{mathouteq}
&BRST0[ A_\alpha{}^a ] \to i g A_{\alpha }{}^{b} c^{c} f^{a}{}_{bc} + \nabla_{\alpha }c^{a} \\
&BRST0[ F_{\alpha\beta}{}^a ] \to -i g c^{b} F_{\alpha\beta}{}^c f^{a}{}_{bc} \\
&BRST0[ \chi^a{}_A ] \to -i g c^{b}\cdot \chi^{c}{}_{A} f^{a}{}_{bc} \\
&BRST0[ \bar{\chi}^{aA} ] \to -i g c^{b}\cdot \bar{\chi}^{cA} f^{a}{}_{bc} \\
&BRST0[ c^a ] \to - \tfrac{1}{2}i g c_{b}\cdot c_{c} f^{abc} \\
&BRST0[ D^a ] \to -i g D^{c} c^{b} f^{a}{}_{bc} \\
&BRST0[ \theta_A ] \to 0 \\
&BRST0[ \bar{\theta}^A ] \to 0 \\
&BRST0[ \alpha^\alpha ] \to 0 \eqend{,}&
\end{mathouteq}
and we see that indeed only the gauge transformations survive with this choice. We can thus define a filtrated differential and check that the gauge transformation of $\mathcal{A}^\text{gauge}$ vanishes:
\begin{longcode}
DefOddDifferential[BRSTgauge]
Filtrate[\{AA $\to$ 0, FF $\to$ 0, chi $\to$ 0, barchi $\to$ 0, cc $\to$ 0, auxD $\to$ 0,
    theta $\to$ 2, bartheta $\to$ 1, alpha $\to$ 2\}, BRST $\to$ BRSTgauge]
BRSTgauge[epsilongg[\_\_]] := 0
BRSTgauge[CD[i\_]@AA[j\_\_]] := CD[i]@BRSTgauge[AA[j]]
mycanon[BRSTgauge[gaugeanomaly] + CD[-$\mu$]@gaugeanomalysurface]
\end{longcode}
\begin{mathouteq}
& 0 \eqend{.}&
\end{mathouteq}
On the other hand, supersymmetry transformations close into a gauge transformation and a translation, such that we can not use filtrations to check the corresponding consistency condition. Instead, we obtain an equivariant differential, for which we define the corresponding transformations directly:
\begin{longcode}
DefOddDifferential[BRSTsusy]
BRSTsusy[expr\_] := BRST[expr] /. \{cc[\_] :> 0, alpha[\_] :> 0\}
erg = mycanon@JoinGammaMatrices@Expand@IrreducibleSpinDecompose[mycanon@
    JoinGammaMatrices@mycanon[BRSTsusy[susyanomaly]
    + CD[-$\mu$]@susyanomalysurface], theta];
erg2 = CollectTensors@FToGradA@JoinGammaMatrices@GammaMatricesToDual@
    IrreducibleSpinDecompose[erg, chi];
erg3 = mycanon@GradAToF@JoinGammaMatrices[erg2]
\end{longcode}
\begin{mathouteq}
&- A^{\alpha a} \epsilon g_{\alpha \mu \nu \rho } F_{\beta }{}^{\mu b} F^{\nu \rho c} d[lie]_{abc} \mathrm{Irr}\mathbb{S} M^{(2,\mathbf{4})}[\theta ]^{\beta } - i g A^{\alpha a} A^{\beta b} A^{\mu c} \epsilon g_{\alpha \beta \mu \rho } F_{\nu }{}^{\rho d} tr[lie]_{abcd} \mathrm{Irr}\mathbb{S} M^{(2,\mathbf{4})}[\theta ]^{\nu } \eqend{.} &
\end{mathouteq}
The result is not yet equal to the gauge anomaly $\mathcal{A}^\text{gauge}$ with the ghost $c^a$ replaced by $- \bar{\theta} \gamma^\mu \theta A_\mu^a$~\cite{guadagninimintchev1985}, which is due to dimension-dependent identities that arise from antisymmetrising an arbitrary expression over five or more indices (which identically vanishes in four dimensions). A convenient way to use these identities is the projection of the corresponding expression onto the Young tableau describing the symmetry, where the dimension-dependent identities correspond to the fact that the projectors for Young tableaux with five or more rows vanish, and which for the $\epsilon$ tensor is implemented using \cmdref{sec_doc_spingamma_epsilongammareduce}{EpsilonYoungProject}:
\begin{longcode}
sn = erg3 - mycanon[mycanon@GradAToF@gaugeanomaly /. \{cc[a\_] :> With[\{$\mu$ =
    DummyIn@TangentM\}, -IrreducibleSpinTensor[theta, 2, "4"][$\mu$] AA[-$\mu$, a]]\}];
CollectTensors[sn /. \{epsilongg[i\_\_] IrreducibleSpinTensor[theta, 2, "4"][j\_]
    :> EpsilonYoungProject[epsilongg[i] IrreducibleSpinTensor[theta, 2, "4"][j],
    gg]\}]
\end{longcode}
\begin{mathouteq}
&0 \eqend{.} &
\end{mathouteq}
Since the consistency conditions are nothing else but the statement that the anomaly is BRST-exact, this shows that $\brst\left( \mathcal{A}^\text{gauge} + \mathcal{A}^\text{susy} \right) = 0$, which of course can also be checked directly with \fix.

In general, to bring expressions into a fully canonical form, one would in principle have to consider the interplay between permutations of the Lie algebra indices and the Lorentz indices (i.e., the composition of the multiterm symmetries described by the associated Young tableaux), a problem related to plethysms in the representation theory of the symmetric group. This is a hard problem in general, but the corresponding results can be emulated by repeatedly applying \cmdref{sec_doc_spingamma_epsilongammareduce}{EpsilonYoungProject} and bringing the result into canonical form using only monoterm symmetries (what \xact's \texttt{ToCanonical} command achieves).\footnote{I thank Igor Khavkine for discussions on this subject.} In the present case, a single use of \texttt{EpsilonYoungProject} was fortunately sufficient, since the expression only contained five different indices, one of which belonged to a tensor that only appeared once in the expression, namely the irreducible spin tensor $\mathrm{Irr}\mathbb{S} M^{(2,\mathbf{4})}[\theta]_\mu$.

Lastly, I want to show how one can also derive the gauge anomaly directly in \fix, at least in the $A$ -- $c$ sector where no canonicalisation of spinors is needed. The reason is that the algorithm implemented in \fix\ is a brute-force one, which simply applies the BRST differential and checks whether the result vanishes. Therefore, the result will only be complete (and thus correct) if a fully canonical form can be found for each term, which is quite involved for spinors. We thus have to make an ansatz for all possible terms that can possibly appear, which is restricted by ghost number and engineering dimension. Furthermore, we do not need to include traces of invariant tensors of the Lie algebra, since they can be expressed using invariant tensors of lower rank~\cite{lie1,lie2}, and need to include at most one totally antisymmetric $\epsilon$ tensor because the product of two can be reduced to products of metrics. We also need ans{\"a}tze for surface terms and BRST-exact terms (and the corresponding surface terms), and define ghost number and dimension of the invariant tensors and $\epsilon$ to ensure this is correctly taken into account:
\begin{longcode}
invtensors = \{lief, lied, Invlie4, Invlie5\};
Map[SetGrading[\#, \{Dimension $\to$ 0, GhostNumber $\to$ 0\}] \&, invtensors];
SetGrading[epsilongg, \{Dimension $\to$ 0, GhostNumber $\to$ 0\}];
SetGrading[CD, \{Dimension $\to$ 1, GhostNumber $\to$ 0\}];
tracelessreps = \{t\_?xTensorQ[i\_\_\_, a\_, j\_\_\_, -a\_, k\_\_\_] :> 0\};
ansatz = GenerateMonomials[\{AA, cc\}, \{invtensors, \{epsilongg\}\}, Constraint $\to$
    (((Dimension[\#]==5) \&\& (GhostNumber[#]==1))\&), Replacements $\to$
    tracelessreps, MaxNumberOfFields $\to$ 5, MaxNumberOfInvTensors $\to$ 1,
    MaxNumberOfDerivatives $\to$ 3];
dansatz = GenerateMonomials[\{AA, cc\}, \{invtensors, \{epsilongg\}\}, Constraint $\to$
    (((Dimension[\#]==5) \&\& (GhostNumber[#]==2))\&), Replacements $\to$
    tracelessreps, MaxNumberOfFields $\to$ 5, MaxNumberOfInvTensors $\to$ 1,
    MaxNumberOfDerivatives $\to$ 3, FreeIndices $\to$ \{$\mu$\}];
csansatz = GenerateMonomials[\{AA, cc\}, \{invtensors, \{epsilongg\}\}, Constraint $\to$
    (((Dimension[\#]==4) \&\& (GhostNumber[#]==0))\&), Replacements $\to$
    tracelessreps, MaxNumberOfFields $\to$ 4, MaxNumberOfInvTensors $\to$ 1,
    MaxNumberOfDerivatives $\to$ 2];
cdansatz = GenerateMonomials[\{AA, cc\}, \{invtensors, \{epsilongg\}\}, Constraint $\to$
    (((Dimension[\#]==4) \&\& (GhostNumber[#]==1))\&), Replacements $\to$
    tracelessreps, MaxNumberOfFields $\to$ 4, MaxNumberOfInvTensors $\to$ 1,
    MaxNumberOfDerivatives $\to$ 2, FreeIndices $\to$ \{$\mu$\}];
\end{longcode}
Since both $A$ and $c$ have engineering dimension 1, at most five (respectively four) fields are needed, and at most three (respectively two) derivatives, since any term with more derivatives would be a total derivative and not contribute anyway to the cohomology. For a more effective generation of monomials, the function \cmdref{sec_doc_contraction_generatemonomialsbygrading}{GenerateMonomialsByGrading} could be used as well. The computation of the relative cohomology then results in
\begin{longcode}
RelativeCohomologyFromAnsatz[BRSTgauge, ansatz, CD[-$\mu$], dansatz, BRSTgauge,
    csansatz, CD[-$\mu$], cdansatz, CanonicalizeMethod $\to$ mycanon]
\end{longcode}
\begin{mathouteq}
&\{\{ c^{a} \epsilon g_{\alpha \beta \gamma \delta } d[lie]_{abc} \nabla^{\beta }A^{\alpha b} \nabla^{\delta }A^{\gamma c} + i g A^{\alpha a} A^{\beta b} c^{c} \epsilon g_{\alpha \beta \gamma \delta } tr[lie]_{abcd} \nabla^{\delta }A^{\gamma d} \\
&+ i g A^{\alpha a} A^{\beta b} c^{c} \epsilon g_{\alpha \beta \gamma \delta } tr[lie]_{abdc} \nabla^{\delta }A^{\gamma d} + i g A^{\alpha a} A^{\beta b} c^{c} \epsilon g_{\alpha \beta \gamma \delta } tr[lie]_{acbd} \nabla^{\delta }A^{\gamma d}, \\
&i g^2 A^{\alpha a} A^{\beta b} A^{\gamma c} c_{d}\cdot c_{e} \epsilon g^{\mu }{}_{\alpha \beta \gamma } tr[lie]_{abc}{}^{de} - i g A^{\alpha a} c_{c}\cdot c_{d} \epsilon g^{\mu }{}_{\alpha \beta \gamma } tr[lie]_{ab}{}^{cd} \nabla^{\gamma }A^{\beta b} \\
&- i g A^{\alpha a} c_{c}\cdot c_{d} \epsilon g^{\mu }{}_{\alpha \beta \gamma } tr[lie]_{a}{}^{c}{}_{b}{}^{d} \nabla^{\gamma }A^{\beta b} - i g A^{\alpha a} c_{c}\cdot c_{d} \epsilon g^{\mu }{}_{\alpha \beta \gamma } tr[lie]_{a}{}^{cd}{}_{b} \nabla^{\gamma }A^{\beta b} \}\} \eqend{,}&
\end{mathouteq}
which agrees with $\mathcal{A}^\text{gauge}$. Denoting the first term by $A$ and the second term in the result by $S^\mu$, $S^\mu$ is the needed surface term such that $\brst A + \nabla_\mu S^\mu = 0$:
\begin{longcode}
mycanon[erg[[1, 1]] - gaugeanomaly]
mycanon[BRSTgauge[erg[[1, 1]]] + CD[-$\mu$]@erg[[1, 2]]]
\end{longcode}
\begin{mathouteq}
& 0 \\
& 0 \eqend{.} &
\end{mathouteq}

\section{Functions in {\sc FieldsX}}
\label{sec_doc}

In this section, I give a complete list of all functions and variables contained in \fix. In most cases the functions are self-explanatory, or their usage has already been demonstrated in the previous section. Other functions are only needed to implement advanced functionality, and will not be used by a normal user of the package. In the remaining cases, I give short descriptions and/or the rationale for the given implementation.

\subsection{Helper functions}
\label{sec_doc_helper}

These are small helper functions that are needed for other parts of the package, but are also useful on their own.

\ssubsection{DummiesIn}
\label{sec_doc_helper_dummiesin}

\begin{usagetable}
	\usageline*{DummiesIn[\textit{bundle},\,\textit{k}]}
		{returns a list of \textit{k} unique abstract dollar-indices on the vector bundle \textit{bundle}, using the last of the user-defined indices.}
\end{usagetable}

This is just a straightforward extension of the \xact\ command \texttt{DummyIn}.

\ssubsection{TensorCount}
\label{sec_doc_helper_tensorcount}

\begin{usagetable}
	\usageline{TensorCount[\textit{expr},\,\textit{T}]}
		{returns the number of tensors \textit{T} occuring in \textit{expr} including covariant derivatives of \textit{T}. \textit{expr} can be given in pseudo index-free notation.}
	\usageline*{TensorCount[\textit{expr},\,\textit{T},\,False]}
        {does not include covariant derivatives of \textit{T}.}
\end{usagetable}

The pseudo index-free notation was introduced in the \textsc{xTras} package, where only heads of tensors are given and not their indices. Expressions in pseudo index-free notation must be wrapped in the head \texttt{IndexFree}. For example, \texttt{IndexFree[CD@RicciCD]} stands for \texttt{CD[-$\mu$][RicciCD[-$\alpha$,-$\beta$]}.
    
\ssubsection{AllTensors}
\label{sec_doc_helper_alltensors}

\begin{usagetable}
	\usageline*{AllTensors[\textit{expr}]}
        {returns a list containing all tensors occuring in \textit{expr} together with their multiplicity. \textit{expr} must be given in pseudo index-free notation.}
\end{usagetable}

To convert an expression with indices to pseudo index-free notation, use the command \texttt{ToIndexFree} of the \textsc{xTras} package.

\subsection{Inner bundles (Lie algebra functions)}
\label{sec_doc_inner}

These are functions to work with inner bundles in general, and Lie-algebra valued fields in particular, including invariant tensors.

\ssubsection{DefVBundleWithMetric}
\label{sec_doc_inner_defvbundlewithmetric}

\begin{usagetable}
	\usageline*{DefVBundleWithMetric[\textit{bundle},\,\textit{M},\,\textit{dim},\,\{\textit{a},\,\textit{b},\,\textit{c},\,\dots\},\,\textit{metric}]}
        {defines \textit{bundle} to be a vector bundle with base manifold \textit{M} and fiber vector space with dimension given by \textit{dim} and represented by the abstract indices \{\textit{a},\,\textit{b},\,\textit{c},\,\dots\}. It defines \textit{metric} to be a metric on the vector bundle \textit{bundle} without associated covariant derivative, which is constant with respect to the covariant derivative on the tangent bundle of \textit{M}.}
\end{usagetable}

By default, it is not possible to define metrics on inner vector bundles with \xact; this command is adapted from code of the \xact\ mailing list and the \textsc{Spinors} package where such a definition was already needed. The dimension \textit{dim} can be an integer or any constant symbol (defined with \texttt{DefConstantSymbol}). Since the metric is defined to be constant with respect to the covariant derivative of the tangent bundle of the manifold, a covariant derivative must be defined on \textit{M} first. The main application of this command is for principal $G$-bundles with $G$ a semisimple Lie group or products thereof, since gauge fields transform in some representation of the associated Lie algebra $\mathfrak{g}$ (in a local trivialisation of the bundle). The metric is then just given by the Cartan--Killing form of the Lie algebra $\mathfrak{g}$, and \textit{dim} is the dimension of the chosen representation.

\ssubsection{\$InvariantTraceTensors}
\label{sec_doc_inner_invarianttracetensors}

\begin{usagetable}
	\usageline*{\$InvariantTraceTensors}
        {is a global variable storing the list of all currently defined invariant tensors on inner bundles.}
\end{usagetable}

This is the analogue of the lists of metrics, etc.

\ssubsection{InvariantTraceTensor}
\label{sec_doc_inner_invarianttracetensor}

\begin{usagetable}
	\usageline*{InvariantTraceTensor[\textit{bundle},\,\textit{n},\,\textit{sym}]}
        {returns the invariant tensor on the inner bundle \textit{bundle} obtained as the trace over \textit{n} basis elements. For $n=3$ \textit{sym} determines whether the antisymmetric \textit{f} tensor or the symmetric \textit{d} tensor is returned.}
\end{usagetable}

By the Chevalley restriction theorem, the $G$-invariant polynomials for a simple Lie algebra $\mathfrak{g}$ are linear combinations of products of traces of the generators $\mathfrak{t}^a$ in some representation of $\mathfrak{g}$. This function returns $\tr\left( \mathfrak{t}^{a_1} \cdots \mathfrak{t}^{a_n} \right)$ (only the head of the tensor, without the abstract indices). For $n = 2$, the invariant tensor is the Cartan--Killing form (the metric on \textit{bundle}). For $n = 3$, specifying \textit{sym} $\to$ \texttt{Antisymmetric} returns the completely antisymmetric $f^{abc}$ (equal to the structure constants of $\mathfrak{g}$), while using \textit{sym} $\to$ \texttt{Symmetric} returns the completely symmetric part $d^{abc}$. In the general case $n \geq 4$ it would in principle be possible to further decompose the traces according to their transformations under the permutation group, but this seems not be useful in applications such that only the cyclic symmetry of the trace is imposed. Note that for any given Lie algebra, invariant tensors of high enough rank are not independent and expressable in terms of tensors of lower rank or vanish. For example, $d^{abc} = 0$ for $\mathfrak{su}(2)$; see~\cite{lie1,lie2} for derivations of such relations. These relations are not taken into account by \fix\ and must be imposed by the user, giving rules for the corresponding tensors.

\ssubsection{Invariant\-TraceTensorQ}
\label{sec_doc_inner_invarianttracetensorq}

\begin{usagetable}
	\usageline*{InvariantTraceTensorQ[\textit{expr}]}
        {gives \texttt{True} if \textit{expr} is an invariant tensor on some inner bundle, and \texttt{False} otherwise.}
\end{usagetable}

This is the analogue of the corresponding functions for metrics, etc.

\ssubsection{\$StructureConstantSign}
\label{sec_doc_inner_structureconstantsign}

\begin{usagetable}
	\usageline*{\$StructureConstantSign}
        {defines the global sign of the structure constants of inner bundles:\linebreak $\left[ \mathfrak{t}_a, \mathfrak{t}_b \right] = \texttt{\$StructureConstantSign} \, f_{ab}{}^c \, \mathfrak{t}_c$. The default value is $\mathi$.}
\end{usagetable}

By default, \fix\ assumes Hermitian generators $\mathfrak{t}$ and structure constants.

\ssubsection{ReduceInvariantTraceTensors}
\label{sec_doc_inner_reduceinvarianttracetensors}

\begin{usagetable}
    \usageline{ReduceInvariantTraceTensors[\textit{expr}]}
        {expands products of the structure constants $f_{ab}{}^c$ with invariant tensors on inner bundles into sums of invariant tensors.}
    \usageline{ReduceInvariantTraceTensors[\textit{expr},\,\textit{tens}]}
        {expands products of $f_{ab}{}^c$ with the invariant tensor \textit{tens} only.}
	\usageline*{ReduceInvariantTraceTensors[\textit{expr},\,\{\textit{tens1},\,\textit{tens2},\,\dots\}]}
        {expands products of $f_{ab}{}^c$ with the invariant tensors \textit{tens1}, \textit{tens2}, \dots}
\end{usagetable}

Since gauge transformations involve the structure constants, this function can be used to put the result back into canonical form.

\subsection{Functions extended to work with more than one bundle}
\label{sec_doc_extend}

These functions are needed for other parts of the package, and extend existing functionality of \xact.

\ssubsection{BundleSymmetryOf}
\label{sec_doc_extend_bundlesymmetryof}

\begin{usagetable}
	\usageline{BundleSymmetryOf[\textit{expr}]}
        {gives a description (a result with head \texttt{Symmetry}) of the symmetry of \textit{expr}. This includes a generating set for that symmetry using \texttt{Cycles} notation on the indices of \textit{expr}. Extending \texttt{SymmetryOf}, the number of indices and the symmetry group are ordered by bundle.}
	\usageline{Sorted}
        {is an option for \texttt{BundleSymmetryOf} that specifies if the replacements should be sorted by slot number. By default, it is \texttt{True}.}
	\usageline*{Offset}
        {is also an option for \texttt{BundleSymmetryOf} that specifies if the generating sets should use offsets for the slot numbers. By default, it is \texttt{False}.}
\end{usagetable}

This is needed for computing cohomologies.

\subsection{Noncommuting product, Grassmann-even and -odd tensors}
\label{sec_doc_grassmann}

Here functions to work with non-commuting objects are listed.

\ssubsection{CenterDot}
\label{sec_doc_grassmann_centerdot}

\begin{usagetable}
	\usageline*{CenterDot}
        {stands for the non-commutative product of Grassmann-odd indexed objects.}
\end{usagetable}

\texttt{CenterDot} ($\cdot$) is an operator in \textsc{Mathematica} with no built-in rules, which makes it ideal to use as the non-commutative product, in contrast to \texttt{NonCommutativeMultiply} (\texttt{**}). Commuting objects are automatically taken out of the non-commutative product, so it is safe to use \texttt{CenterDot} for all objects. \texttt{CenterDot} can be entered as \keys{Esc + . + Esc}. Since \texttt{**} is nevertheless faster to input, to automatically convert \texttt{NonCommutativeMultiply} to \texttt{CenterDot} one can use the definition
\begin{longcode}
Unprotect[NonCommutativeMultiply];
NonCommutativeMultiply /: NonCommutativeMultiply[expr\_\_\_] := CenterDot[expr];
Protect[NonCommutativeMultiply];
\end{longcode}
Because of the automatic rules set up for \texttt{CenterDot}, in replacement rules it is necessary to use the \texttt{Verbatim} or \texttt{HoldPattern} wrappers: \texttt{Verbatim[CenterDot][expr\_\_\_] :> \dots} or \texttt{HoldPattern[ expr1\_ $\cdot$ expr2\_ ] :> \dots}

\ssubsection{Parity}
\label{sec_doc_grassmann_parity}

\begin{usagetable}
	\usageline*{Parity[\textit{expr}]}
        {returns the Grassmann parity of \textit{expr}.}
\end{usagetable}

This gives either 0 or 1.

\ssubsection{TimesToCenterDot}
\label{sec_doc_grassmann_timestocenterdot}

\begin{usagetable}
	\usageline*{TimesToCenterDot[\textit{expr}]}
        {returns \textit{expr} with all products replaced by non-commutative ones.}
\end{usagetable}

This is useful when converting from pseudo index-free notation when non-commutative objects are included, since \texttt{IndexFree} does not work well with the non-commutative product.

\ssubsection{DefEvenTensor}
\label{sec_doc_grassmann_defeventensor}

\begin{usagetable}
	\usageline{DefEvenTensor[\textit{T}[-\textit{a},\,\textit{b},\,\dots],\,\textit{M}]}
        {defines \textit{T} to be a Grassmann-even tensor field on the manifold \textit{M} and the base manifolds associated to the vector bundles of its indices $-\textit{a},\textit{b},\dots$}
	\usageline*{DefEvenTensor[\textit{T}[-\textit{a},\,\textit{b},\,\dots],\,\textit{M},\,\textit{sym}]}
        {defines \textit{T} to be a Grassmann-even tensor field with symmetry \textit{sym}.}
\end{usagetable}

This extends the \xact\ command \texttt{DefTensor} for commuting objects, and takes the same additional arguments (for example, \texttt{PrintAs}).

\ssubsection{DefOddTensor}
\label{sec_doc_grassmann_defoddtensor}

\begin{usagetable}
	\usageline{DefOddTensor[\textit{T}[-\textit{a},\,\textit{b},\,\dots],\,\textit{M}]}
        {defines \textit{T} to be a Grassmann-odd tensor field on the manifold \textit{M} and the base manifolds associated to the vector bundles of its indices $-\textit{a},\textit{b},\dots$}
	\usageline*{DefOddTensor[\textit{T}[-\textit{a},\,\textit{b},\,\dots],\,\textit{M},\,\textit{sym}]}
        {defines \textit{T} to be a Grassmann-odd tensor field with symmetry \textit{sym}.}
\end{usagetable}

This extends the \xact\ command \texttt{DefTensor} for non-commuting objects, and takes the same additional arguments (for example, \texttt{PrintAs}).

\ssubsection{\$CenterDot\-TexSymbol}
\label{sec_doc_grassmann_centerdottexsymbol}

\begin{usagetable}
	\usageline*{\$CenterDotTexSymbol}
        {gives the symbol to use for \TeX\ output of the non-commutative product. The default value is a space.}
\end{usagetable}

\fix\ includes proper \TeX\ output that works together with the \textsc{TexAct} package and can be customised using its commands.

\subsection{Frame bundles and spin connections}
\label{sec_doc_framespin}

These are functions to work with frame bundles on the manifold.

\ssubsection{FrameBundleQ}
\label{sec_doc_framespin_framebundleq}

\begin{usagetable}
	\usageline*{FrameBundleQ[\textit{bundle}]}
        {gives \texttt{True} if \textit{bundle} is a frame bundle, and \texttt{False} otherwise.}
\end{usagetable}

This is the analogue of the corresponding functions for metrics, etc.

\ssubsection{DefFrameBundle}
\label{sec_doc_framespin_defframebundle}

\begin{usagetable}
	\usageline*{DefFrameBundle[\textit{frame}[-$\mu$,\,\textit{a}],\,\textit{eta}[-\textit{a},\,-\textit{b}],\,\{\textit{a},\,\textit{b},\,\dots\}]}
        {defines a frame bundle \textit{FrameM} with abstract indices $\textit{a},\textit{b},\dots$ on the base manifold \textit{M} of the tangent bunde represented by the abstract index $\mu$. It also defines the flat frame bundle metric \textit{eta}[$-\textit{a},\,-\textit{b}$] and the frame field \textit{frame}[$-\mu,\,\textit{a}$] with inverse \textit{frame}[$\mu,\,-\textit{a}$].}
\end{usagetable}

Before using this command, a metric (with its associated covariant derivative) must be defined on $M$. The frame field, its determinant \textit{Detframe} and the frame bundle metric are defined to be constant with respect to this covariant derivative.

\ssubsection{UndefFrameBundle}
\label{sec_doc_framespin_undefframebundle}

\begin{usagetable}
	\usageline*{UndefFrameBundle[\textit{frame}]}
        {undefines the frame bundle whose frame field is \textit{frame}.}
\end{usagetable}

If there are spin connections defined on the bundle, they must be undefined first.

\ssubsection{FrameFieldOfBundle}
\label{sec_doc_framespin_framefieldofbundle}

\begin{usagetable}
	\usageline*{FrameFieldOfBundle[\textit{tb},\,\textit{fb}]}
	{returns the frame field connecting the tangent bundle \textit{tb} and the frame bundle \textit{fb}.}
\end{usagetable}

This is needed for other parts of the package.

\ssubsection{\$SpinConnections}
\label{sec_doc_spingamma_spinconnections}

\begin{usagetable}
	\usageline*{\$SpinConnections}
        {is a global variable storing the list of all currently defined spin connections.}
\end{usagetable}

This is the analogue of the lists of metrics, etc.

\ssubsection{DefSpinConnection}
\label{sec_doc_framespin_defspinconnection}

\begin{usagetable}
	\usageline*{DefSpinConnection[$\omega$[-$\mu$,\,-\textit{a},\,-\textit{b}],\,\textit{CD}]}
	{defines the spin connection $\omega$[-$\mu$,\,-\textit{a},\,-\textit{b}] of the covariant derivative \textit{CD} and the associated curvature and torsion tensors.}
\end{usagetable}

If the covariant derivative is the standard Levi-Civita one, the corresponding torsion and contortion vanishes. The torsion and contortion defined by \fix\ for the spin connection agree with the conventions of~\cite{freedmanvanproeyen}, which differ in the ordering of the indices from the convention used by \xact.

\ssubsection{UndefSpinConnection}
\label{sec_doc_framespin_undefspinconnection}

\begin{usagetable}
	\usageline*{UndefSpinConnection[$\omega$]}
	{undefines the spin connection $\omega$.}
\end{usagetable}

This also removes the associated curvature and torsion tensors.

\ssubsection{CovDOfSpinCon\-nection}
\label{sec_doc_framespin_covdofspinconnection}

\begin{usagetable}
	\usageline*{CovDOfSpinConnection[$\omega$]}
	{returns the covariant derivative corresponding to the spin connection $\omega$.}
\end{usagetable}

This is needed for other parts of the package.

\ssubsection{SpinConnectionToFrame}
\label{sec_doc_framespin_spinconnectiontoframe}

\begin{usagetable}
	\usageline*{SpinConnectionToFrame[\textit{expr},\,$\omega$]}
	{expands all occurrences of the spin connection $\omega$ inside \textit{expr} in terms of derivatives of the frame field and the torsion. It also expands perturbations of the spin connection in terms of covariant derivatives of the perturbation of the frame field and the torsion. The second argument can be a list of spin connections.}
\end{usagetable}

This is useful in a second-order formalism, where one only needs to specify the perturbation of the frame field.

\ssubsection{ContortionToTorsion}
\label{sec_doc_framespin_contortiontotorsion}

\begin{usagetable}
	\usageline{ContortionToTorsion[\textit{expr},\,\textit{CD}]}
	{expresses the contortion tensor \textit{ContortionCD} of the covariant derivative \textit{CD} occurring inside \textit{expr} by the torsion tensor \textit{TorsionCD}.}
	\usageline{ContortionToTorsion[\textit{expr},\,$\omega$]}
	{expresses the contortion tensor \textit{Contortion}$\omega$ of the spin connection $\omega$ occurring inside \textit{expr} by the torsion tensor \textit{Torsion}$\omega$.}
	\usageline*{\vspace{-1.5em}}
	{In both cases, the second argument can be a list of covariant derivatives or spin connections.}
\end{usagetable}

The contortion defined by \fix\ for a spin connection agrees with the conventions of~\cite{freedmanvanproeyen}, while the one defined for the covariant derivative differs in the ordering of the indices to match the torsion convention used by \xact.

\subsection{Spin structure, \texorpdfstring{$\gamma$}{\textgamma} matrices}
\label{sec_doc_spingamma}

These are functions to define a spin structure on the manifold and work with the corresponding curved-space $\gamma$ matrices.

\ssubsection{GammaMatrixQ,\,GammaStarQ,\,GammaZeroQ}
\label{sec_doc_spingamma_gammamatrixq}

\begin{usagetable}
	\usageline{GammaMatrixQ[\textit{expr}]}
        {gives \texttt{True} if \textit{expr} is a $\gamma$ matrix, and \texttt{False} otherwise.}
	\usageline{GammaStarQ[\textit{expr}]}
        {gives \texttt{True} if \textit{expr} is the $\gamma_*$ (chiral) matrix, and \texttt{False} otherwise.}
	\usageline*{GammaZeroQ[\textit{expr}]}
        {gives \texttt{True} if \textit{expr} is the $\gamma^0$ matrix, and \texttt{False} otherwise.}
\end{usagetable}

This is the analogue of the corresponding functions for metrics, etc. The chiral $\gamma_*$ matrix is only defined for even dimensions of the manifold, and $\gamma^0$ exists only in Lorentzian signature; both of them are fixed (numerical) matrices.

\ssubsection{\$GammaStarSign}
\label{sec_doc_spingamma_gammastarsign}

\begin{usagetable}
	\usageline*{\$GammaStarSign}
        {defines the global sign of the $\gamma_*$ (chiral) matrix. By default it is 1.}
\end{usagetable}

By default, \fix\ assumes the conventions of~\cite{freedmanvanproeyen}, such that
\begin{equation*}
\gamma_* = \frac{(-\mathi)^{d/2+3}}{d!} \epsilon_{\mu_1 \cdots \mu_d} \gamma^{\mu_1} \cdots \gamma^{\mu_d}
\end{equation*}
in even dimensions $d$ and $\gamma_*^2 = \unitmatrix$. \texttt{\$GammaStarSign} can be used to insert an additional overall factor. In odd dimensions $d$, the $(d-1)$-dimensional chiral matrix is used as the last basis element of the Clifford algebra, and \texttt{\$GammaStarSign} is the sign in eq.~(3.40) of~\cite{freedmanvanproeyen}.

\ssubsection{GammaMatrix}
\label{sec_doc_spingamma_gammamatrix}

\begin{usagetable}
	\usageline{GammaMatrix[\textit{metric},\,\textit{n}]}
        {returns the generalized (totally antisymmetric) $\gamma$ matrix of order $n$ of the Clifford algebra associated to the metric \textit{metric}.}
	\usageline{GammaMatrix[\textit{metric},\,Star]}
        {returns the $\gamma_*$ (chiral) matrix of the Clifford algebra associated to the metric \textit{metric}.}
	\usageline*{GammaMatrix[\textit{metric},\,Zero]}
        {returns the $\gamma^0$ matrix of the Clifford algebra associated to the metric \textit{metric}.}
\end{usagetable}

The generalised $\gamma$ matrices are normalised with unit strength: $\gamma^{\mu_1 \cdots \mu_n} = \gamma^{[\mu_1} \cdots \gamma^{\mu_n]}$. The chiral $\gamma_*$ matrix is only defined for even dimensions of the manifold, and $\gamma^0$ exists only in Lorentzian signature. These functions only return the head of the tensor, without the abstract (spin bundle) indices.

\ssubsection{Metric\-OfGammaMatrix}
\label{sec_doc_spingamma_metricofgammamatrix}

\begin{usagetable}
	\usageline*{MetricOfGammaMatrix[$\gamma$]}
        {returns the metric associated to the Clifford algebra of $\gamma$.}
\end{usagetable}

This is needed for other parts of the package.

\ssubsection{\$GammaMatrices}
\label{sec_doc_spingamma_gammamatrices}

\begin{usagetable}
	\usageline*{\$GammaMatrices}
        {is a global variable storing the list of all currently defined $\gamma$ matrices.}
\end{usagetable}

This is the analogue of the lists of metrics, etc.

\ssubsection{\$PrecomputeGammaMatrixProducts}
\label{sec_doc_spingamma_precomputegammamatrixproducts}

\begin{usagetable}
	\usageline*{\$PrecomputeGammaMatrixProducts}
        {controls whether products of generalised $\gamma$ matrices are precomputed when a spin structure is defined, to speed up later computations. By default, it is \texttt{True}.}
\end{usagetable}

In low dimensions, the precomputation only takes a few seconds, but in higher dimensions it might be preferable to switch it off.

\ssubsection{DefSpinStruc\-ture}
\label{sec_doc_spingamma_defspinstructure}

\begin{usagetable}
	\usageline*{DefSpinStructure[\textit{metric},\,\{\textit{A},\,\textit{B},\,\textit{C},\,\dots\}]}
        {defines a spin structure on the base manifold \textit{M} of the metric \textit{metric}. This includes the $\gamma$ matrices of the Clifford algebra associated to \textit{metric} and a spin bundle \textit{SpinM} with abstract indices \{\textit{A},\,\textit{B},\,\textit{C},\,\dots\}, whose covariant derivative is induced from the one of \textit{metric}, with the same name.}
\end{usagetable}

Before using this command, a metric and (Levi-Civita) covariant derivative must be defined on $M$, whose dimension must be a positive integer. The curvature $\left( \mathcal{R}_{\mu\nu} \right)_A{}^B$ of the induced spin connection is given in terms of the Riemann tensor $R_{\mu\nu\rho\sigma}$ of the metric by the well-known formula
\begin{equation}
\left( \mathcal{R}_{\mu\nu} \right)_A{}^B = \frac{1}{4} R_{\mu\nu\rho\sigma} \left( \gamma^{\rho\sigma} \right)_A{}^B \eqend{.}
\end{equation}
If \cmdref{sec_doc_spingamma_precomputegammamatrixproducts}{$PrecomputeGammaMatrixProducts} is \texttt{True}, \texttt{DefSpinStructure} also precomputes products of generalised $\gamma$ matrices to speed up later computations. If a frame bundle has been defined, \fix\ also defines the corresponding frame $\gamma$ matrices.

\ssubsection{UndefSpinStructure}
\label{sec_doc_spingamma_undefspinstructure}

\begin{usagetable}
	\usageline*{UndefSpinStructure[\textit{metric}]}
        {undefines the spin structure on the base manifold of the metric \textit{metric}.}
\end{usagetable}

This also removes all related definitions of $\gamma$ matrices etc.

\ssubsection{SpinBundleQ}
\label{sec_doc_spingamma_spinbundleq}

\begin{usagetable}
	\usageline*{SpinBundleQ[\textit{bundle}]}
        {gives \texttt{True} if \textit{bundle} is a spin bundle, and \texttt{False} otherwise.}
\end{usagetable}

This is the analogue of the corresponding functions for metrics, etc.

\ssubsection{SplitGammaMatrix,\,SplitGammaMatrices}
\label{sec_doc_spingamma_splitgammamatrix}

\begin{usagetable}
	\usageline{SplitGammaMatrix[$\gamma$,\,\textit{keep}]}
        {decomposes the generalized $\gamma$ matrix $\gamma$ into an antisymmetrized product of individual $\gamma$ matrices. If $\textit{keep}=\texttt{True}$, the $\gamma_*$ (chiral) matrix is kept.}
    \usageline*{SplitGammaMatrices[\textit{expr},\,\textit{keep}]}
        {decomposes all the generalized $\gamma$ matrix appearing within \textit{expr} into an antisymmetrized product of individual $\gamma$ matrices. If $\textit{keep}=\texttt{True}$, the $\gamma_*$ (chiral) matrix is kept.}
\end{usagetable}

For example, this is needed for computing perturbations of curved-space $\gamma$ matrices from the perturbation of the individual matrices.

\ssubsection{JoinGammaMatrices}
\label{sec_doc_spingamma_joingammamatrices}

\begin{usagetable}
    \usageline*{JoinGammaMatrices[\textit{expr},\,\textit{keep}]}
        {replaces products of $\gamma$ matrices within \textit{expr} by generalized $\gamma$ matrices. If \textit{keep} = \texttt{True}, products of a single $\gamma_*$ (chiral) matrix and other generalized $\gamma$ matrices are kept.}
\end{usagetable}

Since supersymmetry transformations involve generalised $\gamma$ matrices, this function can be used to put the result back into canonical form. \texttt{JoinGammaMatrices} always simplifies expressions with products of generalised $\gamma$ matrices and at least two chiral $\gamma_*$ matrices, but keeps products of generalised $\gamma$ matrices and a single $\gamma_*$ if the second parameter is \texttt{True}.

The formula for the product of two generalised $\gamma$ matrices is~\cite[eq.~(3.41)]{hep-th/9910030}
\begin{equation*}
\gamma_{a_1 \cdots a_i} \gamma^{b_1 \cdots b_j} = \sum_{s = 0}^{\min(i,j)} \frac{i! j!}{s! (i-s)! (j-s)!} \delta^{[b_1}_{[a_i} \cdots \delta^{b_s}_{a_{i-s+1}} \gamma_{a_1 \cdots a_{i-s}]}{}^{b_{s+1} \cdots b_j]} \eqend{.}
\end{equation*}

\ssubsection{DualGammaMatrix}
\label{sec_doc_spingamma_dualgammamatrix}

\begin{usagetable}
    \usageline*{DualGammaMatrix[$\gamma$[$\mu$,\,\dots,\,-\textit{A},\,\textit{B}]]}
        {returns the dual matrix of the generalized $\gamma$ matrix $\gamma^{\mu \cdots }{}_A{}^B$.}
\end{usagetable}

The duality relations are the ones given in~\cite{freedmanvanproeyen}, eqs.~(3.41) and (3.42). In odd dimensions, \cmdref{sec_doc_spingamma_gammastarsign}{$GammaStarSign} gives the sign used in eq.~(3.41).

\ssubsection{GammaMatricesToDual}
\label{sec_doc_spingamma_gammamatricestodual}

\begin{usagetable}
    \usageline{GammaMatricesToDual[\textit{expr}]}
        {replaces generalized $\gamma$ matrices with more than $\frac{d}{2}$ indices within \textit{expr} by their dual, where $d$ is the dimension of the manifold.}
    \usageline{GammaMatricesToDual[\textit{expr},\,\texttt{All}]}
        {replaces all generalized $\gamma$ matrices within \textit{expr} by their dual.}
    \usageline*{GammaMatricesToDual[\textit{expr},\,$\gamma$]}
        {replaces only the matrix $\gamma$ within \textit{expr} by its dual.}
\end{usagetable}

The dual $\gamma$ matrices are given by \cmdref{sec_doc_spingamma_dualgammamatrix}{DualGammaMatrix}. It is often useful to perform computations with $\gamma$ matrices with less indices, since canonicalisation is faster.

\ssubsection{ChangeGamma\-Matrices}
\label{sec_doc_spingamma_changegammamatrices}

\begin{usagetable}
	\usageline*{ChangeGammaMatrices[\textit{expr},\,\textit{met1},\,\textit{met2}]}
	{converts all generalized $\gamma$ matrices within \textit{expr} depending on the metric \textit{met1} to the corresponding ones depending on the metric \textit{met2}. One of the metrics must be the metric of a tangent bundle, and the other one of the associated frame bundle.}
\end{usagetable}

To convert between curved-space $\gamma$ matrices and the ones in a frame, a frame bundle must be defined using \cmdref{sec_doc_framespin_defframebundle}{DefFrameBundle}.

\ssubsection{EpsilonGammaReduce, EpsilonYoung\-Project}
\label{sec_doc_spingamma_epsilongammareduce}

\begin{usagetable}
    \usageline{EpsilonGammaReduce[\textit{expr},\,\textit{metric}]}
        {replaces products of the totally antisymmetric $\epsilon$ tensor and generalized $\gamma$ matrices associated to the metric \textit{metric} within \textit{expr} by suitably contracted ones.}
    \usageline*{EpsilonYoungProject[\textit{expr},\,\textit{metric}]}
        {replaces products of the totally antisymmetric $\epsilon$ tensor associated to the metric \textit{metric} and other tensors within \textit{expr} onto the corresponding Young tableaux.}
\end{usagetable}

This can also be used to put expressions into canonical form. The list of replacements is given in appendix~\ref{app_youngproj}.

\subsection{Spinors}
\label{sec_doc_spinors}

These are functions to define spinors of various types.

\ssubsection{MajoranaQ,\,DiracQ,\,SpinorQ,\,SpinorUnbarQ,\,SpinorBarQ}
\label{sec_doc_spinors_majoranaq}

\begin{usagetable}
	\usageline{MajoranaQ[\textit{expr}]}
        {gives \texttt{True} if \textit{expr} is a Majorana spinor, and \texttt{False} otherwise.}
	\usageline{DiracQ[\textit{expr}]}
        {gives \texttt{True} if \textit{expr} is a Dirac spinor, and \texttt{False} otherwise.}
	\usageline{SpinorQ[\textit{expr}]}
        {gives \texttt{True} if \textit{expr} is a spinor, and \texttt{False} otherwise.}
	\usageline{SpinorUnbarQ[\textit{expr}]}
        {gives \texttt{True} if \textit{expr} is a spinor but not a conjugate one, and \texttt{False} otherwise.}
	\usageline*{SpinorBarQ[\textit{expr}]}
        {gives \texttt{True} if \textit{expr} is a conjugate spinor, and \texttt{False} otherwise.}
\end{usagetable}

This is the analogue of the corresponding functions for metrics, etc. \fix\ does not (yet) support neither 2- nor 4-component Weyl spinors. Since 4-component Weyl spinors can be viewed as Dirac spinors $\psi_A$ that satisfy
\begin{equation*}
\psi_a = \left( \mathcal{P}_\pm \psi \right)_A = \frac{1}{2} \left( \unitmatrix \pm \gamma_* \right)_A{}^B \psi_B = \frac{1}{2} \psi_A \pm \frac{1}{2} \left( \gamma_* \right)_A{}^B \psi_B \eqend{,}
\end{equation*}
they can be emulated by defining rules that replace contractions of $\gamma_*$ with $\psi$.

\ssubsection{DefSpinor}
\label{sec_doc_spinors_defspinor}

\begin{usagetable}
	\usageline{DefSpinor[$\psi$[-$A$],\,\textit{M}]}
        {defines $\psi$ to be a spinor field on the manifold \textit{M} and the spin bundle associated to the index $-A$. The conjugate spinor $\bar{\psi}$ is automatically defined, with name \textit{bar}$\psi$.}
	\usageline{DefSpinor[$\psi$[-$A$],\,\textit{M},\,\textit{sym}]}
        {defines $\psi$ to be a spinor field with symmetry \textit{sym}.}
	\usageline{DefSpinor[$\psi$[$b$,-$A$],\,\textit{M}]}
        {defines $\psi$ to be a spinor field valued in the inner bundle associated to the index $b$.}
	\usageline{SpinorType}
        {is an option for \texttt{DefSpinor} that specifies the type of spinor. By default, it is \texttt{Majorana}.}
	\usageline{Majorana}
        {is a value for the option \texttt{SpinorType} of \texttt{DefSpinor}.}
	\usageline{Dirac}
        {is a value for the option \texttt{SpinorType} of \texttt{DefSpinor}.}
	\usageline*{Conjugate}
        {is also an option for \texttt{DefSpinor} that specifies if the roles of spinor and conjugate spinor should be switched.}
\end{usagetable}

\texttt{DefSpinor} is the general function (extending \texttt{DefTensor} from \xact, and taking thus the same additional arguments such as \texttt{PrintAs}), but for applications it is probably more useful to use \cmdref{sec_doc_spinors_defevenspinor}{DefEvenSpinor} and \cmdref{sec_doc_spinors_defoddspinor}{DefOddSpinor} to define spinors with given Grassmann parity. The \texttt{Conjugate} option is useful in the BV formalism, where the antifield $\psi^\dagger$ for a spinor $\psi$ is really a conjugate spinor, while the antifield $\bar{\psi}^\dagger$ of the conjugate $\bar{\psi}$ is not a conjugate one, such that their respective products are scalars.

\ssubsection{UndefSpinor}
\label{sec_doc_spinors_undefspinor}

\begin{usagetable}
	\usageline*{UndefSpinor[$\psi$]}
        {undefines $\psi$ and the conjugate spinor $\bar{\psi}$.}
\end{usagetable}

This ensures that both $\psi$ and its conjugate are removed.

\ssubsection{DefEvenSpinor}
\label{sec_doc_spinors_defevenspinor}

\begin{usagetable}
	\usageline{DefEvenSpinor[$\psi$[-$A$],\,\textit{M}]}
        {defines $\psi$ to be a Grassmann-even spinor field on the manifold \textit{M} and the spin bundle associated to the index $-A$. The conjugate spinor $\bar{\psi}$ is automatically defined, with name \textit{bar}$\psi$.}
	\usageline{DefEvenSpinor[$\psi$[-$A$],\,\textit{M},\,\textit{sym}]}
        {defines $\psi$ to be a Grassmann-even spinor field with symmetry \textit{sym}.}
	\usageline*{DefEvenSpinor[$\psi$[$b$,-$A$],\,\textit{M}]}
        {defines $\psi$ to be a Grassmann-even spinor field valued in the inner bundle associated to the index $b$.}
\end{usagetable}

This extends \cmdref{sec_doc_spinors_defspinor}{DefSpinor}, such that the options given there (and the ones of the \xact\ command \texttt{DefTensor}) are also valid for \texttt{DefEvenSpinor}. Lie-algebra valued spinors are defined with the last variant; the inner bundle must be defined first with \cmdref{sec_doc_inner_defvbundlewithmetric}{DefVBundleWithMetric}.

\ssubsection{DefOddSpinor}
\label{sec_doc_spinors_defoddspinor}

\begin{usagetable}
	\usageline{DefOddSpinor[$\psi$[-$A$],\,\textit{M}]}
        {defines $\psi$ to be a Grassmann-odd spinor field on the manifold \textit{M} and the spin bundle associated to the index $-A$. The conjugate spinor $\bar{\psi}$ is automatically defined, with name \textit{bar}$\psi$.}
	\usageline{DefOddSpinor[$\psi$[-$A$],\,\textit{M},\,\textit{sym}]}
        {defines $\psi$ to be a Grassmann-odd spinor field with symmetry \textit{sym}.}
	\usageline*{DefOddSpinor[$\psi$[$b$,-$A$],\,\textit{M}]}
        {defines $\psi$ to be a Grassmann-odd spinor field valued in the inner bundle associated to the index $b$.}
\end{usagetable}

This extends \cmdref{sec_doc_spinors_defspinor}{DefSpinor}, such that the options given there (and the ones of the \xact\ command \texttt{DefTensor}) are also valid for \texttt{DefOddSpinor}. Lie-algebra valued spinors are defined with the last variant; the inner bundle must be defined first with \cmdref{sec_doc_inner_defvbundlewithmetric}{DefVBundleWithMetric}.

\ssubsection{ConjugateSpinor}
\label{sec_doc_spinors_conjugatespinor}

\begin{usagetable}
    \usageline{ConjugateSpinor[$\psi$]}
        {returns $\bar{\psi}$.}
    \usageline*{ConjugateSpinor[$\bar{\psi}$]}
        {returns $\psi$.}
\end{usagetable}

Note that the conjugate is the charge conjugate for Majorana spinors and the Dirac conjugate $\bar{\psi} = \mathi \psi^\dagger \gamma^0$ for Dirac spinors.

\ssubsection{SpinScalar}
\label{sec_doc_spinors_spinscalar}

\begin{usagetable}
    \usageline*{SpinScalar[\textit{expr}]}
        {gives \texttt{True} if \textit{expr} (given in pseudo index-free notation) is a spin bundle scalar (i.e., all indices with values in a spin bundle can be contracted), and \texttt{False} otherwise.}
\end{usagetable}

This is needed for computing cohomologies. The expression \textit{expr} must be wrapped in the head \texttt{IndexFree} of the \textsc{xTras} package.

\subsection{Flip and Fierz relations}
\label{sec_doc_flipfierz}

These are functions for flip relations (valid for Majorana spinors) and Fierz identities and rearrangements (valid for all spinors).

\ssubsection{\$SpinorFlipSigns}
\label{sec_doc_flipfierz_spinorflipsigns}

\begin{usagetable}
    \usageline*{\$SpinorFlipSigns}
        {is a table of signs appearing in the Majorana flip relations.}
\end{usagetable}

The signs are taken from~\cite{freedmanvanproeyen}, table 3.1 (the choices in boldface).

\ssubsection{SignOfGammaMatrix}
\label{sec_doc_flipfierz_signofgammamatrix}

\begin{usagetable}
    \usageline*{SignOfGammaMatrix[$\gamma$]}
        {returns the sign needed for the generalized $\gamma$ matrix $\gamma$ appearing in the Majorana flip relations.}
\end{usagetable}

This is the sign appearing in the Majorana flip relation (3.51) of~\cite{freedmanvanproeyen}.

\ssubsection{FindSpinChain}
\label{sec_doc_flipfierz_findspinchain}

\begin{usagetable}
    \usageline*{FindSpinChain[\textit{expr},\,\textit{start}[\textit{inds}]]}
        {returns a list of spinors and $\gamma$ matrices appearing within \textit{expr} whose indices are contracted with each other (spin chain). The chain starts with \textit{start}[\textit{inds}], which must be a (conjugate) spinor appearing in \textit{expr}.}
\end{usagetable}

This functions extracts a spinor bilinear which can be flipped using the Majorana flip relations. The functions \cmdref{sec_doc_flipfierz_flipspinor}{FlipSpinor}, \cmdref{sec_doc_flipfierz_flipspinorstoconjugateamount}{FlipSpinorsToConjugateAmount}, and \cmdref{sec_doc_flipfierz_spinorflipsymmetrize}{SpinorFlipSymme-!trize} might be more useful for applications.

\ssubsection{FlipSpinChain}
\label{sec_doc_flipfierz_flipspinchain}

\begin{usagetable}
    \usageline*{FlipSpinChain[\textit{expr},\,\textit{chain}]}
        {returns \textit{expr} with the spin chain \textit{chain} flipped using the Majorana flip relations.}
\end{usagetable}

The spin chain \textit{chain} can be found using \cmdref{sec_doc_flipfierz_findspinchain}{FindSpinChain}. The functions \cmdref{sec_doc_flipfierz_flipspinor}{FlipSpinor}, \cmdref{sec_doc_flipfierz_flipspinorstoconjugateamount}{FlipSpinorsToConjugateAmount}, and \cmdref{sec_doc_flipfierz_spinorflipsymmetrize}{SpinorFlipSymmetrize} might be more useful for applications.

\ssubsection{FlipSpinor}
\label{sec_doc_flipfierz_flipspinor}

\begin{usagetable}
    \usageline{FlipSpinor[\textit{expr}]}
        {returns \textit{expr} with a spinor bilinear flipped using the Majorana flip relations. \textit{expr} must contain a single spinor bilinear.}
    \usageline{FlipSpinor[\textit{expr},\,$\psi$]}
        {returns \textit{expr} with a spinor bilinear flipped using the Majorana flip relations. \textit{expr} must contain a single bilinear formed with the spinor $\psi$.}
    \usageline*{FlipSpinor[\textit{expr},\,$\psi_1$,\,$\psi_2$]}
        {returns \textit{expr} with a spinor bilinear flipped using the Majorana flip relations. \textit{expr} must contain a single bilinear formed with the spinors $\psi_1$ and $\psi_2$.}
\end{usagetable}

This function uses the Majorana flip relations on a single bilinear, with the variants controlling which bilinear is flipped.

\ssubsection{FlipSpinorsToConjugateAmount}
\label{sec_doc_flipfierz_flipspinorstoconjugateamount}

\begin{usagetable}
    \usageline*{FlipSpinorsToConjugateAmount[\textit{expr},\,$\psi$,\,\textit{count}]}
        {returns \textit{expr} with spinor bilinears formed with the spinor $\psi$ flipped using the Majorana flip relations until \textit{count} conjugate spinors $\bar{\psi}$ remain.}
\end{usagetable}

This function is useful to ensure that all terms of a long expression have the same form, in order to exhibit cancellations.

\ssubsection{SortSpinor}
\label{sec_doc_flipfierz_sortspinor}

\begin{usagetable}
    \usageline{SortSpinor[\textit{expr},\,$\psi \to \bar{\psi}$]}
        {returns \textit{expr} with all bilinears formed with the spinor $\psi$ flipped using the Majorana flip relations. The spinor $\psi$ and its conjugate $\bar{\psi}$ can be exchanged.}
    \usageline*{SortSpinor[\textit{expr},\,\{$\psi_1 \to \bar{\psi}_1$,\,\dots\}]}
        {returns \textit{expr} with all bilinears formed with the spinors $\psi_1,\dots$ flipped using the Majorana flip relations. The spinors $\psi_i$ and their conjugates $\bar{\psi}_i$ can be exchanged.}
\end{usagetable}

Also this function is useful to bring all terms of a long expression into the same form.

\ssubsection{SpinorFlipSymme\-trize}
\label{sec_doc_flipfierz_spinorflipsymmetrize}

\begin{usagetable}
    \usageline*{SpinorFlipSymmetrize[\textit{expr}]}
        {returns \textit{expr} with all spinor bilinears symmetrized using the Majorana flip relations.}
\end{usagetable}

This function replaces all bilinears in \textit{expr} with half of the sum of the bilinear and its flipped version.

\ssubsection{FierzExpand}
\label{sec_doc_flipfierz_fierzexpand}

\begin{usagetable}
    \usageline*{FierzExpand[$\bar{\psi}_1[\textit{inds}_1]$,\,$\psi_2[\textit{inds}_2]$]}
        {expands the tensor product $\bar{\psi}_1[\textit{inds}_1] \psi_2[\textit{inds}_2]$ in the basis of generalized $\gamma$ matrices (Fierz rearrangement). Both spinors may have covariant derivatives acting on them.}
\end{usagetable}

This is the basic Fierz rearrangement formula
\begin{equation*}
\bar{\psi}^A \chi_B = 2^{-[d/2]} \sum_{k=0}^d \frac{1}{k!} (-1)^{k (k-1)/2} \gamma_{\mu_1 \cdots \mu_k}{}_B{}^A \left( \bar{\psi}^C \gamma^{\mu_1 \cdots \mu_k}{}_C{}^D \chi_D \right) \eqend{,}
\end{equation*}
which is valid in the given form for both commuting and anticommuting spinors. From this, various rearrangement formulas for three, four and more spinors can be obtained by contracting this equation with more $\gamma$ matrices and spinors.

\subsection{Irreducible spin tensors}
\label{sec_doc_irreducible}

These functions implement the approach to Fierz identities of d'Auria, Fr{\'e}, Maina and Regge~\cite{dauriafremainaregge1982} based on the decomposition of the product of group representations into irreducible components. Currently, the corresponding decompositions are worked out for the tensor product of up to four Majorana spinors in four dimensions, which can have an additional inner bundle index.

\ssubsection{IrreducibleSpinTensor}
\label{sec_doc_irreducible_irreduciblespintensor}

\begin{usagetable}
    \usageline*{IrreducibleSpinTensor[$\psi$,\,\textit{n},\,\textit{rep}]}
        {returns the irreducible tensor of the Lorentz group corresponding to the representation \textit{rep} appearing in the decomposition of the tensor product of \textit{n} copies of the spinor $\psi$.}
\end{usagetable}

This function returns the invariant tensor $\Psi^{(n,\textbf{\textit{rep}})}$ (only the head of the tensor, without the abstract indices) that transforms in the irreducible representation $\textit{rep}$ (labeled by its dimension) of the Lorentz group (and the symmetric group for spinors with inner bundle indices) contained in the product of \textit{n} spinors $\psi$. The list of valid representations is given in appendix~\ref{app_decomp}.

\ssubsection{IrreducibleSpinTensorQ}
\label{sec_doc_irreducible_irreduciblespintensorq}

\begin{usagetable}
    \usageline*{IrreducibleSpinTensorQ[\textit{expr}]}
        {gives \texttt{True} if \textit{expr} is an irreducible tensor of the Lorentz group for some representation, and \texttt{False} otherwise.}
\end{usagetable}

This is the analogue of the corresponding functions for metrics, etc.

\ssubsection{ExpandIrreducibleSpinTensor, ExpandIrreducibleSpinTensors}
\label{sec_doc_irreducible_expandirreduciblespintensor}

\begin{usagetable}
    \usageline{ExpandIrreducibleSpinTensor[\textit{tens}[$\psi$,\,\dots][\textit{inds}]]}
        {expands the irreducible tensor \textit{tens} into a sum of products of spinors $\psi$ with the right symmetries.}
    \usageline*{ExpandIrreducibleSpinTensors[\textit{expr}]}
        {expands all irreducible tensors of the Lorentz group within \textit{expr} into a sum of products of spinors with the right symmetries.}
\end{usagetable}

These functions give the explicit form of the invariant tensors in terms of the spinor $\psi$ and $\gamma$ matrices. It is the inverse function to \cmdref{sec_doc_irreducible_irreduciblespindecompose}{IrreducibleSpinDecompose}.

\ssubsection{IrreducibleSpinDecompose}
\label{sec_doc_irreducible_irreduciblespindecompose}

\begin{usagetable}
    \usageline*{IrreducibleSpinDecompose[\textit{expr},\,$\psi$]}
        {decomposes all products of spinors $\psi$ within \textit{expr} into sums of irreducible representations of the Lorentz group.}
\end{usagetable}

This function performs the decomposition of products of the spinor $\psi$ into invariant tensors transforming in irreducible representations of the Lorentz group (and the symmetric group for spinors with inner bundle indices). It is the inverse function to \cmdref{sec_doc_irreducible_expandirreduciblespintensor}{ExpandIrreducibleSpinTensors}.

\ssubsection{IrreducibleSpinProject}
\label{sec_doc_irreducible_irreduciblespinproject}

\begin{usagetable}
    \usageline*{IrreducibleSpinProject[\textit{expr},\,$\psi$]}
        {projects all irreducible tensors of the Lorentz group within \textit{expr} depending on the spinor $\psi$ onto their Young tableaux. This includes their products with $\gamma$ matrices, and induced symmetries if the spinor $\psi$ depends on inner bundles.}
\end{usagetable}

This function ensures that multiterm symmetries are correctly taken into account when bringing terms into canonical form, using Young projectors. The list of applied projections is given in appendix~\ref{app_youngproj}.

\subsection{Gradings}
\label{sec_doc_gradings}

These are functions to work with arbitrary gradings, for example ghost number, antifield number or engineering dimension.

\ssubsection{\$Gradings}
\label{sec_doc_gradings_gradings}

\begin{usagetable}
	\usageline*{\$Gradings}
        {is a global variable storing the list of all currently defined gradings.}
\end{usagetable}

This is the analogue of the lists of metrics, etc.

\ssubsection{DefGrading}
\label{sec_doc_gradings_defgrading}

\begin{usagetable}
	\usageline{DefGrading[\textit{grad}]}
        {defines the grading \textit{grad}.}
	\usageline{DefGrading[\{$\textit{grad}_1$,\,\dots\}]}
        {defines the gradings $\textit{grad}_1,\dots$}
	\usageline{SumGrading}
        {is an option for \texttt{DefGrading} that specifies a function that determines the grading of a sum. By default it is \texttt{Undefined\&}.}
	\usageline*{ZeroGrading}
        {is an option for \texttt{DefGrading} that specifies the grading of 0. By default it is \texttt{Undefined}.}
\end{usagetable}

Numeric expressions are not taken into account when computing the grading of an expression, i.e. $\textit{grad}[2 \textit{expr}] = \textit{grad}[\textit{expr}]$. The grading of a product is the sum of the gradings of the individual terms, and the grading of a covariant derivative of an expression is the sum of the gradings of the expression and the derivative. Gradings work properly with the symmetrised covariant derivatives of the \textsc{xTras} package, where the grading of the derivative gets multiplied by the number of indices it contains. In some cases, it is meaningful to assign a definite grading to a sum or to $0$, for example if all the terms in the sum have the same ghost number.

\ssubsection{UndefGrading}
\label{sec_doc_gradings_undefgrading}

\begin{usagetable}
	\usageline*{UndefGrading[\textit{grad}]}
        {undefines the grading \textit{grad}.}
\end{usagetable}

This function does not remove gradings set for individual tensors.

\ssubsection{SetGrading}
\label{sec_doc_gradings_setgrading}

\begin{usagetable}
	\usageline{SetGrading[\textit{T},\,$\textit{grad}\to\textit{val}$]}
        {sets the grading \textit{grad} of the tensor or spinor \textit{T} to the value \textit{val}.}
	\usageline{SetGrading[\{$\textit{T}_1$,\,$\textit{T}_2$,\,\dots\},\,$\textit{grad}\to\textit{val}$]}
        {sets the grading of all the tensors $\textit{T}_i$.}
	\usageline*{SetGrading[\textit{T},\,\{$\textit{grad}_1\to\textit{val}_1$,\,$\textit{grad}_2\to\textit{val}_2$,\,\dots\}]}
        {sets all gradings $\textit{grad}_i$ to their respective values $\textit{val}_i$.}
\end{usagetable}

This function sets the grading of individual tensors or spinors, which must be defined first with \cmdref{sec_doc_gradings_defgrading}{DefGrading}. Setting the grading for a spinor automatically sets the same grading for the conjugate spinor; if the spinor and its conjugate have different gradings the one of the spinor must be set first. Gradings can also be set for covariant derivatives, for example an engineering dimension, and for the symmetrised covariant derivatives of the \textsc{xTras} package get multiplied by the number of indices.

\subsection{Left and right variational derivatives}
\label{sec_doc_varder}

For non-commuting objects, left and right variational derivatives can differ by signs, and the functions here handle non-commutative products correctly.

\ssubsection{LeftVarD}
\label{sec_doc_varder_leftvard}

\begin{usagetable}
	\usageline{LeftVarD[\textit{T}[\textit{inds}]][\textit{expr}]}
        {returns the left variational derivative of \textit{expr} with respect to the tensor or spinor field \textit{T}[\textit{inds}].}
	\usageline*{LeftVarD[\textit{T}[\textit{inds}],\,\textit{covd}][\textit{expr}]}
        {returns the left variational derivative of \textit{expr} with respect to the tensor or spinor field \textit{T}[\textit{inds}]. Integration by parts uses the covariant derivative \textit{covd} instead of the partial derivative \texttt{PD}.}
\end{usagetable}

The left variational or functional derivative $\delta_\text{L}$ is defined such that the Leibniz (product) rule holds from the left:
\begin{equation*}
\frac{\delta_\text{L}}{\delta T(x)} \left( A B \right) = \frac{\delta_\text{L} A}{\delta T(x)} B + (-1)^{\epsilon_T \epsilon_A} A \frac{\delta_\text{L} B}{\delta T(x)} \eqend{,}
\end{equation*}
where $\epsilon_T$ is the Grassmann parity of \textit{T}. \textit{expr} must be a scalar (or scalar density) and is implicitly assumed to be integrated over, such that integration by parts can be performed, as for the \texttt{VarD} command of \xact. If a tensor \textit{T} implicitly depends on another tensor \textit{U} and the functional derivative of $\textit{T}^{\textit{it}}$ with respect to $\textit{U}^{\textit{iu}}$ is given by the tensor $\textit{R}^{\textit{it}, \textit{iu}}$, this can be defined using
\begin{longcode}
T /: ImplicitTensorDepQ[T, U] = True;
Unprotect[LeftVarD];
LeftVarD[U[iu\_\_], cd\_][T[it\_\_], l\_, r\_] := CenterDot[l, R[it,iu], r];
Protect[LeftVarD];
\end{longcode}
For Majorana spinors where the spinor and its (charge) conjugate are not independent, the appropriate definitions are automatically set up by \cmdref{sec_doc_spinors_defspinor}{DefSpinor}, \cmdref{sec_doc_spinors_defevenspinor}{DefEvenSpinor}, and \cmdref{sec_doc_spinors_defoddspinor}{DefOddSpinor}.

\ssubsection{RightVarD}
\label{sec_doc_varder_rightvard}

\begin{usagetable}
	\usageline{RightVarD[\textit{T}[\textit{inds}]][\textit{expr}]}
        {returns the right variational derivative of \textit{expr} with respect to the tensor or spinor field \textit{T}[\textit{inds}].}
	\usageline*{RightVarD[\textit{T}[\textit{inds}],\,\textit{covd}][\textit{expr}]}
        {returns the right variational derivative of \textit{expr} with respect to the tensor or spinor field \textit{T}[\textit{inds}]. Integration by parts uses the covariant derivative \textit{covd} instead of the partial derivative \texttt{PD}.}
\end{usagetable}

The same comments as for the left variational derivative \cmdref{sec_doc_varder_leftvard}{LeftVarD} apply, except that for the right variational derivative $\delta_\text{R}$ the Leibniz rule holds from the right:
\begin{equation*}
\frac{\delta_\text{R}}{\delta T(x)} \left( A B \right) = A \frac{\delta_\text{R} B}{\delta T(x)} + (-1)^{\epsilon_T \epsilon_B} \frac{\delta_\text{R} A}{\delta T(x)} B \eqend{.}
\end{equation*}

\subsection{Contractions and Monomials}
\label{sec_doc_contraction}

These are functions to generate all possible terms with given tensor content or grading, extending functionality from the \textsc{xTras} package, and can be used to find actions and ans{\"a}tze for the computation of cohomologies.

\ssubsection{FindAllContractions}
\label{sec_doc_contraction_findallcontractions}

\begin{usagetable}
	\usageline{FindAllContractions[\textit{expr}]}
        {returns a list of all possible full contractions of \textit{expr} over its free indices. Extending \texttt{AllContractions}, this function also works if the tensors within \textit{expr} depend on more than one bundle. \textit{expr} can be given in pseudo index-free notation.}
	\usageline{FindAllContractions[\textit{expr},\,\{$a$,\,$b$,\,\dots\}]}
        {returns a list of all possible full contractions of \textit{expr} that have $a,b,\dots$ as free indices.}
	\usageline{FindAllContractions[\textit{expr},\,\{$a$,\,$b$,\,\dots\},\,\textit{sym}]}
        {returns a list of all possible full contractions of \textit{expr} with the symmetry \textit{sym} imposed on the free indices $a,b,\dots$}
	\usageline{SymmetrizeMethod}
        {is also an option for \texttt{FindAllContractions} that specifies a function to symmetrize the free indices. By default, it is \texttt{ImposeSymmetry}.}
	\usageline{AuxiliaryTensor}
        {is also an option for \texttt{FindAllContractions} that specifies the name of the auxiliary tensor used for the free indices.}
	\usageline*{Parallelization}
        {is also an option for \texttt{FindAllContractions} that specifies whether contractions should be calculated in parallel.}
\end{usagetable}

Given a product of tensors (a monomial), this function finds all possible contractions of them with a specified set of free indices. \texttt{FindAllContractions} extends the corresponding functionality of \texttt{AllContractions} from the \textsc{xTras} package (using some of the internal \textsc{xTras} routines) to also allow spinors and tensor with indices in inner bundles (for example, Lie-algebra valued ones). The resulting list can be used with \texttt{MakeAnsatz} from the \textsc{xTras} package to determine a complete ansatz with undetermined constants. If one is not interested in a single monomial, but needs all monomials containing a given set of tensors, the functions \cmdref{sec_doc_contraction_generatemonomialsbygrading}{GenerateMonomialsByGrading} or \cmdref{sec_doc_contraction_generatemonomialsbygrading}{GenerateMonomialsByGrading} are more useful.

\ssubsection{GenerateMonomials}
\label{sec_doc_contraction_generatemonomials}

\begin{usagetable}
	\usageline{GenerateMonomials[\textit{fields},\,\textit{invtens}]}
        {returns a list of all monomials that can be formed from the fields \textit{fields}, their covariant derivatives, and the invariant tensors \textit{invtens}.}
	\usageline{FreeIndices}
        {is an option for \texttt{GenerateMonomials} and \cmdref{sec_doc_contraction_generatemonomialsbygrading}{GenerateMonomialsByGrading} that specifies a list of free (uncontracted) indices that the returned monomials should have. By default, it is an empty list.}
	\usageline{Constraint}
        {is an option for \texttt{GenerateMonomials} and \cmdref{sec_doc_contraction_generatemonomialsbygrading}{GenerateMonomialsByGrading} that specifies the constraint function, a function returning \texttt{True} if its argument should be added to the list of monomials and \texttt{False} otherwise. By default, it is given by \texttt{True\&} (i.e., no constraint).}
	\usageline{Replacements}
        {is an option for \texttt{GenerateMonomials} and \cmdref{sec_doc_contraction_generatemonomialsbygrading}{GenerateMonomialsByGrading} that specifies a list of replacements to be made after contracting free indices in each monomial. By default, it is an empty list.}
	\usageline{MaxNumberOfFields}
        {is an option for \texttt{GenerateMonomials} and \cmdref{sec_doc_contraction_generatemonomialsbygrading}{GenerateMonomialsByGrading} that specifies the maximum number of fields of each type that can appear in an monomial. It can be an integer, or a list of integers specifying the maximum number for each field. By default, it is \texttt{Infinity} for \cmdref{sec_doc_contraction_generatemonomialsbygrading}{GenerateMonomialsByGrading} and 5 for \texttt{GenerateMonomials}.}
	\usageline{MaxNumberOfInvTensors}
        {is an option for \texttt{GenerateMonomials} and \cmdref{sec_doc_contraction_generatemonomialsbygrading}{GenerateMonomialsByGrading} that specifies the maximum number of invariant tensors of each type that can appear in an monomial. It can be an integer, or a list of integers specifying the maximum number for each tensor. By default, it is \texttt{Infinity} for \cmdref{sec_doc_contraction_generatemonomialsbygrading}{GenerateMonomialsByGrading} and 1 for \texttt{GenerateMonomials}.}
	\usageline{MaxNumberOfDerivatives}
        {is an option for \texttt{GenerateMonomials} and \cmdref{sec_doc_contraction_generatemonomialsbygrading}{GenerateMonomialsByGrading} that specifies the maximum number of derivatives that can be applied to a field. It can be an integer, or a list of integers specifying the maximum number for each field. By default, it is \texttt{Infinity} for \cmdref{sec_doc_contraction_generatemonomialsbygrading}{GenerateMonomialsByGrading} and 3 for \texttt{GenerateMonomials}.}
	\usageline*{IndexFree}
        {is also an option for \texttt{GenerateMonomials} and \cmdref{sec_doc_contraction_generatemonomialsbygrading}{GenerateMonomialsByGrading} that specifies if the list of monomials should returned in pseudo index-free notation. By default, it is \texttt{False}.}
\end{usagetable}

This function generates a list of all products of tensors (monomials) that can be obtained with a maximum number of a given tensor or spinor (\texttt{MaxNumberOfFields}), with a maximum number of derivatives acting on them (\texttt{MaxNumberOfDerivatives}) and a maximum number of invariant tensors (\texttt{MaxNumberOfInvTensors}), for example metrics, $\gamma$ matrices, or the tensors returned from \cmdref{sec_doc_inner_invarianttracetensor}{InvariantTraceTensor}. For an efficient generation, the \textsc{Multisets} package is used. The resulting list is then passed to \cmdref{sec_doc_contraction_findallcontractions}{FindAllContractions} to suitably contract free indices, except if the option \texttt{IndexFree} is given. The only difference between the \textit{fields} and \textit{invtens} arguments is that covariant derivatives are only generated on \textit{fields}, but otherwise arbitrary tensors or spinors can be used for both. A constraint function is applied on each monomial before it is added to the list, and can be used to filter terms (for example, to impose a certain grading: engineering dimension, ghost number, \dots). In many cases, one needs a list of monomials with a fixed grading, and \cmdref{sec_doc_contraction_generatemonomialsbygrading}{GenerateMonomialsByGrading} can be more efficient.

\ssubsection{GenerateMonomialsByGrading}
\label{sec_doc_contraction_generatemonomialsbygrading}

\begin{usagetable}
	\usageline{GenerateMonomialsByGrading[\textit{fields},\,\textit{invtens},\,$\textit{grad}\to\textit{n}$]}
        {returns a list of all monomials that can be formed from the fields \textit{fields}, their covariant derivatives, and the invariant tensors \textit{invtens}, restricted to the value \textit{n} for the grading \textit{grad}.}
	\usageline*{FilterGammaMatrices}
        {is an option for \texttt{GenerateMonomialsByGrading} that specifies whether spinor bilinears containing more than one generalized $\gamma$ matrix should be dropped. By default, it is \texttt{True}.}
\end{usagetable}

All options of the \cmdref{sec_doc_contraction_generatemonomials}{GenerateMonomials} command can also be used for \texttt{GenerateMonomials\-ByGrading}. This function uses a more efficient algorithm to generate monomials than \cmdref{sec_doc_contraction_generatemonomials}{GenerateMonomials}, based on integer partitions, and does not rely on the \textsc{Multisets} package. However, this leads to limitations (which are fulfilled in many appplications): the grading $\textit{grad}$ must be non-negative and rational for all tensors in \textit{fields} and \textit{invtens}. For fields with zero grading, a maximum number must be specified, as well as for derivatives whch have zero grading. Moreover, because of the way that fields with zero grading are incorporated, they cannot have derivatives acting on them. The \texttt{FilterGammaMatrices} option is used to filter terms of the form $\bar{\psi} \gamma^\mu \gamma^\nu \chi$, which can be expressed using $\gamma^{\mu\nu}$ and $g^{\mu\nu}$, and similar ones. (To obtain a complete ansatz, one of course needs to include all $\gamma$ matrices.)

\subsection{BRST operator and filtrations}
\label{sec_doc_brst}

These are functions to work with the BV/BRST formalism for gauge theories.

\ssubsection{DefOddDifferential}
\label{sec_doc_brst_defodddifferential}

\begin{usagetable}
	\usageline*{DefOddDifferential[\textit{brst}]}
        {defines a Grassmann-odd differential \textit{brst} that commutes with covariant derivatives. The BRST differential BRST is predefined.}
\end{usagetable}

This function defines linear fermionic differentials, such as the BRST differential. It can be used for filtrations of the predefined \cmdref{sec_doc_brst_brst}{BRST} differential using the \cmdref{sec_doc_brst_filtrate}{Filtrate} command.

\ssubsection{BRST}
\label{sec_doc_brst_brst}

\begin{usagetable}
	\usageline*{BRST[\textit{expr}]}
        {returns the BRST differential applied to \textit{expr}. BRST transformations can be defined using \texttt{F /: BRST[F[inds]] \textasciicircum:= G[inds]}.}
\end{usagetable}

The BRST differential is set up as a linear fermionic operator commuting with covariant derivatives. However, nilpotency is not (and cannot be) enforced, but depends on the transformations defined by the user.

\ssubsection{BRSTWeightInequalities}
\label{sec_doc_brst_brstweightinequalities}

\begin{usagetable}
	\usageline{BRSTWeightInequalities[\{$\textit{field}_1$,\,$\textit{field}_2$,\,\dots\},\,\textit{brst},\,\textit{weight}]}
        {returns a list with inequalities that the weight function \textit{weight} has to fulfill to be an admissible filtration for the BRST differential \textit{brst} applied to the fields $\textit{field}_1, \textit{field}_2, \dots$}
	\usageline*{BRSTWeightInequalities[\{$\textit{field}_1$,\,$\textit{field}_2$,\,\dots\},\,\textit{brst},\,\textit{weight},\texttt{False}]}
        {does not apply the Reduce function to the obtained system of inequalities before returning it.}
\end{usagetable}

Filtrations, where each field is assigned a (non-negative) weight, can be used to simplify the computation of BRST cohomologies by first computing the cohomology at lowest weight~\cite{piguetsorella}. To obtain a consistent filtration, the assignment of weights is restricted, and \texttt{BRSTWeightInequalities} gives a list of inequalities that the weight function \textit{weight} has to fulfill. If some assignments of weights are already determined, this function can be used to obtain conditions on the remaining assignments.

\ssubsection{FindBRSTWeights}
\label{sec_doc_brst_findbrstweights}

\begin{usagetable}
	\usageline*{FindBRSTWeights[\{$\textit{field}_1$,\,$\textit{field}_2$,\,\dots\},\,\textit{brst},\,\textit{maxweight}]}
        {returns a list of all weights admissible for filtrations of the BRST differential \textit{brst} applied to the fields $\textit{field}_1, \textit{field}_2, \dots$, with the maximum weight of each field restricted to be $\leq \textit{maxweight}$.}
\end{usagetable}

This function can be used to find all permissible weights, with \textit{maxweight} a positive integer. \texttt{FindBRSTWeights} uses a brute-force algorithm, such that for large \textit{maxweight}'s the computation may take quite long.

\ssubsection{CheckFiltration}
\label{sec_doc_brst_checkfiltration}

\begin{usagetable}
	\usageline{CheckFiltration[\{$\textit{field}_1$,\,$\textit{field}_2$,\,\dots\},\,\{$w_1$,\,$w_2$,\,\dots\},\,\textit{brst}]}
        {displays a table of the lowest-order terms of the BRST differential \textit{brst} applied to the fields $\textit{field}_1, \textit{field}_2, \dots$ and filtrated according to the weights $w_1, w_2, \dots$}
	\usageline{CheckFiltration[\{$\textit{field}_1 \to w_1$,\,$\textit{field}_2 \to w_2$,\,\dots\},\,\textit{brst}]}
        {displays a table of the lowest-order terms of the BRST differential \textit{brst} applied to the fields $\textit{field}_1, \textit{field}_2, \dots$ and filtrated according to the weights $w_1, w_2, \dots$}
    \usageline*{Display}
        {is also an option for \texttt{CheckFiltration} with values \texttt{IndexFree} and \texttt{Full} that specifies how the table entries should be displayed.}
\end{usagetable}

This function allows to quickly check how the filtrated BRST differential (at lowest weight) looks like for a given assignment of weights.

\ssubsection{Filtrate}
\label{sec_doc_brst_filtrate}

\begin{usagetable}
	\usageline{Filtrate[\{$\textit{field}_1$,\,$\textit{field}_2$,\,\dots\},\,\{$w_1$,\,$w_2$,\,\dots\},\,$\textit{brst} \to \textit{brst}_0$]}
        {defines rules for the operator $\textit{brst}_0$ such that it acts as the lowest-order terms of the BRST differential \textit{brst} applied to the fields $\textit{field}_1, \textit{field}_2, \dots$ and filtrated according to the weights $w_1, w_2, \dots$}
	\usageline*{Filtrate[\{$\textit{field}_1 \to w_1$,\,$\textit{field}_2 \to w_2$,\,\dots\},\,$\textit{brst} \to \textit{brst}_0$]}
        {defines rules for the operator $\textit{brst}_0$ such that it acts as the lowest-order terms of the BRST differential \textit{brst} applied to the fields $\textit{field}_1, \textit{field}_2, \dots$ and filtrated according to the weights $w_1, w_2, \dots$}
\end{usagetable}

This function performs the actual filtration. To remove transformations that are already defined for the lowest-weight BRST differential $\textit{brst}_0$, \cmdref{sec_doc_brst_removefiltration}{RemoveFiltration} can be used.

\ssubsection{RemoveFiltration}
\label{sec_doc_brst_removefiltration}

\begin{usagetable}
	\usageline*{RemoveFiltration[$\textit{brst}_0$]}
        {removes the rules defined for the differential $\textit{brst}_0$ defined by Filtrate.}
\end{usagetable}

This function clears all user-defined rules for the differential $\textit{brst}_0$.

\subsection{Cohomology}
\label{sec_doc_cohom}

\ssubsection{CohomologyFromAnsatz}
\label{sec_doc_cohom_cohomologyfromansatz}

\begin{usagetable}
	\usageline{CohomologyFromAnsatz[\textit{brst},\,\textit{ansatzcc},\,\textit{brstb},\,\textit{ansatzcb}]}
        {returns a list of representatives of elements of the cohomology Ker(\textit{brst})/Im(\textit{brstb}). For the BRST cohomology H(\textsf{s}), one needs to take $\textit{brst} = \textit{brstb} = \texttt{BRST}$. A list of possible elements (cocycles) must be given as \textit{ansatzcc}, and a list of possible exact elements (coboundaries) as \textit{ansatzcb}. Both lists could be calculated using \cmdref{sec_doc_contraction_generatemonomials}{GenerateMonomials} or \cmdref{sec_doc_contraction_generatemonomialsbygrading}{GenerateMonomialsByGrading}.}
	\usageline{CanonicalizeMethod}
        {is an option for \texttt{CohomologyFromAnsatz} and \cmdref{sec_doc_cohom_relativecohomologyfromansatz}{RelativeCohomologyFromAnsatz} that specifies the function applied to an expression after the differential has acted, to obtain a canonical form. By default, it is given by \texttt{CollectTensors[ReduceInvariantTraceTensors[ContractMetric[SymmetrizeCov\-Ds[Expand[\#]]]\&}.}
	\usageline*{SimplifyMethod}
        {is an option for \texttt{CohomologyFromAnsatz} and \cmdref{sec_doc_cohom_relativecohomologyfromansatz}{RelativeCohomologyFromAnsatz} that specifies the function applied to representatives of elements of the cohomology before they are returned. By default, it is given by \texttt{Identity} (i.e., no transformation).}
\end{usagetable}

This function computes (representatives of) the (co-)homology of a pair of differentials, which can be elements of a (co-)chain complex. It uses a simple straightforward algorithm, evaluating the differential \textit{brst} on each of the elements of the ansatz \textit{ansatzcc} and checking whether the result can be written as a linear combination of the boundary differential \textit{brstb} evaluated on the elements of the ansatz \textit{ansatzcb}. Since this function is completely agnostic about the nature of the differentials \textit{brst} and \textit{brstb}, it can also compute homologies. The choice of \texttt{CanonicalizeMethod} is extremely important, since to check whether an element of the ansatz \textit{ansatzcc} lies in the (co-)homology, it must be compared with zero (after application of the differential \textit{brst}), and checking whether some expression is zero is in general a hard problem. \texttt{CanonicalizeMethod} must thus be chosen such that a canonical form is obtained, which for example can include the use of Young projectors to take into account multiterm symmetries; see the \textsc{xTras} package documentation for examples involving the Riemann tensor, which satisfies the multiterm Bianchi identities. The default value for \texttt{CanonicalizeMethod} is suitable for theories without fermions, while for theories with fermions one must also use \cmdref{sec_doc_spingamma_joingammamatrices}{JoinGammaMatrices}, \cmdref{sec_doc_spingamma_epsilongammareduce}{EpsilonGammaReduce}, \cmdref{sec_doc_irreducible_irreduciblespindecompose}{IrreducibleSpinDecompose}, \cmdref{sec_doc_irreducible_irreduciblespinproject}{IrreducibleSpinProject}, ... \texttt{CohomologyFromAnsatz} can be used to determine local gauge-invariant operators.

\ssubsection{RelativeCohomologyFromAnsatz}
\label{sec_doc_cohom_relativecohomologyfromansatz}

\begin{usagetable}
	\usageline*{RelativeCohomologyFromAnsatz[\textit{brst},\,\textit{ansatzcc},\,\textit{d},\,\textit{ansatzd},\,\textit{brstb},\,\textit{ansatzcb},\,\textit{db},\,\textit{ansatzdb}]}
        {returns a list of representatives of elements of the relative cohomology Ker(\textit{brst}$\vert$\textit{d})/Im(\textit{brstb}$\vert$\textit{db}). For the relative BRST cohomology H(\textsf{s}$\vert$\textsf{d}), one needs to take $\textit{brst} = \textit{brstb} = \texttt{BRST}$, and $\textit{d} = \textit{db} = \texttt{CD[-$\mu$]}$. A list of possible elements (cocycles) must be given as \textit{ansatzcc} and \textit{ansatzd}, and a list of possible exact elements (coboundaries) as \textit{ansatzcb} and \textit{ansatzdb}. All lists could be calculated using \cmdref{sec_doc_contraction_generatemonomials}{GenerateMonomials} or \cmdref{sec_doc_contraction_generatemonomialsbygrading}{GenerateMonomialsByGrading}.}
\end{usagetable}

This is the analogue of \cmdref{sec_doc_cohom_cohomologyfromansatz}{CohomologyFromAnsatz} for relative (co-)homologies, and can be used to determine possible anomalies and invariant actions, which both are only defined up to surface terms (total derivatives).

\begin{acknowledgments}
This work has been funded by the Deutsche Forschungsgemeinschaft (DFG, German Research Foundation) --- project nos. 415803368 and 406116891 within the Research Training Group RTG 2522/1. I thank Thomas B{\"a}ckdahl, Camillo Imbimbo, Ioannis Papadimitriou and Tung Tran for comments.
\end{acknowledgments}

\appendix

\section{Decomposition of tensor products}
\label{app_decomp}

Here I give the full list of decompositions used by \fix, valid for commuting and anticommuting Majorana spinors in four dimensions. These decompositions were obtained following the algorithm of~\cite{dauriafremainaregge1982}, to which the reader is referred for details.

\subsection{Commuting spinors without an inner bundle index}

For two commuting spinors, Majorana flip relations show that all bilinears except the ones with one or two $\gamma$ matrices vanish, such that 
\begin{equation}
\bar{\psi}^A \psi_B = \frac{1}{4} \gamma^\mu{}_B{}^A \Psi^{(2,\repdim{4})}_\mu - \frac{1}{8} \gamma^{\mu\nu}{}_B{}^A \Psi^{(2,\repdim{6})}_{\mu\nu}
\end{equation}
with the invariants
\begin{equations}
\Psi^{(2,\repdim{4})}_\mu &\equiv \bar{\psi}^C \gamma_{\mu C}{}^D \psi_D \eqend{,} \\
\Psi^{(2,\repdim{6})}_{\mu\nu} &\equiv \bar{\psi}^C \gamma_{\mu\nu C}{}^D \psi_D \eqend{.}
\end{equations}
For three spinors, it follows that
\begin{equation}
\psi_A \bar{\psi}^B \psi_C = - \frac{1}{8} \gamma^{\mu\nu}{}_{(A}{}^B \Psi^{(3,\repdim{8})}_{\mu\nu C)} + \frac{1}{8} \left( 2 \gamma^\nu{}_{(A}{}^B \delta_{C)}^D + \gamma^{\mu\nu}{}_{(A}{}^B \gamma_{\mu C)}{}^D \right) \Psi^{(3,\repdim{12})}_{\nu D}
\end{equation}
with the invariants
\begin{equations}
\Psi^{(3,\repdim{12})}_{\mu A} &\equiv \frac{3}{4} \Psi^{(2,\repdim{4})}_\mu \psi_A - \frac{1}{4} \gamma_\mu{}^\nu{}_A{}^B \Psi^{(2,\repdim{4})}_\nu \psi_B \eqend{,} \\
\Psi^{(3,\repdim{8})}_{\mu\nu A} &\equiv \frac{1}{3} \Psi^{(2,\repdim{6})}_{\mu\nu} \psi_A - \frac{1}{6} \gamma_{\mu\nu}{}^{\rho\sigma}{}_A{}^B \Psi^{(2,\repdim{6})}_{\rho\sigma} \psi_B + \frac{1}{3} \gamma_{[\mu}{}^\rho{}_A{}^B \Psi^{(2,\repdim{6})}_{\nu]\rho} \psi_B \eqend{.}
\end{equations}
For four spinors one obtains
\begin{splitequation}
\psi_A \bar{\psi}^B \psi_C \bar{\psi}^D &= \frac{1}{16} \left( 2 \gamma^\mu{}_{(A}{}^B \gamma^{\nu)}{}_C{}^{D} + \gamma^{\mu\rho}{}_{(A}{}^B \gamma^\nu{}_\rho{}_{C)}{}^D \right) \Psi^{(4,\repdim{9})}_{\mu\nu} \\
&\quad- \frac{1}{6} \gamma^\mu{}_{(A}{}^B \gamma^{\rho\sigma}{}_{C)}{}^D \Psi^{(4,\repdim{16})}_{\mu\rho\sigma} + \frac{1}{32} \gamma^{\mu\nu}{}_{(A}{}^B \gamma^{\rho\sigma}{}_{C)}{}^D \Psi^{(4,\repdim{10})}_{\mu\nu\rho\sigma}
\end{splitequation}
with the invariants
\begin{equations}
\Psi^{(4,\repdim{9})}_{\mu\nu} &\equiv \Psi^{(2,\repdim{4})}_\mu \Psi^{(2,\repdim{4})}_\nu \eqend{,} \\
\Psi^{(4,\repdim{16})}_{\mu\rho\sigma} &\equiv \Psi^{(2,\repdim{4})}_\mu \Psi^{(2,\repdim{6})}_{\rho\sigma} \eqend{,} \\
\Psi^{(4,\repdim{10})}_{\mu\nu\rho\sigma} &\equiv \Psi^{(2,\repdim{6})}_{\mu\nu} \Psi^{(2,\repdim{6})}_{\rho\sigma} - 2 \Psi^{(2,\repdim{6})}_{\alpha[\mu} g_{\nu][\rho} \Psi^{(2,\repdim{6})}_{\sigma]}{}^\alpha + \frac{1}{3} g_{\mu[\rho} g_{\sigma]\nu} \Psi^{(2,\repdim{6})}_{\alpha\beta} \Psi^{(2,\repdim{6})}{}^{\alpha\beta} \eqend{,}
\end{equations}
where the relations (obtained from Fierz rearrangements)
\begin{equations}
\Psi^{(2,\repdim{4})}_{[\mu} \Psi^{(2,\repdim{6})}_{\nu]\rho} &= - \frac{1}{2} \Psi^{(2,\repdim{4})}_\rho \Psi^{(2,\repdim{6})}_{\mu\nu} \eqend{,} \\
\Psi^{(2,\repdim{6})}_{\mu[\alpha} \Psi^{(2,\repdim{6})}_{\beta]\nu} &= - \frac{1}{2} \Psi^{(2,\repdim{6})}_{\mu\nu} \Psi^{(2,\repdim{6})}_{\alpha\beta}
\end{equations}
are needed to derive the above expressions.

\subsection{Commuting spinors with an inner bundle index}

For spinors with an additional inner bundle index (Lie-algebra valued Majorana spinors), the Majorana flip relations instead determine whether the symmetric or antisymmetric representation of the symmetric group (for permutations of the Lie algebra index) contributes. For two spinors, it follows that
\begin{splitequation}
\bar{\psi}^{aA} \psi^b_B &= \frac{1}{4} \delta_B^A \Psi^{(2,\repdim{d(d-1)/2})}{}^{ab} + \frac{1}{4} \gamma_{\mu B}{}^A \Psi^{(2,\repdim{2d(d+1)})}{}^{ab\mu} - \frac{1}{8} \gamma_{\mu\nu B}{}^A \Psi^{(2,\repdim{3d(d+1)})}{}^{ab\mu\nu} \\
&\quad- \frac{1}{24} \epsilon^{\mu\nu\rho\sigma} \gamma_{\mu\nu\rho}{}_B{}^A \Psi^{(2,\repdim{2d(d-1)})}{}^{ab}_\sigma - \frac{1}{96} \epsilon^{\mu\nu\rho\sigma} \gamma_{\mu\nu\rho\sigma}{}_B{}^A \Psi^{(2,\repdim{d(d-1)/2}')}{}^{ab}
\end{splitequation}
with the invariants
\begin{equations}
\Psi^{(2,\repdim{d(d-1)/2})}{}^{ab} &\equiv \bar{\psi}^{[aC} \psi^{b]}_C \eqend{,} \\
\Psi^{(2,\repdim{2d(d+1)})}{}^{ab}_\mu &\equiv \bar{\psi}^{(aC} \gamma_{\mu C}{}^D \psi^{b)}_D \eqend{,} \\
\Psi^{(2,\repdim{3d(d+1)})}{}^{ab}_{\mu\nu} &\equiv \bar{\psi}^{(aC} \gamma_{\mu\nu C}{}^D \psi^{b)}_D \eqend{,} \\
\Psi^{(2,\repdim{2d(d-1)})}{}^{ab}_\mu &\equiv \frac{1}{6} \epsilon_{\mu\nu\rho\sigma} \bar{\psi}^{[aC} \gamma^{\nu\rho\sigma}{}_C{}^D \psi^{b]}_D \eqend{,} \\
\Psi^{(2,\repdim{d(d-1)/2}')}{}^{ab} &\equiv \frac{1}{24} \epsilon^{\mu\nu\rho\sigma} \bar{\psi}^{[aC} \gamma_{\mu\nu\rho\sigma C}{}^D \psi^{b]}_D \eqend{,}
\end{equations}
whose dimension now depends on the dimension $d$ of the Lie algebra representation. In principle, the fully irreducible tensors are traceless also on the Lie algebra indices (and one thus should subtract the traces for the symmetric tensors $\Psi^{(2,\repdim{d(d+1)/2})}{}^{ab}$, \dots), but this does not seem to be useful for practical applications. Since a Lie-algebra valued Majorana spinor has $4d$ degrees of freedom, the counting of degrees of freedom also works out:
\begin{equation}
\frac{4d(4d+1)}{2} = \frac{d(d-1)}{2} + 2d(d+1) + 3d(d+1) + 2d(d-1) + \frac{d(d-1)}{2} \eqend{.}
\end{equation}
In the case $d = 1$, one naturally recovers the previous result.

For three spinors, the decomposition is already quite complicated:
\begin{splitequation}
\psi^a_A \bar{\psi}^{bB} \psi^c_C &= - \frac{1}{96} \left( 24 \delta_{[A}^B \delta_{C]}^D + 4 \gamma_{\mu\nu\rho}{}_{[A}{}^B \gamma^{\mu\nu\rho}{}_{C]}{}^D - \gamma_{\mu\nu\rho\sigma}{}_{[A}{}^B \gamma^{\mu\nu\rho\sigma}{}_{C]}{}^D \right) \\
&\qquad\quad\times \Psi^{(3,\repdim{2d(d-1)(d-2)/3})}{}\ytableausmall{1,2,3}{}^{abc}_D \\
&\quad- \frac{1}{192} \left( 24 \delta_A^B \delta_C^D + 6 \gamma^\mu{}_C{}^B \gamma_{\mu A}{}^D - 2 \gamma^{\mu\nu}{}_C{}^B \gamma_{\mu\nu A}{}^D - \gamma^{\mu\nu\rho}{}_A{}^B \gamma_{\mu\nu\rho C}{}^D \right) \\
&\qquad\quad\times \Psi^{(3,\repdim{4d(d+1)(d-1)/3})}{}\ytableausmall{13,2}{}^{abc}_D \\
&\quad- \frac{1}{192} \left( 24 \delta_C^B \delta_A^D + 6 \gamma^\mu{}_A{}^B \gamma_{\mu C}{}^D - 2 \gamma^{\mu\nu}{}_A{}^B \gamma_{\mu\nu C}{}^D - \gamma^{\mu\nu\rho}{}_C{}^B \gamma_{\mu\nu\rho A}{}^D \right) \\
&\qquad\quad\times \Psi^{(3,\repdim{4d(d+1)(d-1)/3})}{}\ytableausmall{13,2}{}^{cba}_D \\
&\quad- \frac{1}{48} \left( 3 \gamma^\mu{}_{[A}{}^B \gamma_{\mu C]}{}^D - \gamma^{\mu\nu}{}_{[A}{}^B \gamma_{\mu\nu C]}{}^D \right) \Psi^{(3,\repdim{4d(d+1)(d-1)/3})}{}\ytableausmall{13,2}{}^{cab}_D \\
&\quad- \frac{1}{192} \left( 6 \gamma^\mu{}_C{}^B \gamma_{\mu A}{}^D + 2 \gamma^{\mu\nu}{}_C{}^B \gamma_{\mu\nu A}{}^D - \gamma^{\mu\nu\rho}{}_A{}^B \gamma_{\mu\nu\rho C}{}^D - \gamma^{\mu\nu\rho\sigma}{}_A{}^B \gamma_{\mu\nu\rho\sigma C}{}^D \right) \\
&\qquad\quad\times \Psi^{(3,\repdim{4d(d+1)(d-1)/3}*)}{}\ytableausmall{13,2}{}^{abc}_D \\
&\quad- \frac{1}{192} \left( 6 \gamma^\mu{}_A{}^B \gamma_{\mu C}{}^D + 2 \gamma^{\mu\nu}{}_A{}^B \gamma_{\mu\nu C}{}^D - \gamma^{\mu\nu\rho}{}_C{}^B \gamma_{\mu\nu\rho A}{}^D - \gamma^{\mu\nu\rho\sigma}{}_C{}^B \gamma_{\mu\nu\rho\sigma A}{}^D \right) \\
&\qquad\quad\times \Psi^{(3,\repdim{4d(d+1)(d-1)/3}*)}{}\ytableausmall{13,2}{}^{cba}_D \\
&\quad- \frac{1}{48} \left( 3 \gamma^\mu{}_{[A}{}^B \gamma_{\mu C]}{}^D + \gamma^{\mu\nu}{}_{[A}{}^B \gamma_{\mu\nu C]}{}^D \right) \Psi^{(3,\repdim{4d(d+1)(d-1)/3}*)}{}\ytableausmall{13,2}{}^{cab}_D \\
&\quad+ \frac{1}{8} \left( 2 \gamma^\nu{}_{(A}{}^B \delta_{C)}^D + \gamma^{\mu\nu}{}_{(A}{}^B \gamma_{\mu C)}{}^D \right) \Psi^{(3,\repdim{2d(d+1)(d+2)})}{}\ytableausmall{123}{}^{abc}_{\nu D} \\
&\quad+ \frac{1}{16} \left( 2 \gamma^\nu{}_A{}^B \delta_C^D - 2 \gamma^{\mu\nu}{}_A{}^B \gamma_{\mu C}{}^D + \gamma^{\mu\nu\rho}{}_C{}^B \gamma_{\mu\rho A}{}^D \right) \Psi^{(3,\repdim{4d(d+1)(d-1)})}{}\ytableausmall{12,3}{}^{abc}_{\nu D} \\
&\quad+ \frac{1}{16} \left( 2 \gamma^\nu{}_C{}^B \delta_A^D - 2 \gamma^{\mu\nu}{}_C{}^B \gamma_{\mu A}{}^D + \gamma^{\mu\nu\rho}{}_A{}^B \gamma_{\mu\rho C}{}^D \right) \Psi^{(3,\repdim{4d(d+1)(d-1)})}{}\ytableausmall{12,3}{}^{cba}_{\nu D} \\
&\quad+ \frac{1}{8} \gamma^{\mu\nu\rho}{}_{(A}{}^B \gamma_{\mu\nu C)}{}^D \Psi^{(3,\repdim{4d(d+1)(d-1)})}{}\ytableausmall{12,3}{}^{cab}_{\rho D} \\
&\quad- \frac{1}{8} \gamma_{\mu\nu (A}{}^B \Psi^{(3,\repdim{4d(d+1)(d+2)/3})}{}\ytableausmall{123}{}^{abc}_{\mu\nu C)} \eqend{.}
\end{splitequation}
The symmetry of the invariants with respect to permutations of the Lie algebra indices is now more complicated than total (anti-)symmetry, and described in the standard way by Young tableaux.\footnote{See for example~\cite{xact7} or the very readable lecture notes~\cite{alcockzeilinger} and references therein.} The invariants are given by
\begin{equations}
\Psi^{(3,\repdim{2d(d-1)(d-2)/3})}{}\ytableausmall{1,2,3}{}^{abc}_A &\equiv \Psi^{(2,\repdim{d(d-1)/2})}{}^{[ab} \psi^{c]}_A \eqend{,} \\
\Psi^{(3,\repdim{4d(d+1)(d-1)/3})}{}\ytableausmall{13,2}{}^{abc}_A &\equiv \frac{2}{3} \Psi^{(2,\repdim{d(d-1)/2})}{}^{ab} \psi^c_A - \frac{2}{3} \Psi^{(2,\repdim{d(d-1)/2})}{}^{c[a} \psi^{b]}_A \eqend{,} \\
\Psi^{(3,\repdim{4d(d+1)(d-1)/3}*)}{}\ytableausmall{13,2}{}^{abc}_A &\equiv \frac{2 \mathi}{3} \gamma_*{}_A{}^B \left[ \Psi^{(2,\repdim{d(d-1)/2}')}{}^{ab} \psi^c_B - \Psi^{(2,\repdim{d(d-1)/2}')}{}^{c[a} \psi^{b]}_B \right] \eqend{,} \\
\Psi^{(3,\repdim{2d(d+1)(d+2)})}{}\ytableausmall{123}{}^{abc}_{\mu A} &\equiv \frac{3}{4} \Psi^{(2,\repdim{2d(d+1)})}{}^{(ab}_\mu \psi^{c)}_A - \frac{1}{4} \gamma_\mu{}^\nu{}_A{}^B \Psi^{(2,\repdim{2d(d+1)})}{}^{(ab}_\nu \psi^{c)}_B \eqend{,} \\
\begin{split}
\Psi^{(3,\repdim{4d(d+1)(d-1)})}{}\ytableausmall{12,3}{}^{abc}_{\mu A} &\equiv \frac{1}{2} \Psi^{(2,\repdim{2d(d+1)})}{}^{ab}_\mu \psi^c_A - \frac{1}{6} \gamma_\mu{}^\nu{}_A{}^B \Psi^{(2,\repdim{2d(d+1)})}{}^{ab}_\nu \psi^{c}_B \\
&\quad- \frac{1}{2} \Psi^{(2,\repdim{2d(d+1)})}{}^{c(a}_\mu \psi^{b)}_A + \frac{1}{6} \gamma_\mu{}^\nu{}_A{}^B \Psi^{(2,\repdim{2d(d+1)})}{}^{c(a}_\nu \psi^{b)}_B \eqend{,}
\end{split} \\
\begin{split}
\Psi^{(3,\repdim{4d(d+1)(d+2)/3})}{}\ytableausmall{123}{}^{abc}_{\mu\nu A} &\equiv \frac{1}{3} \Psi^{(2,\repdim{3d(d+1)})}{}^{(ab}_{\mu\nu} \psi^{c)}_A - \frac{1}{3} \gamma^\rho{}_{[\mu A}{}^B \Psi^{(2,\repdim{3d(d+1)})}{}^{(ab}_{\nu]\rho} \psi^{c)}_B \\
&\quad- \frac{1}{6} \gamma_{\mu\nu}{}^{\rho\sigma}{}_A{}^B \Psi^{(2,\repdim{3d(d+1)})}{}^{(ab}_{\rho\sigma} \psi^{c)}_B \eqend{,}
\end{split}
\end{equations}
and the counting of degrees of freedom also works out:
\begin{splitequation}
\frac{4d(4d+1)(4d+2)}{2\cdot 3} &= \frac{2d(d-1)(d-2)}{3} + \frac{4d(d+1)(d-1)}{3} + \frac{4d(d+1)(d-1)}{3} \\
&\quad+ 2d(d+1)(d+2) + 4d(d+1)(d-1) + \frac{4d(d+1)(d+2)}{3} \eqend{.}
\end{splitequation}

These decompositions assume that the Lie algebra and the chosen representation are sufficiently generic (with high enough dimension $d$), since otherwise it is possible that some of the invariant tensors vanish. For example, if $d = 2$ one obviously has
\begin{equation}
\Psi^{(2,\repdim{2d(d-1)/2})}{}^{ab} = \frac{1}{2} \epsilon^{ab} \epsilon_{cd} \Psi^{(2,\repdim{2d(d-1)/2})}{}^{cd}
\end{equation}
with the antisymmetric tensor $\epsilon_{ab}$ normalised as $\epsilon_{12} = 1$, and total antisymmetrisation over more than two indices leads to a vanishing result: $\Psi^{(3,\repdim{2d(d-1)(d-2)/3})}{}\ytableausmall{1,2,3}{}^{abc}_A = 0$. If applicable, such relations must be specified by the user.

\subsection{Anticommuting spinors without an inner bundle index}

For two anticommuting spinors, Majorana flip relations show that the bilinears with one or two $\gamma$ matrices vanish, and one obtains
\begin{equation}
\bar{\psi}^A \psi_B = \frac{1}{4} \delta_B^A \Psi^{(2,\repdim{1})} - \frac{1}{24} \gamma^{\mu\nu\rho}{}_B{}^A \epsilon_{\mu\nu\rho\sigma} \Psi^{(2,\repdim{4})}{}^\sigma - \frac{1}{96} \gamma^{\mu\nu\rho\sigma}{}_B{}^A \epsilon_{\mu\nu\rho\sigma} \Psi^{(2,\repdim{1}')}
\end{equation}
with the invariants
\begin{equations}
\Psi^{(2,\repdim{1})} &\equiv \bar{\psi}^C \psi_C \eqend{,} \\
\Psi^{(2,\repdim{4})}_\mu &\equiv \frac{1}{6} \epsilon_{\mu\nu\rho\sigma} \bar{\psi}^C \gamma^{\nu\rho\sigma}{}_C{}^D \psi_D \eqend{,} \\
\Psi^{(2,\repdim{1}')} &\equiv \frac{1}{24} \epsilon^{\mu\nu\rho\sigma} \bar{\psi}^C \gamma_{\mu\nu\rho\sigma C}{}^D \psi_D \eqend{.}
\end{equations}
For three anticommuting spinors, one has
\begin{equation}
\psi_A \bar{\psi}^B \psi_C = - \frac{1}{4} \left[ \delta_{[A}^B \delta_{C]}^D + \frac{1}{6} \gamma_{\mu\nu\rho [A}{}^B \gamma^{\mu\nu\rho}{}_{C]}{}^D - \frac{1}{24} \gamma_{\mu\nu\rho\sigma [A}{}^B \gamma^{\mu\nu\rho\sigma}_{C]}{}^D \right] \Psi^{(3,\repdim{4})}_D \eqend{,}
\end{equation}
with the single independent invariant
\begin{equation}
\Psi^{(3,\repdim{4})}_D \equiv \psi_A \Psi^{(2,\repdim{1})} \eqend{,}
\end{equation}
which agrees with the counting of degrees of freedom $4 (4-1) (4-2)/(2\cdot 3) = 4$. Lastly, for four spinors one has the decomposition
\begin{equation}
\psi_A \bar{\psi}^B \psi_C \bar{\psi}^D = - \frac{1}{16} \left[ \delta_{[A}^B \delta_{C]}^D + \frac{1}{6} \gamma_{\mu\nu\rho [A}{}^B \gamma^{\mu\nu\rho}{}_{C]}{}^D - \frac{1}{24} \gamma_{\mu\nu\rho\sigma [A}{}^B \gamma^{\mu\nu\rho\sigma}_{C]}{}^D \right] \Psi^{(4,\repdim{1})}
\end{equation}
with the invariant
\begin{equation}
\Psi^{(4,\repdim{1})} = \left( \bar{\psi}^A \psi_A \right)^2 \eqend{,}
\end{equation}
and the product of five or more anticommuting spinors vanishes since at least two components will be equal.

\subsection{Anticommuting spinors with an inner bundle index}

For Lie-algebra valued Majorana spinors, the Majorana flip relations again determine whether the symmetric or antisymmetric representation of the symmetric group (for permutations of the Lie algebra index) contributes. For two spinors, it follows that
\begin{splitequation}
\bar{\psi}^{aA} \psi^b_B &= \frac{1}{4} \delta_B^A \Psi^{(2,\repdim{d(d+1)/2})}{}^{ab} + \frac{1}{4} \gamma_{\mu B}{}^A \Psi^{(2,\repdim{2d(d-1)})}{}^{ab\mu} - \frac{1}{8} \gamma_{\mu\nu B}{}^A \Psi^{(2,\repdim{3d(d-1)})}{}^{ab\mu\nu} \\
&\quad- \frac{1}{24} \epsilon^{\mu\nu\rho\sigma} \gamma_{\mu\nu\rho}{}_B{}^A \Psi^{(2,\repdim{2d(d+1)})}{}^{ab}_\sigma - \frac{1}{96} \epsilon^{\mu\nu\rho\sigma} \gamma_{\mu\nu\rho\sigma}{}_B{}^A \Psi^{(2,\repdim{d(d+1)/2}')}{}^{ab}
\end{splitequation}
with the invariants
\begin{equations}
\Psi^{(2,\repdim{d(d+1)/2})}{}^{ab} &\equiv \bar{\psi}^{(aC} \psi^{b)}_C \eqend{,} \\
\Psi^{(2,\repdim{2d(d-1)})}{}^{ab}_\mu &\equiv \bar{\psi}^{[aC} \gamma_{\mu C}{}^D \psi^{b]}_D \eqend{,} \\
\Psi^{(2,\repdim{3d(d-1)})}{}^{ab}_{\mu\nu} &\equiv \bar{\psi}^{[aC} \gamma_{\mu\nu C}{}^D \psi^{b]}_D \eqend{,} \\
\Psi^{(2,\repdim{2d(d+1)})}{}^{ab}_\mu &\equiv \frac{1}{6} \epsilon_{\mu\alpha\beta\gamma} \bar{\psi}^{(aC} \gamma^{\alpha\beta\gamma}{}_C{}^D \psi^{b)}_D \eqend{,} \\
\Psi^{(2,\repdim{d(d+1)/2}')}{}^{ab} &\equiv \frac{1}{24} \epsilon^{\mu\nu\rho\sigma} \bar{\psi}^{(aC} \gamma_{\mu\nu\rho\sigma C}{}^D \psi^{b)}_D \eqend{,}
\end{equations}
whose dimension again depends on the dimension $d$ of the Lie algebra representation, and the counting of degrees of freedom works out to be
\begin{equation}
\frac{4d(4d-1)}{2} = \frac{d(d+1)}{2} + 2d(d-1) + 3d(d-1) + 2d(d+1) + \frac{d(d+1)}{2} \eqend{.}
\end{equation}

For three anticommuting spinors, one obtains again a quite complicated result
\begin{splitequation}
\psi^a_A \bar{\psi}^{bB} \psi^c_C &= - \frac{1}{96} \left( 24 \delta_{[A}^B \delta_{C]}^D + 4 \gamma_{\mu\nu\rho [A}{}^B \gamma^{\mu\nu\rho}{}_{C]}{}^D - \gamma_{\mu\nu\rho\sigma [A}{}^B \gamma^{\mu\nu\rho\sigma}_{C]}{}^D \right) \\
&\qquad\quad\times \Psi^{(3,\repdim{2d(d+1)(d+2)/3})}{}\ytableausmall{123}{}^{abc}_D \\
&\quad- \frac{1}{192} \left( 24 \delta_A^B \delta_C^D + 6 \gamma^\mu{}_C{}^B \gamma_{\mu A}{}^D - 2 \gamma^{\mu\nu}{}_C{}^B \gamma_{\mu\nu A}{}^D - \gamma^{\mu\nu\rho}{}_A{}^B \gamma_{\mu\nu\rho C}{}^D \right) \\
&\qquad\quad\times \Psi^{(3,\repdim{4d(d+1)(d-1)/3})}{}\ytableausmall{12,3}{}^{abc}_D \\
&\quad+ \frac{1}{192} \left( 24 \delta_C^B \delta_A^D + 6 \gamma^\mu{}_A{}^B \gamma_{\mu C}{}^D - 2 \gamma^{\mu\nu}{}_A{}^B \gamma_{\mu\nu C}{}^D - \gamma^{\mu\nu\rho}{}_C{}^B \gamma_{\mu\nu\rho A}{}^D \right) \\
&\qquad\quad\times \Psi^{(3,\repdim{4d(d+1)(d-1)/3})}{}\ytableausmall{12,3}{}^{cba}_D \\
&\quad- \frac{1}{48} \left( 3 \gamma^\mu{}_{[A}{}^B \gamma_{\mu C]}{}^D - \gamma^{\mu\nu}{}_{[A}{}^B \gamma_{\mu\nu C]}{}^D \right) \Psi^{(3,\repdim{4d(d+1)(d-1)/3})}{}\ytableausmall{12,3}{}^{cab}_D \\
&\quad- \frac{1}{192} \left( 6 \gamma_{\mu C}{}^B \gamma^\mu{}_A{}^D + 2 \gamma_{\mu\nu C}{}^B \gamma^{\mu\nu}{}_A{}^D - \gamma_{\mu\nu\rho}{}_A{}^B \gamma^{\mu\nu\rho}{}_C{}^D - \gamma_{\mu\nu\rho\sigma}{}_A{}^B \gamma^{\mu\nu\rho\sigma}{}_C{}^D \right) \\
&\qquad\quad\times \Psi^{(3,\repdim{4d(d+1)(d-1)/3}*)}{}\ytableausmall{12,3}{}^{abc}_D \\
&\quad+ \frac{1}{192} \left( 6 \gamma_{\mu A}{}^B \gamma^\mu{}_C{}^D + 2 \gamma_{\mu\nu A}{}^B \gamma^{\mu\nu}{}_C{}^D - \gamma_{\mu\nu\rho}{}_C{}^B \gamma^{\mu\nu\rho}{}_A{}^D - \gamma_{\mu\nu\rho\sigma}{}_C{}^B \gamma^{\mu\nu\rho\sigma}{}_A{}^D \right) \\
&\qquad\quad\times \Psi^{(3,\repdim{4d(d+1)(d-1)/3}*)}{}\ytableausmall{12,3}{}^{cba}_D \\
&\quad- \frac{1}{48} \left( 3 \gamma_{\mu [A}{}^B \gamma^\mu{}_{C]}{}^D + \gamma_{\mu\nu [A}{}^B \gamma^{\mu\nu}{}_{C]}{}^D \right) \Psi^{(3,\repdim{4d(d+1)(d-1)/3}*)}{}\ytableausmall{12,3}{}^{cab}_D \\
&\quad+ \frac{1}{8} \left( 2 \gamma^\nu{}_{(A}{}^B \delta_{C)}^D + \gamma^{\mu\nu}{}_{(A}{}^B \gamma_{\mu C)}{}^D \right) \Psi^{(3,\repdim{2d(d-1)(d-2)})}{}\ytableausmall{1,2,3}{}^{abc}_{\nu D} \\
&\quad+ \frac{1}{16} \left( 2 \gamma^\nu{}_A{}^B \delta_C^D - 2 \gamma^{\mu\nu}{}_A{}^B \gamma_{\mu C}{}^D + \gamma^{\mu\nu\rho}{}_C{}^B \gamma_{\mu\rho A}{}^D \right) \Psi^{(3,\repdim{4d(d+1)(d-1)})}{}\ytableausmall{13,2}{}^{abc}_{\nu D} \\
&\quad- \frac{1}{16} \left( 2 \gamma^\nu{}_C{}^B \delta_A^D - 2 \gamma^{\mu\nu}{}_C{}^B \gamma_{\mu A}{}^D + \gamma^{\mu\nu\rho}{}_A{}^B \gamma_{\mu\rho C}{}^D \right) \Psi^{(3,\repdim{4d(d+1)(d-1)})}{}\ytableausmall{13,2}{}^{cba}_{\nu D} \\
&\quad+ \frac{1}{8} \gamma^{\mu\nu\rho}{}_{(A}{}^B \gamma_{\mu\nu C)}{}^D \Psi^{(3,\repdim{4d(d+1)(d-1)})}{}\ytableausmall{13,2}{}^{cab}_{\rho D} \\
&\quad- \frac{1}{8} \gamma^{\mu\nu}{}_{(A}{}^B \Psi^{(3,\repdim{4d(d-1)(d-2)/3})}{}\ytableausmall{1,2,3}{}^{abc}_{\mu\nu C)}
\end{splitequation}
with the invariants
\begin{equations}
\Psi^{(3,\repdim{2d(d+1)(d+2)/3})}{}\ytableausmall{123}{}^{abc}_A &\equiv \Psi^{(2,\repdim{d(d+1)/2})}{}^{(ab} \psi^{c)}_A \eqend{,} \\
\Psi^{(3,\repdim{4d(d+1)(d-1)/3})}{}\ytableausmall{12,3}{}^{abc}_A &\equiv \frac{2}{3} \Psi^{(2,\repdim{d(d+1)/2})}{}^{ab} \psi^c_A - \frac{2}{3} \Psi^{(2,\repdim{d(d+1)/2})}{}^{c(a} \psi^{b)}_A \eqend{,} \\
\Psi^{(3,\repdim{4d(d+1)(d-1)/3}*)}{}\ytableausmall{12,3}{}^{abc}_A &\equiv \frac{2 \mathi}{3} \gamma_*{}_A{}^B \left[ \Psi^{(2,\repdim{d(d+1)/2}')}{}^{ab} \psi^c_B - \Psi^{(2,\repdim{d(d+1)/2}')}{}^{c(a} \psi^{b)}_B \right] \eqend{,} \\
\Psi^{(3,\repdim{2d(d-1)(d-2)})}{}\ytableausmall{1,2,3}{}^{abc}_{\mu A} &\equiv \frac{3}{4} \Psi^{(2,\repdim{2d(d-1)})}{}^{[ab}_\mu \psi^{c]}_A - \frac{1}{4} \gamma_\mu{}^\nu{}_A{}^B \Psi^{(2,\repdim{2d(d-1)})}{}^{[ab}_\nu \psi^{c]}_B \eqend{,} \\
\begin{split}
\Psi^{(3,\repdim{4d(d+1)(d-1)})}{}\ytableausmall{13,2}{}^{abc}_{\mu A} &\equiv \frac{1}{2} \Psi^{(2,\repdim{2d(d-1)})}{}^{ab}_\mu \psi^c_A - \frac{1}{2} \Psi^{(2,\repdim{2d(d-1)})}{}^{c[a}_\mu \psi^{b]}_A \\
&\quad- \frac{1}{6} \gamma_\mu{}^\nu{}_A{}^B \Psi^{(2,\repdim{2d(d-1)})}{}^{ab}_\nu \psi^c_B + \frac{1}{6} \gamma_\mu{}^\nu{}_A{}^B \Psi^{(2,\repdim{2d(d-1)})}{}^{c[a}_\nu \psi^{b]}_B \eqend{,}
\end{split} \\
\begin{split}
\Psi^{(3,\repdim{4d(d-1)(d-2)/3})}{}\ytableausmall{1,2,3}{}^{abc}_{\mu\nu A} &\equiv \frac{1}{3} \Psi^{(2,\repdim{3d(d-1)})}{}^{[ab}_{\mu\nu} \psi^{c]}_A - \frac{1}{3} \gamma^\rho{}_{[\mu A}{}^B \Psi^{(2,\repdim{3d(d-1)})}{}^{[ab}_{\nu]\rho} \psi^{c]}_B \\
&\quad- \frac{1}{6} \gamma_{\mu\nu}{}^{\rho\sigma}{}_A{}^B \Psi^{(2,\repdim{3d(d-1)})}{}^{[ab}_{\rho\sigma} \psi^{c]}_B \eqend{,}
\end{split}
\end{equations}
and the counting of degrees of freedom
\begin{splitequation}
\frac{4d (4d-1) (4d-2)}{6} &= \frac{2d(d+1)(d+2)}{3} + \frac{4d(d+1)(d-1)}{3} + \frac{4d(d+1)(d-1)}{3} \\
&\quad+ 2d(d-1)(d-2) + 4d(d+1)(d-1) + \frac{4d(d-1)(d-2)}{3} \eqend{.}
\end{splitequation}

Again we note that for $d = 1$ we recover the previous result, and that for Lie algebra representations of low dimensionality some of these invariants can vanish.

\section{Young projectors}
\label{app_youngproj}

To deal with multiterm symmetries (such as the Bianchi identities for the Riemann tensor), we use Young projectors $P$ associated to the corresponding Young tableaux, applied to the symmetric group describing permutations of the Lorentz indices.\footnote{See for example~\cite{xact7} or the very readable lecture notes~\cite{alcockzeilinger} and references therein.}

Let there be given tensor-spinors $t_{\mu A}$ and $t_{\mu\nu A} = t_{[\mu\nu]A}$ such that
\begin{equation}
\gamma^\mu{}_A{}^B t_{\mu B} = 0 = \gamma^\mu{}_A{}^B t_{\mu\nu B} \eqend{.}
\end{equation}
By multiplying with $\delta_C^A = \gamma_*{}_C{}^D \gamma_*{}_D{}^A$ and expanding $\gamma_*{}_D{}^A \gamma_{\mu\cdots}{}_A{}^B = - \tfrac{\mathi}{24} \epsilon_{\alpha\beta\gamma\delta} \gamma^{\alpha\beta\gamma\delta}{}_D{}^A \gamma_{\mu\cdots}{}_A{}^B$ in generalised $\gamma$ matrices in the following expressions, it follows that
\begin{equations}
P_\ytableausmall{\mu,\nu,\rho,\sigma} \gamma_{\mu\nu\rho A}{}^B t_{\sigma B} &= \gamma_{[\mu\nu\rho A}{}^B t_{\sigma] B} = 0 \eqend{,} \qquad P_\ytableausmall{\mu,\nu,\rho} \gamma_{\mu A}{}^B t_{\nu\rho B} = \gamma_{[\mu A}{}^B t_{\nu\rho] B} = 0 \eqend{,} \\
P_\ytableausmall{\mu,\nu,\rho,\sigma} \gamma_{\mu\nu A}{}^B t_{\rho\sigma B} &= \gamma_{[\mu\nu A}{}^B t_{\rho\sigma] B} = 0 \eqend{,} \qquad P_\ytableausmall{\alpha\rho,\beta,\mu,\nu} \gamma_{\mu\nu\rho A}{}^B t_{\alpha\beta B} = 0 \eqend{,}
\end{equations}
and obviously all projectors which antisymmetrise five or more indices vanish in four dimensions. It follows that
\begin{equations}
\begin{split}
\gamma^{\mu\nu\rho}{}_A{}^B t^\alpha_B &= P_{\ytableausmall{\mu,\nu,\rho} \otimes \ytableausmall{\alpha}} \gamma^{\mu\nu\rho}{}_A{}^B t^\alpha_B = \left( P_\ytableausmall{\mu\alpha,\nu,\rho} + P_\ytableausmall{\mu,\nu,\rho,\alpha} \right) \gamma^{\mu\nu\rho}{}_A{}^B t^\alpha_B \\
&= P_\ytableausmall{\mu\alpha,\nu,\rho} \gamma^{\mu\nu\rho}{}_A{}^B t^\alpha_B = \frac{3}{4} \gamma^{\mu\nu\rho}{}_A{}^B t^\alpha_B + \frac{3}{4} \gamma^{\alpha[\mu\nu}{}_A{}^B t^{\rho]}_B \eqend{,}
\end{split} \\
\gamma^\mu{}_A{}^B t^{\alpha\beta}_B &= P_\ytableausmall{\alpha\mu,\beta} \gamma^\mu{}_A{}^B t^{\alpha\beta}_B = \frac{2}{3} \gamma^\mu{}_A{}^B t^{\alpha\beta}_B - \frac{2}{3} \gamma^{[\alpha}{}_A{}^B t^{\beta]\mu}_B \eqend{,} \\
\begin{split}
\gamma^{\mu\nu}{}_A{}^B t^{\alpha\beta}_B &= \left( P_\ytableausmall{\alpha\mu,\beta\nu} + P_\ytableausmall{\alpha\mu,\beta,\nu} + P_\ytableausmall{\alpha\nu,\beta,\mu} \right) \gamma^{\mu\nu}{}_A{}^B t^{\alpha\beta}_B \\
&= \frac{5}{6} \gamma^{\mu\nu}{}_A{}^B t^{\alpha\beta}_B - \frac{1}{3} \gamma^{\mu[\alpha}{}_A{}^B t^{\beta]\nu}_B + \frac{1}{3} \gamma^{\nu[\alpha}{}_A{}^B t^{\beta]\mu}_B - \frac{1}{6} \gamma^{\alpha\beta}{}_A{}^B t^{\mu\nu}_B \eqend{,}
\end{split} \\
\begin{split}
\gamma^{\mu\nu\rho}{}_A{}^B t^{\alpha\beta}_B &= \left( P_\ytableausmall{\alpha\mu,\beta\nu,\rho} + P_\ytableausmall{\alpha\mu,\beta\rho,\nu} + P_\ytableausmall{\alpha\nu,\beta\rho,\mu} \right) \gamma^{\mu\nu\rho}{}_A{}^B t^{\alpha\beta}_B \\
&= \frac{1}{2} \gamma^{\mu\nu\rho}{}_A{}^B t^{\alpha\beta}_B + \frac{1}{2} \gamma^{\alpha\beta[\mu}{}_A{}^B t^{\nu\rho]}_B + \frac{1}{2} \gamma^{\alpha[\mu\nu}{}_A{}^B t^{\rho]\beta}_B - \frac{1}{2} \gamma^{\beta[\mu\nu}{}_A{}^B t^{\rho]\alpha}_B \eqend{,}
\end{split} \\
\begin{split}
\gamma^{\mu\nu\rho\sigma}{}_A{}^B t^{\alpha\beta}_B &= \left( P_\ytableausmall{\alpha\mu,\beta\nu,\rho,\sigma} + P_\ytableausmall{\alpha\mu,\beta\rho,\nu,\sigma} + P_\ytableausmall{\alpha\nu,\beta\rho,\mu,\sigma} + P_\ytableausmall{\alpha\mu,\beta\sigma,\nu,\rho} + P_\ytableausmall{\alpha\nu,\beta\sigma,\mu,\rho} + P_\ytableausmall{\alpha\rho,\beta\sigma,\mu,\nu} \right) \gamma^{\mu\nu\rho\sigma}{}_A{}^B t^{\alpha\beta}_B \\
&= \frac{3}{5} \gamma^{\mu\nu\rho\sigma}{}_A{}^B t^{\alpha\beta}_B - \frac{3}{5} \gamma^{\alpha[\mu\nu\rho}{}_A{}^B t^{\sigma]\beta}_B + \frac{3}{5} \gamma^{\beta[\mu\nu\rho}{}_A{}^B t^{\sigma]\alpha}_B + \frac{3}{5} \gamma^{\alpha\beta[\mu\nu}{}_A{}^B t^{\rho\sigma]}_B \eqend{.}
\end{split}
\end{equations}

\fix\ applies these identities to the invariant tensors
\begin{splitequation}
\Psi^{(3,\repdim{2d(d+1)(d+2)})}{}\ytableausmall{123}{}^{abc}_{\nu A} \eqend{,} \quad \Psi^{(3,\repdim{4d(d+1)(d-1)})}{}\ytableausmall{12,3}{}^{abc}_{\nu A} \eqend{,} \quad \Psi^{(3,\repdim{4d(d+1)(d+2)/3})}{}\ytableausmall{123}{}^{abc}_{\mu\nu A}
\end{splitequation}
for commuting spinors and
\begin{splitequation}
\Psi^{(3,\repdim{2d(d-1)(d-2)})}{}\ytableausmall{1,2,3}{}^{abc}_{\nu A} \eqend{,} \quad \Psi^{(3,\repdim{4d(d+1)(d-1)})}{}\ytableausmall{13,2}{}^{abc}_{\nu A} \eqend{,} \quad \Psi^{(3,\repdim{4d(d-1)(d-2)/3})}{}\ytableausmall{1,2,3}{}^{abc}_{\mu\nu A}
\end{splitequation}
for anticommuting spinors.

In the same way, the products of the totally antisymmetric $\epsilon$ tensor and generalised $\gamma$ matrices can be replaced by simpler expressions, by multiplying $\gamma^{\mu \cdots}{}_A{}^B$ with $\delta_C^A = \gamma_*{}_C{}^D \gamma_*{}_D{}^A$ and expanding $\gamma_*{}_D{}^A \gamma_{\mu\cdots}{}_A{}^B = - \tfrac{\mathi}{24} \epsilon_{\alpha\beta\gamma\delta} \gamma^{\alpha\beta\gamma\delta}{}_D{}^A \gamma_{\mu\cdots}{}_A{}^B$ in generalised $\gamma$ matrices, or alternatively (in odd dimensions) using that the antisymmetrisation over more than $d$ indices vanishes in $d$ dimensions. This results in the identities
\begin{equation}
\epsilon_{\mu_1 \cdots \mu_d} \gamma^{\nu_1 \cdots \nu_k} = (-1)^{k (d+1)} \frac{d!}{k! (d-k)!} \delta^{\nu_1}_{[\mu_1} \cdots \delta^{\nu_k}_{\mu_k} \epsilon_{\mu_{k+1} \cdots \mu_d] \rho_{d-k+1} \cdots \rho_d} \gamma^{\rho_{d-k+1} \cdots \rho_d} \eqend{,}
\end{equation}
where now all indices of the generalised $\gamma$ matrices are contracted with the $\epsilon$ tensor. In particular in $d = 4$ dimensions, we have
\begin{equations}
\epsilon_{\mu\nu\rho\sigma} \gamma^{\alpha\beta\gamma\delta} &= \delta^\alpha_{[\mu} \delta^\beta_\nu \delta^\gamma_\rho \delta^\delta_{\sigma]} \epsilon_{\kappa\lambda\upsilon\tau} \gamma^{\kappa\lambda\upsilon\tau} \eqend{,} \\
\epsilon_{\mu\nu\rho\sigma} \gamma^{\alpha\beta\gamma} &= - 4 \delta^\alpha_{[\mu} \delta^\beta_\nu \delta^\gamma_\rho \epsilon_{\sigma]\kappa\lambda\upsilon} \gamma^{\kappa\lambda\upsilon} \eqend{,} \\
\epsilon_{\mu\nu\rho\sigma} \gamma^{\alpha\beta} &= 6 \delta^\alpha_{[\mu} \delta^\beta_\nu \epsilon_{\rho\sigma]\kappa\lambda} \gamma^{\kappa\lambda} \eqend{,} \\
\epsilon_{\mu\nu\rho\sigma} \gamma^\alpha &= - 4 \delta^\alpha_{[\mu} \epsilon_{\nu\rho\sigma]\kappa} \gamma^\kappa \eqend{.}
\end{equations}

Lastly, also the product of the totally antisymmetric $\epsilon$ tensor and an arbitrary number of other tensors can be projected onto the corresponding Young tableaux. \fix\ implements the following projections:
\begin{equations}
\epsilon_{\mu\nu\rho\sigma} t_\alpha &= P_\ytableausmall{\mu\alpha,\nu,\rho,\sigma} \epsilon_{\mu\nu\rho\sigma} t_\alpha = \frac{4}{5} \epsilon_{\mu\nu\rho\sigma} t_\alpha - \frac{4}{5} \epsilon_{\alpha[\mu\nu\rho} t_{\sigma]} \eqend{,} \\
\begin{split}
\epsilon_{\mu\nu\rho\sigma} t_{\alpha\beta} &= \left( P_\ytableausmall{\mu\alpha\beta,\nu,\rho,\sigma} + P_\ytableausmall{\mu\alpha,\nu\beta,\rho,\sigma} \right) \epsilon_{\mu\nu\rho\sigma} t_{\alpha\beta} = \frac{3}{5} \epsilon_{\alpha\beta[\mu\nu} t_{\rho\sigma]} + \frac{1}{30} \epsilon_{\mu\nu\rho\sigma} t_{\beta\alpha} - \frac{1}{30} \epsilon_{\beta[\mu\nu\rho} t_{\sigma]\alpha} \\
&\qquad- \frac{1}{30} \epsilon_{\alpha[\mu\nu\rho} t_{|\beta|\sigma]} + \frac{19}{30} \epsilon_{\mu\nu\rho\sigma} t_{\alpha\beta} - \frac{19}{30} \epsilon_{\alpha[\mu\nu\rho} t_{\sigma]\beta} - \frac{19}{30} \epsilon_{\beta[\mu\nu\rho} t_{|\alpha|\sigma]} \eqend{.}
\end{split}
\end{equations}

\addcontentsline{toc}{section}{References}
\bibliography{literature}

\end{document}